\documentclass[preprint, notitlepage, nofootinbib, amsmath,amssymb,aps,floatfix]{revtex4-2}

\usepackage{graphicx}
\usepackage{dcolumn}
\usepackage{bm}
\usepackage[hidelinks]{hyperref}

\usepackage{xcolor}
\usepackage{cleveref}
\usepackage{footmisc}

\newcommand{\be}{\begin{equation}}
\newcommand{\ee}{\end{equation}}
\newcommand{\bea}{\begin{eqnarray}}
\newcommand{\eea}{\end{eqnarray}}

\newcommand{\eq}[2]{\be\begin{aligned}#1 \label{#2}\end{aligned}\ee}
\newcommand{\OO}{\mathcal{O}}
\newcommand{\Fig}[1]{Fig.~\ref{#1}}

\newcommand{\Eq}[1]{Eq.~(\ref{#1})}
\newcommand{\Eqs}[2]{Eqs.~(\ref{#1}) and (\ref{#2})}
\newcommand{\Sec}[1]{Sec.~\ref{#1}}

\newcommand{\mL}{m_L}
\newcommand{\mH}{m_H}
\newcommand{\eH}{\eta_H}
\newcommand{\eL}{\eta_L}

\newcommand{\vL}{u_L}
\newcommand{\vH}{u_H}

\newcommand{\rs}{r_S}

\newcommand{\dotbarxL}{\dot{\bar{x}}_L}

\newcommand{\ddotbarxH}{\ddot{\bar{x}}_H}

\newcommand{\barxL}{\bar{x}_L}
\newcommand{\barxH}{\bar{x}_H}

\newcommand{\dx}{\delta x}
\newcommand{\dA}{\delta A}
\newcommand{\dF}{\delta F}

\newcommand{\dg}{\delta g}
\newcommand{\dG}{\delta \Gamma}
\newcommand{\dE}{\delta E}

\newcommand{\del}[1]{\hat{\delta}(#1)}

\usepackage[compat=1.1.0]{tikz-feynman}
\usepackage{tikz}
\usetikzlibrary{shapes.geometric}

\tikzset{
bgraviton/.style={decorate, decoration={snake, amplitude=0.5mm, segment length=2mm, pre length=.0mm, post length=.2mm}, double, thick},
source/.style={decorate, thick, blob, scale = 0.37},
hsource/.style={decorate, thick, blob, scale = 0.5},
graviton/.style={decorate, decoration={complete sines, amplitude=1mm, segment length=2mm, pre length=0mm, post length=.2mm}, thick},
sourceone/.style={decorate, thick, dot, fill=white, scale=1.5, inner sep = 1pt},
bscalar/.style={decorate, double, thick},
myphoton/.style={decorate, thick},
myscalar/.style={decorate, dashed, thick},
binsertion/.style={decorate, thick, crossed dot},
gvert/.style={
},
vert/.style={decorate, dot, scale=0.75},
recoil/.style={decorate, thick, empty dot, scale=1.5, inner sep = 1pt}
}

\newcommand{\flatprop}{
\begin{tikzpicture}
        \begin{feynman}
          \vertex (i) at (-1, 0);
          \vertex (f) at (1, 0);
          \diagram*{
            (i) -- [graviton]  (f) 
          };
        \end{feynman}
    \end{tikzpicture}
}

\newcommand{\flatscalar}{
\begin{tikzpicture}
        \begin{feynman}
          \vertex (i) at (-1, 0);
          \vertex (f) at (1, 0);
          \diagram*{
            (i) -- [myphoton]  (f) 
          };
        \end{feynman}
    \end{tikzpicture}
}

\newcommand{\photonsource}{
\begin{tikzpicture}
        \begin{feynman}
          \vertex (f) at (1, 0);
          \vertex[sourceone] (c) at (0,0) {\tiny{\(L\)}};
          \diagram*{
            (c) -- [myphoton] (f) 
          };
        \end{feynman}
    \end{tikzpicture}
}

\newcommand{\gravitonsource}{
\begin{tikzpicture}
        \begin{feynman}
          \vertex (f) at (1, 0);
          \vertex[sourceone] (c) at (0,0) {\tiny{\(L\)}};
          \diagram*{
            (c) -- [graviton] (f) 
          };
        \end{feynman}
    \end{tikzpicture}
}

\newcommand{\recoilvertGG}{
\begin{tikzpicture}
        \begin{feynman}
          \vertex (i) at (-1, 0);
          \vertex (f) at (1, 0);
          \vertex[recoil] (c) at (0,0) {\tiny{\(H\)}};
          \diagram*{
            (c) -- [graviton] (i),
            (c) -- [graviton] (f) 
          };
        \end{feynman}
    \end{tikzpicture}
}

\newcommand{\recoilvertAG}{
\begin{tikzpicture}
        \begin{feynman}
          \vertex (i) at (-1, 0);
          \vertex (f) at (1, 0);
          \vertex[recoil] (c) at (0,0) {\tiny{\(H\)}};
          \diagram*{
            (c) -- [myphoton] (i),
            (c) -- [graviton] (f) 
          };
        \end{feynman}
    \end{tikzpicture}
}

\newcommand{\recoilvertAA}{
\begin{tikzpicture}
        \begin{feynman}
          \vertex (i) at (-1, 0);
          \vertex (f) at (1, 0);
          \vertex[recoil] (c) at (0,0) {\tiny{\(H\)}};
          \diagram*{
            (c) -- [myphoton] (i),
            (c) -- [myphoton] (f) 
          };
        \end{feynman}
    \end{tikzpicture}
}

\newcommand{\backgroundvertAAone}{
\begin{tikzpicture}
        \begin{feynman}
          \vertex (i) at (-1, 0);
          \vertex (f) at (1, 0);
          \vertex[recoil] (c) at (0,0) {\tiny{\(1\)}};
          \diagram*{
            (c) -- [myphoton] (i),
            (c) -- [myphoton] (f) 
          };
        \end{feynman}
    \end{tikzpicture}
}

\newcommand{\backgroundvertAGone}{
\begin{tikzpicture}
        \begin{feynman}
          \vertex (i) at (-1, 0);
          \vertex (f) at (1, 0);
          \vertex[recoil] (c) at (0,0) {\tiny{\(1\)}};
          \diagram*{
            (c) -- [myphoton] (i),
            (c) -- [graviton] (f) 
          };
        \end{feynman}
    \end{tikzpicture}
}

\newcommand{\backgroundvertGGone}{
\begin{tikzpicture}
        \begin{feynman}
          \vertex (i) at (-1, 0);
          \vertex (f) at (1, 0);
          \vertex[recoil] (c) at (0,0) {\tiny{\(1\)}};
          \diagram*{
            (c) -- [graviton] (i),
            (c) -- [graviton] (f) 
          };
        \end{feynman}
    \end{tikzpicture}
}

\newcommand{\backgroundvertAAtwo}{
\begin{tikzpicture}
        \begin{feynman}
          \vertex (i) at (-1, 0);
          \vertex (f) at (1, 0);
          \vertex[recoil] (c) at (0,0) {\tiny{\(2\)}};
          \diagram*{
            (c) -- [myphoton] (i),
            (c) -- [myphoton] (f) 
          };
        \end{feynman}
    \end{tikzpicture}
}

\newcommand{\backgroundvertAGtwo}{
\begin{tikzpicture}
        \begin{feynman}
          \vertex (i) at (-1, 0);
          \vertex (f) at (1, 0);
          \vertex[recoil] (c) at (0,0) {\tiny{\(2\)}};
          \diagram*{
            (c) -- [myphoton] (i),
            (c) -- [graviton] (f) 
          };
        \end{feynman}
    \end{tikzpicture}
}

\newcommand{\backgroundvertGGtwo}{
\begin{tikzpicture}
        \begin{feynman}
          \vertex (i) at (-1, 0);
          \vertex (f) at (1, 0);
          \vertex[recoil] (c) at (0,0) {\tiny{\(2\)}};
          \diagram*{
            (c) -- [graviton] (i),
            (c) -- [graviton] (f) 
          };
        \end{feynman}
    \end{tikzpicture}
}

\newcommand{\HtwoPM}{
\begin{tikzpicture}
        \begin{feynman}
          \vertex[recoil] (c) at (0,1.0) {\tiny{\(H\)}};
          \vertex[sourceone] (i) at (-1, 0 ) {\tiny{\(L_{0}\)}};
          \vertex[sourceone] (f) at (1, 0) {\tiny{\(L_{0}\)}};
          \diagram*{
            (c) -- [myscalar, bend right=30] (i), (c) --  [myscalar, bend left=30] (f),
          };
        \end{feynman}
    \end{tikzpicture}
    }

    \newcommand{\VtwoPM}{
\begin{tikzpicture}
        \begin{feynman}
          \vertex[recoil] (c) at (0,1.0) {\scriptsize{\(1\)}};
          \vertex[sourceone] (i) at (-1, 0 ) {\tiny{\(L_{0}\)}};
          \vertex[sourceone] (f) at (1, 0) {\tiny{\(L_{0}\)}};
          \diagram*{
            (c) -- [myscalar, bend right=30] (i), (c) --  [myscalar, bend left=30] (f),
          };
        \end{feynman}
    \end{tikzpicture}
    }

\newcommand{\LzeroHLone}{
\begin{tikzpicture}
        \begin{feynman}
          \vertex[recoil] (c) at (0,1.0) {\tiny{\(H\)}};
          \vertex[sourceone] (i) at (-1, 0 ) {\tiny{\(L_{0}\)}};
          \vertex[sourceone] (f) at (1, 0) {\tiny{\(L_{1}\)}};
          \diagram*{
            (c) -- [myscalar, bend right=30] (i), (c) --  [myscalar, bend left=30] (f),
          };
        \end{feynman}
    \end{tikzpicture}
    }

    \newcommand{\LzeroVLone}{
\begin{tikzpicture}
        \begin{feynman}
          \vertex[recoil] (c) at (0,1.0) {\scriptsize{\(1\)}};
          \vertex[sourceone] (i) at (-1, 0 ) {\tiny{\(L_{0}\)}};
          \vertex[sourceone] (f) at (1, 0) {\tiny{\(L_{1}\)}};
          \diagram*{
            (c) -- [myscalar, bend right=30] (i), (c) --  [myscalar, bend left=30] (f),
          };
        \end{feynman}
    \end{tikzpicture}
    }

        \newcommand{\HVthreePM}{
\begin{tikzpicture}
        \begin{feynman}
          \vertex[recoil] (c) at (-0.5,0.87) {\tiny{\(H\)}};
          \vertex[recoil] (d) at (0.5,0.87) {\scriptsize{\(1\)}};
          \vertex[sourceone] (i) at (-1, 0 ) {\tiny{\(L_{0}\)}};
          \vertex[sourceone] (f) at (1, 0) {\tiny{\(L_{0}\)}};
          \diagram*{
            (c) -- [myscalar, bend right=30] (i), (d) --  [myscalar, bend left=30] (f), (c) --  [myscalar] (d),
          };
        \end{feynman}
    \end{tikzpicture}
    }

            \newcommand{\VVthreePM}{
\begin{tikzpicture}
        \begin{feynman}
          \vertex[recoil] (c) at (-0.5,0.87) {\scriptsize{\(1\)}};
          \vertex[recoil] (d) at (0.5,0.87) {\scriptsize{\(1\)}};
          \vertex[sourceone] (i) at (-1, 0 ) {\tiny{\(L_{0}\)}};
          \vertex[sourceone] (f) at (1, 0) {\tiny{\(L_{0}\)}};
          \diagram*{
            (c) -- [myscalar, bend right=30] (i), (d) --  [myscalar, bend left=30] (f), (c) --  [myscalar] (d),
          };
        \end{feynman}
    \end{tikzpicture}
    }

                \newcommand{\VtwothreePM}{
\begin{tikzpicture}
        \begin{feynman}
          \vertex[recoil] (c) at (0,1.0) {\scriptsize{\(2\)}};
          \vertex[sourceone] (i) at (-1, 0 ) {\tiny{\(L_{0}\)}};
          \vertex[sourceone] (f) at (1, 0) {\tiny{\(L_{0}\)}};
          \diagram*{
            (c) -- [myscalar, bend right=30] (i), (c) --  [myscalar, bend left=30] (f),
          };
        \end{feynman}
    \end{tikzpicture}
    }

\begin{document}
\count\footins = 1000

\preprint{CALT-TH 2023-041}

\title{Conservative Scattering of Reissner-Nordstr\"{o}m Black Holes at Third Post-Minkowskian Order}

\author{Jordan Wilson-Gerow}
\email{wilsonjs@caltech.edu}
\affiliation{%
 Walter Burke Institute for Theoretical Physics\\
 California Institute of Technology, Pasadena, California 91125
}


\begin{abstract}
Using a recently developed effective field theory formalism for extreme mass ratios [\href{https://arxiv.org/abs/2308.14832}{2308.14832}], we present a calculation of charged black hole scattering at third post-Minkowskian order. The charges and masses are kept arbitrary, and the result interpolates from the scattering of Schwarzschild to extremal charged black holes, and beyond to charged particles in electrodynamics---agreeing with previously reported results in all such limits. The computation of the radial action is neatly organized in powers of the mass ratio. The probe (0SF) contributions are readily computed by direct integration of the radial momentum, and we use the effective field theory to compute the subleading (1SF) contributions via background-field Feynman rules supplemented by an operator encoding recoil of the background. Together these contributions completely determine the conservative physics at order~$\mathcal{O}(G^{3})$.

\end{abstract}

\maketitle
\newpage

\tableofcontents

\section{Introduction}

Gravitational wave science has proven to be an incredibly successful endeavour, as epitomized by the detections of the LIGO/Virgo collaboration~\cite{LIGO} and of NANOGrav~\cite{NANOGrav}. Precision measurements have necessitated precision theoretical calculations,  leading to a variety of powerful and complementary approaches to the computation of gravitational waveforms from orbiting binary systems. While numerical relativity~\cite{Pretorius:2005gq, Lehner_2014,Cardoso_2015} and gravitational self-force~\cite{PoissonReview, PoundReview, BarackReview} successfully describe strong-field dynamics, a practical route to covering the parameter space requires complementary weak-field methods---ideally re-summed via e.g., effective one-body theory~\cite{Buonanno:1998gg}. Such weak-field approximations include the post-Newtonian (PN)~\cite{Blanchet2014, NRGR} and the post-Minkowskian (PM) expansions~\cite{Bertotti1960, Westpfahl1979, Westpfahl1985, Damour:2016gwp, Damour2018-pmeob2, Cheung2018-2PM}. The different approaches have overlapped, cross-fertilized, and complimented each other well, e.g., \cite{Barack:2010ny, Nagar:2022fep, Detweiler:2008ft, Barack:2011ed, Bini:2013zaa, 
Damour:2014afa,
vandeMeent:2016hel, 
Antonelli:2019ytb,
Bini:2019nra, 
Bini:2020wpo, Bini:2020rzn, Gralla:2021qaf, Long:2021ufh, 
Khalil:2022ylj,
Barack:2022pde, Barack:2023oqp, Whittall:2023xjp,  Galley_2009}.

In recent years the PM approach has seen rapid development due to the import of tools from the modern scattering amplitudes program (e.g. \cite{ElvangHuang, Dixon:2013uaa, Cheung:2017pzi,  Chetyrkin:1981qh, Henn:2013pwa, Kosower:2018adc}). These ideas, which include but are not limited to: effective field theory~\cite{NRGR}, on-shell constructions of loop integrands~\cite{Bern:1994zx, Bern:1994cg}, and dimensional regularization with its accompanying Feynman integral calculus~\cite{tHooft:1972tcz,Bollini:1972ui, Smirnov:2012gma}, all originate in quantum theory yet have proven to be highly efficient for classical computations. Indeed in just a few years the state of the art for scattering dynamics has been pushed from $\mathcal{O}(G^{2})$ (2PM) through to 4PM~(see e.g.~\cite{Smatrix, Bjerrum-Bohr:2018xdl, Cheung2018-2PM, Bern:2019nnu, Bern:2019crd, Bjerrum-Bohr:2019kec, Kalin:2019rwq, Kalin:2019inp, Kalin:2020fhe, Cheung:2020gyp, Bjerrum-Bohr:2021din, Bern:2021dqo, Bern:2021yeh, Herrmann:2021tct, DiVecchia:2021bdo, Herrmann:2021lqe, DiVecchia:2021ndb, Jakobsen:2021smu, Mougiakakos:2021ckm,  Brandhuber:2021eyq, Manohar:2022dea, Bjerrum-Bohr:2022blt, Kalin:2020mvi, Dlapa:2021npj, Dlapa:2021vgp, Jakobsen:2022fcj, Jakobsen:2023ndj,  Dlapa:2022lmu,  dlapa2023-letter, Mogull:2020sak, jakobsen2023dissipative, DiVecchia:2023frv, Damgaard:2023ttc}). The relationship between classical gravitation and scattering theory has even been symbiotic, with exact solutions in General Relativity being utilized via the background field method (and its systematic corrections) to efficiently compute scattering amplitudes~\cite{Mason_2009, Cheung_2021, bautista2023scattering-1, bautista2023scattering-2, Adamo_2023, Adamo:2023cfp, cheung2023effective, long_paper, kosmopoulos2023gravitational}.

These theoretical developments have primarily been in service of a practical experimentally focused goal, however it is interesting to pause and ask whether these general tools can be utilized for other classical systems. The answer is the affirmative; with these new techniques one can more deeply study the two-body dynamics in a variety of theories of both phenomenological and formal interest, eg. the Abelian Higgs Model~\cite{jones2023classical}, electrodynamics~\cite{Bern_2022, bern2023conservative}, and toy self-force models~\cite{long_paper, Barack:2023oqp}. In this paper we will continue this effort and use these modern techniques to study the dynamics of electrically charged black holes in  Einstein-Maxwell theory.

While the electric charge of a black hole is expected to be astrophysically irrelevant~\cite{gibbons1975vacuum, 10.1093/mnras/179.3.433}, charged solutions are of widespread formal interest. Black hole solutions carrying a variety of types of charge appear ubiquitously in supergravity and Kaluza-Klein theories, and consequently in low-energy string models. Extremal charged black holes in particular have played a central role in our understanding of black hole microstate counting and the weak gravity conjecture~(e.g., \cite{Strominger_1996,Arkani-Hamed_2007,Cheung_2018}). These remarkable objects have surprising properties, including invariance under supersymmetry and the related existence of exact multi-black-hole solutions~\cite{GIBBONS1982190, PhysRevD.46.5278}. 

This is true even for the familiar Reissner-Nordstr\"{o}m black hole. One can understand Einstein-Maxwell theory as the bosonic sector of $\mathcal{N}=2$ supergravity, and the Reissner-Nordstr\"{o}m black hole continues to exist as a solution which even preserves supersymmetry in the extremal limit. Consequently, like-charged extremal black holes can be arranged together as in the Majumdar-Papapetrou multi-centered solutions~\cite{papapetrou_1945, PhysRev.72.390, hartle1972solutions}. It is interesting to ask whether or not this implies any special properties for the binary dynamics of Reissner-Nordstr\"{o}m black holes such that they may serve as a toy model for purely gravitational black holes (as do black holes in maximal supergravity~\cite{Caron-Huot:2018ape}). 

For obvious phenomenological reasons the binary dynamics of charged black holes have been studied much less than Schwarzschild or Kerr, however there has been considerable progress from multiple directions. For general values of the charge, leading radiation effects on Newtonian orbits and relativistic test-bodies have been studied~\cite{PhysRevD.102.103520, PhysRevD.107.023005}, as well as their accompanying waveforms~\cite{Benavides_Gallego_2023}. Both conservative and dissipative dynamics, as well as quantum corrections, have been computed at 1PN order~\cite{Juli__2018, Khalil_2018, Bjerrum-Bohr:2002aqa, Butt:2006gv, Faller:2007sy, Holstein:2008sy}, with conservative dynamics computed further to 2PN order~\cite{gupta2022binary}. Numerical simulations have been performed, initially for head-on collisions~\cite{PhysRevD.85.124062}, and recently for more generic configurations~\cite{PhysRevD.99.104044, PhysRevD.104.044004}.  Modern on-shell amplitudes methods have been used to compute the fully relativistic scattering impulse and radiated momentum of extremal black holes at the 2PM and 3PM orders~\cite{Caron-Huot:2018ape, Parra-Martinez:2020dzs, DiVecchia:2020ymx, Herrmann:2021tct, DiVecchia:2021bdo, Bjerrum-Bohr:2021vuf} in the context of $\mathcal{N}=8$ supergravity. Amplitudes techniques have also been applied in Einstein-Maxwell theory to compute the two-body and three-body potentials~\cite{Cheung_2021, Jones_2023} at 2PM order. 

For supersymmetric (BPS) extremal black holes one can write down exact solutions to the field equations. Beyond the static multi-centered solutions there are also stationary but non-static solutions corresponding to bound electric and magnetic black holes~\cite{Denef_2000, denef2000correspondence, Bates_2011}. These solutions carry angular momentum, but this is more akin to the intrinsic angular momentum carried by a resting electron-monopole system in Maxwell theory than it is to a orbiting binary system. The static multi-center solutions are also built upon by the \textit{moduli space approximation} (MSA) which analytically describes their leading post-static, $\OO(v^{2}G^{\infty})$, dynamics~\cite{Manton:1981mp, PhysRevLett.57.1492, Ferrell:1987gf, Camps:2017gxz}.

In this paper we extend the efforts in Einstein-Maxwell theory and study the conservative scattering of black holes with arbitrary charge and mass, at 3PM order. Specifically we compute the radial action, from which one can straightforwardly determine the scattering angle, periapsis precession, or effective Hamiltonian~\cite{Bjerrum-Bohr:2019kec, Kalin:2019inp, Kalin:2019rwq}. We verify consistency of our result in different limits with known PM results in electromagnetism, GR, and self-force toy models, and also against a computation for slowly moving extremal black holes using the moduli space approximation. We then focus on the extremal limit and combine our novel PM results with results from both the probe limit and MSA to determine all but one of the a priori 33 independent coefficients in the conservative scattering angle through \textit{fifth} post-Newtonian order. 

The calculation is done using a recently developed worldline EFT which perturbatively expands about an exact curved spacetime solution and its exact geodesics, rather than expanding about the free particle solution in flat spacetime~\cite{cheung2023effective}. In this approach one expands in powers of the mass ratio, as in the self-force program. The test-mass (aka probe or 0SF) limit is understood by elementary mechanics, and corrections in powers of the mass-ratio are systematically computed in the EFT using curved spacetime Feynman rules that contain information to all orders $G$.\footnote{A more detailed discussion will be given in \cite{long_paper}. See also~\cite{kosmopoulos2023gravitational} for an analogous field theory formalism.} Further expanding these curved spacetime Feynman diagrams in powers of $G$ efficiently produces PM \textit{loop-integrands}. Using the known exact solutions effectively performs partial resummation and tensor reduction on classes of sub-diagrams at arbitrary loop order. The two-loop Feynman integrals which arise in the present computation are defined in dimensional regularization and evaluated using integration-by-parts reduction and canonical differential equations~\cite{Chetyrkin:1981qh,Henn:2013pwa}. The purely conservative physics is extracted at this loop order by restricting the loop momenta to the potential region.

The paper is organized as follows. In \cref{sec:setup} we describe the point particle effective field theory. We review how the radial action depends on the particles' masses, and elaborate on how it can be efficiently computed using a mass ratio expansion---even for charged black holes. We then take the probe limit and then use elementary mechanics to compute the probe radial action in a PM expansion. In \cref{sec:1SF} we discuss the EFT for extreme mass ratios and its application to Einstein-Maxwell theory. We then present the main result of the paper, the complete radial action at 3PM order. Section \ref{sec:extremal} focuses on the extremal limit and its prospects for high order post-Newtonian computations. Our Feynman rules are given in \cref{app:feynmanrules}.

Throughout we use mostly-minus signature. While certain expressions in the paper are written explicitly in four spacetime dimensions, the loop-integration is performed in a general dimension $D=4-2\epsilon$ for the purposes of dimensional regularization. There are no ambiguities with $\epsilon/\epsilon$ cancellations at the order to which we compute. We use the shorthand notation, $\int_{\ell}=\int\frac{d^{D}\ell}{(2\pi)^{D}}$ and $\del{x}=2\pi \delta(x)$ respectively.  We use Gaussian units with $c=1$, ie. $\mu_{0}=(\varepsilon_{0})^{-1}=4\pi$. Planck's constant $\hbar$ is also set to unity in our Feynman rules, however our final results are all classical and independent of $\hbar$.

\medskip

\section{Set-up and Background}\label{sec:setup}
\medskip
We will start from a worldline action for a pair of electrically charged, massive, spinless particles, coupled to Einstein-Maxwell theory,
\eq{S=-\frac{1}{16\pi G}\int d^{4}x\sqrt{-g}\left(R+F_{\mu\nu}F^{\mu\nu}\right)-\sum_{j=L,H}m_{j}\int d\tau\left(\frac{1}{2}\dot{x}_{j}^{\mu}\dot{x}_{j}^{\nu}g_{\mu\nu}(x_{j})+\eta_{j}\dot{x}_{j}^{\mu}A_{\mu}(x_{j})\right) \, ,}
{eq:action}
where we've gauge fixed the einbein such that $\dot{x}_{j}^{2}=1$. The charges are measured  by the dimensionless charge-to-mass ratio
\eq{\eta_{j}\equiv \frac{q_{j}}{G^{1/2}m_{i}}\,,}
{}
with the charges $q_{j}$ in Gaussian units. With this choice, and the non-canonical normalization of the photon field, an extremal charged object has $\eta_{j}=1$. We will hereafter refer to $\eta_{j}$ as the charge of particle $j$. This point particle effective theory is valid for separations larger than the curvature scales, $\frac{Gm_{j}}{r}\ll 1$, and we will use it to study the dynamics of these bodies up to third order in these ratios, \textit{ie.} third Post-Minkowskian order. In this work we'll focus on just the conservative scattering dynamics, which are neatly encoded in the radial on-shell action.

\subsection{Mass Ratio Expansion of the Radial Action}\label{sec:massratioexpansion}
We'll consider the scattering of particles with asymptotic four-velocities $\vH^{\mu},\vL^{\mu}$ and impact parameter $b$. To clarify our approach to the problem, we'll first discuss pure gravity.

In gravity, Lorentz invariance and dimensional analysis dictate that the PM expanded radial action for scattering has the form
\eq{
i_r=Gm_{H}m_{L}\sum_{m=1}\left(\frac{G}{b}\right)^{m-1}\tilde{I}_{m}(\mL,\mH,\sigma)\,,
}{eq:PMradialaction}
where the $m$PM contribution, $\tilde{I}_{m}$, is a symmetric homogeneous polynomial of degree $m-1$ in the masses, with coefficients which are functions only of the relative boost factor, $\sigma=\vH^{\mu}\vL{}_{\mu}$. Alternatively one could reshuffle the contributions in the radial action into an expansion of the form
\eq{
i_r=Gm_{H}m_{L}\sum_{n=0}\lambda^{n}I_{n}(G\mH,b,\sigma)\,,
}{eq:radialActionSFexpansion}
where we've defined the mass ratio
\eq{
\lambda=\frac{\mL}{\mH}\,,
}
{}
and each $I_{n}$ contains all orders in $G$. We refer to this expansion as an expansion in SF orders (ie. ``self-force''). At each SF order one can then PM expand $I_{n}$ as a polynomial in just the heavy mass such that the total radial action has the form
\eq{
i_r=Gm_{H}m_{L}\sum_{n=0}\sum_{m=1}\lambda^{n}\left(\frac{G\mH}{b}\right)^{m-1}I_{n,m}(\sigma)\,,
}{eq:SFradialaction}
where $n$ denotes the SF order and $m$ denotes the PM order. As we'll discuss in later sections, the 0SF radial action is elementary to compute and a recently developed effective field theory for extreme mass ratios can be used to efficiently compute higher $n$SF corrections. 

The total radial action is invariant under relabelling the masses $H\leftrightarrow L$, hence the symmetric polynomials  $\tilde{I}_{m}(\mL,\mH,\sigma)$. Despite the choice to treat $H$ as special when we write the expansion as a mass ratio expansion (\ref{eq:SFradialaction}), this symmetry still applies and now manifests as simple relationships such as
\eq{
I_{0,2}&=I_{1,2}, \qquad I_{0,3}&=I_{2,3},\qquad I_{0,4}&=I_{3,4}, \qquad I_{1,4}&=I_{2,4}\,,
}{}
etc. This relabelling symmetry ensures that at both and $1$PM and $2$PM order the radial action has only one independent function of $\sigma$, and it is determined entirely by probe motion. At both 3PM and 4PM one must supplement the probe result with one additional function of velocity, the 1SF contribution.

To spell this out explicitly we'll use 3PM as an example.\footnote{See \cite{Damour:2019lcq} for a general treatment.} By the $H,L$ relabelling symmetry the radial action has the form
\eq{
i_r&=Gm_{H}m_{L}\bigg(\log(b)I_{0,1}+\frac{G}{b}(\mH +\mL)I_{0,2}+\frac{G^{2}}{b^{2}}\left((\mH^{2}+\mL^{2}) I_{0,3}+\mH\mL I_{1,3}\right)\bigg)\\
&+\OO(G^4)\, ,
}{}
manifesting the fact that only 0SF and 1SF physics is needed at this PM order. The strategy of separately computing 0SF and 1SF contributions to the radial action has been recently applied to the computation of the total 3PM radial action for pure GR as well as toy self-force models~\cite{cheung2023effective}, yielding complete agreement with existing results in the literature where results overlapped. The strategy is referred to as the  ``Tutti-Frutti'' method in the post-Newtonian expansion~\cite{Bini:2019nra, Bini:2020rzn}.

In the case of Einstein-Maxwell theory, one can repeat the logic above and utilize the same organization of the mass ratios, however the functions $I_{n,m}(\sigma)$ are now generalized to (not necessarily symmetric) polynomials of degree $(m,m)$ in the charges $(\eta_{L},\eta_{H})$. We'll call these $I_{n,m}^{H,L}(\sigma)$.  The general form of the 3PM radial action in Einstein-Maxwell theory is then 
\eq{
\frac{i_r}{G\mH\mL}=&\,\,\log(b)I_{0,1}^{H,L}+\frac{G}{b}\big(\mH I_{0,2}^{H,L}+\mL I_{1,2}^{H,L}\big)+\frac{G^{2}}{b^{2}}(\mH^{2} I_{0,3}^{H,L}+\mH\mL I_{1,3}^{H,L}+\mL^{2} I_{2,3}^{H,L})\,. 
}{}
The $H,L$ relabelling symmetry of the total radial action now implies $I_{1,2}^{H,L}=I_{0,2}^{L,H}$ and $I_{2,3}^{H,L}=I_{0,3}^{L,H}$, where swapping $H,L$ to $L,H$ indicates a swap of the labels on the variables $(\eta_{L},\eta_{H})$. We can then write the radial action as
\eq{
\frac{i_r}{G\mH\mL}=&\,\,\log(b)I_{0,1}^{H,L}+\frac{G}{b}\big(\mH I_{0,2}^{H,L}+\mL I_{0,2}^{L,H}\big)+\frac{G^{2}}{b^{2}}(\mH^{2} I_{0,3}^{H,L}+\mL^{2} I_{0,3}^{L,H}+\mH\mL I_{1,3}^{H,L})\, .
}{}
Thus despite having extra fields it is still true that 1PM and 2PM radial actions are determined by 0SF physics, 3PM and 4PM are determined by 0SF and 1SF physics, etc. one just needs to also relabel the charges. The upshot is that we can still utilize the EFT for extreme mass ratios~\cite{cheung2023effective} to efficiently compute in Einstein-Maxwell theory. 

To be less abstract about what we're computing we'll also explicitly write the center of mass frame scattering angle. To do so we first write the impact parameter in terms of the angular momentum, 
\eq{
b=\frac{J\sqrt{1+2\nu(\sigma-1)}}{M\nu\sqrt{\sigma^{2}-1}}\,,
}{}
where $M=\mH+\mL$ and $\nu=\mH\mL/M^{2}$ are the total mass and symmetric mass ratio. Differentiating the radial action with respect to $J$ immediately yields the scattering angle,
\eq{
\chi^{\textrm{1PM}}&=-\frac{G   M^2 \nu}{J}I_{0,1}^{H,L}\\
\chi^{\textrm{2PM}}&= \frac{G^2  M^4\nu ^2 \sqrt{1-4 \nu } \sqrt{\sigma ^2-1} }{2 J^2
   \sqrt{2 \nu  (\sigma -1)+1}}(I_{0,2}^{H,L}-I_{0,2}^{L,H}) +\frac{G^2M^{4} \nu ^2 \sqrt{\sigma ^2-1}
   }{2 J^2 \sqrt{2 \nu  (\sigma -1)+1}}(I_{0,2}^{H,L}+I_{0,2}^{L,H})\\
\chi^{\textrm{3PM}}&=\frac{G^3 \sqrt{1-4 \nu } \nu ^3 M^6 \left(\sigma ^2-1\right) }{J^3
   (2 \nu  (\sigma -1)+1)}(I_{0,3}^{H,L}-I_{0,3}^{L,H}) \\
   &-\frac{G^3 \nu ^3 M^6 \left(\sigma ^2-1\right) }{J^3 (2 \nu  (\sigma -1)+1)}\left((2 \nu -1)(I_{0,3}^{H,L}+I_{0,3}^{L,H})-2 \nu I_{1,3}^{H,L}\right)\,.
}{eq:scatteringangle}
The probe limit determines most of the scattering angle through 3PM, and it is elementary to compute. The main computation in this paper is of the 3PM 1SF function $I_{1,3}^{H,L}(\sigma)$.

\medskip
\subsection{0SF: Background Solutions}\label{sec:0SFsolutions}
The equations of motion which follow from \Eq{eq:action} are
\eq{
\ddot{x}_{j}^{\mu}=-\Gamma^{\mu}_{\,\,\alpha\beta}(x_{j})\dot{x}_{j}^{\alpha}\dot{x}_{j}^{\beta}+\eta_{j}F^{\mu}_{\,\,\,\,\alpha}(x_{j})\dot{x}_{j}^{\alpha}\,,
}{}
\eq{\nabla_{\mu}F^{\mu\nu}=8\pi G \sum_{j}m_{j}\eta_{j}\int\,d\tau \dot{x}_{j}^{\nu}\,\frac{\delta^{4}(x-x_{j}(\tau))}{\sqrt{-g}}
\,,
}{}
and
\eq{
G^{\mu\nu}(x)=8\pi G\sum_{j}m_{j}\int\,d\tau \dot{x}_{j}^{\mu}\dot{x}_{j}^{\nu}\,\frac{\delta^{4}(x-x_{j}(\tau))}{\sqrt{-g}}-2F^{\mu}_{\,\,\,\,\alpha}F^{\nu\alpha}+\frac{1}{2}g^{\mu\nu}F_{\alpha\beta}F^{\alpha\beta}\,.
}{}
One could study the perturbative scattering dynamics of charged black holes by iteratively solving these equations of motion in powers of $G$, however there is a more economical approach based on a expansion in powers of the mass ratio. The power of this approach derives from the fact that at 0SF order we neglect terms of order $m_{L}$ and  can write down exact solutions.

The heavy body's equation of motion trivializes in the 0SF limit\footnote{We use dimensional regularization to treat point-particle singularities. One could either say that we've dropped scaleless integrals, or alternatively, that we've solved the 0SF equations of motion in general dimension $D$ and used the fact that the Reissner-Nordstr\"{o}m-Tangherlini solution is flat as $r\rightarrow0$ for sufficiently negative $D-4$.}, $\ddotbarxH^{\mu}=0$, and we chose the solution
\eq{
\barxH^{\mu}(\tau)=\vH^{\mu}\tau.
}{}
In the rest frame of the heavy body the background field configuration, to all orders in $Gm_{H}$, is just the Reissner-Nordstr\"{o}m solution\footnote{This is a solution to the exact field equations without matter, however when the exact solution is expanded perturbatively to \textit{arbitrary} order in $G$ it requires the matter source.}
\eq{
\bar{ds^{2}}&=\frac{\Delta(r)}{r^{2}}dt^{2}-\frac{r^{2}}{\Delta(r)}-r^{2}d\Omega^{2}\,,\\
\bar{A}&=\frac{G\mH\eH}{r}dt\,,
}{}
with
\eq{
\Delta(r)=\left(1-\frac{2G\mH}{r}+\frac{G^{2}\mH^{2}\eH^{2}}{r^{2}}\right)\,.
}
{}
Here and throughout we'll use an overbar to denote background field configurations. In isotropic coordinates we can boost to a general inertial frame,
\eq{
\bar{g}_{\mu\nu}&=\left(f_{+}f_{+}-\zeta\eH\right)^{2}\eta_{\mu\nu}+\left[\left(\frac{f_{-}f_{+}+\zeta\eH}{f_{+}f_{+}-\zeta\eH}\right)^{2}
-\left(f_{+}f_{+}-\zeta\eH\right)^{2}\right]\vH{}_{\mu}\vH{}_{\nu}\,\\
\bar{A}_{\mu}&=2\eH\zeta\left[\frac{1}{f_{+}f_{+}-\zeta\eH}\right]\vH{}_{\mu}\,,
}{eq:boostedRN}
with $\zeta=\frac{G\mH}{2r}$, $f_{\pm}=1\pm\zeta$, and $r=\sqrt{(\vH x)^{2}-x^{2}}$.

In later sections we will make use of the Fourier transform of these solutions, as they will appear as insertions in Feynman diagrams. Defining $\bar{g}_{\mu\nu}=\eta_{\mu\nu}+\bar{\gamma}_{\mu\nu}$ we write the background fields in isotropic coordinates and Lorenz gauge, 
\eq{
\bar{\gamma}_{\mu\nu}(q)&=\del{\vH q}\frac{8\pi G\mH}{-q^{2}}(\eta_{\mu\nu}-2\vH{}_{\mu}\vH{}_{\nu}) \\
&+\del{\vH q}\frac{\pi^{2} G^{2}\mH^{2}}{\sqrt{-q^{2}}}((3-\eH^{2})\eta_{\mu\nu}+(1+3\eH^{2})\vH{}_{\mu}\vH{}_{\nu})+\cdots\, \\
\bar{A}_{\mu}(q)&=\del{\vH q}\frac{4\pi G\mH\eH}{-q^{2}}\vH{}_{\mu}-\del{\vH q}\frac{2\pi^{2}G^{2}\mH^{2}\eH}{\sqrt{-q^{2}}}\vH{}_{\mu}+\cdots\,.
}{eq:RNisotropic}
The probe particle's motion is given by $\barxL^{\mu}(\tau)=\vL^{\mu}\tau+b^{\mu}+x_{1}^{\mu}(\tau)+\cdots$, with
\eq{
x_{1}^{\mu}(\omega)=&4i\pi G\mH\int_{q}\,e^{iqb}\frac{\del{\vH q}\del{\vL q+\omega}}{q^{2}\omega^{2}} \\
&\times\left(\omega(2\vL^{\mu}+\vH^{\mu}(\eH\eL-4\sigma))+q^{\mu}(1+\eH\eL\sigma-2\sigma^{2})\right)\,.
}{eq:probesoln}
Rather than iteratively solving for the trajectory in a perturbative series, one could leverage the integrability of the test-body motion to obtain simpler expressions~\cite{ChandrasekharBook}. The advantage of such an approach begins at the next perturbative order, but for $x^{\mu}_{1}$ there is no considerable simplification. As for the background fields we've already utilized the analogous simplification; one could build up $\bar{\gamma}_{\mu\nu}(q)$ and $\bar{A}_{\mu}(q)$ perturbatively by evaluating multi-loop Feynman diagrams for one-point functions as originally demonstrated by Duff~\cite{duff}\footnote{See also  \cite{DdimSchw, DOnofrio:2022cvn} for further work.}, or do as we've done and simply expand and Fourier transform the exact solution in \Eq{eq:boostedRN}. 

Here, and throughout the rest of the paper, we are omit the $i\epsilon$ pole prescriptions in the $1/\omega^{2}$ particle Green's function as well as in the photon and graviton propagators. Relatedly, we have not specified whether $\vL, \vH, b$ are defined in the far future or past. We can do so because we are computing the conservative contributions to a time-symmetric observable. The $i\epsilon$, which changes under time-reversal, necessarily drops out of our results. For concreteness though one can safely assume that we are defining the dimensionless energy $\sigma=\vH^{\mu}\vL{}_{\mu}$, and the angular momentum $J$, in the center-of-mass frame of the total system in the asymptotic past. While there are corrections at finite $\tau$ to both the heavy and light worldlines away from inertial trajectories, the asymptotically defined energy and angular momentum are still constants of motion.

\subsection{0SF: Radial Action}
The 0SF radial action can be computed perturbatively via Feynman diagrams in a similar fashion to the higher SF orders, however it is far more efficient to simply identify the radial momentum and integrate. This manifests the symmetries of the background sourced by the heavy body, eg. staticity and isotropy for Reissner-Nordstr\"{o}m black holes, and utilizes the integrability of the trajectories.  

The radial action for probe motion is given by the one-dimensional integral
\eq{
G\mH\mL I_{0}=\pi\mL b(\sigma^{2}-1)^{1/2}+2\int_{r_{\textrm{min}}}^{\infty}\,dr\,p_{r}(r)\,,
}{eq:radialmomentumintegral}
where $p_{r}(r)$ is the radial component of the canonical momentum, and $r_{\textrm{min}}$ its largest real zero. For a particle with mass $\mL$ and charge $\eL$ the ``on-shell'' constraint is 
\eq{
g^{\mu\nu}(p_{\mu}-\mL\eL \bar{A}_{\mu})(p_{\nu}-\mL\eL \bar{A}_{\nu})=\mL^{2}.
}
{eq:onshellcondition}
Using conservation of energy and angular momentum we may write $p_{0}=\mL\sigma,\;p_{\phi}=\mL b(\sigma^{2}-1)^{1/2}$, and solve \Eq{eq:onshellcondition} for $p_{r}$ as a function of $r,b,\sigma$.  In isotropic coordinates the radial momentum has simple dependence on the angular momentum and all non-trivialities depend only on functions of $\sigma$,
\eq{
p^{2}_{r}(r)=\mL^{2}(\sigma^{2}-1)\left(1-\frac{b^{2}}{r^{2}}\right)+\mL^{2}\sum_{k=1}\frac{(G\mH)^{k}N_{k}(\sigma)}{r^{k}}\,.
}{eq:generalradialmomentum}
The coefficients $N_{k}(\sigma)$ can be easily solved for order by order from \Eq{eq:onshellcondition}. 

The radial action integral for a general radial momentum of the form \Eq{eq:generalradialmomentum} can be evaluated by residues if one carefully manipulates the integration contour. Details can be found in the appendix of Damour and Schaefer~\cite{Damour:1988mr}. The general result is
\eq{
I_{0}=\frac{1}{2}\sum_{k=1}^{\infty}\sum_{q=1}^{k}(G\mH)^{k-1}c_{k,q}\frac{(\sigma^{2}-1)^{1/2-q}}{b^{k-1}}B(\tfrac{1}{2}(k-1),\tfrac{3}{2}-q)\,,
}{eq:generalradialaction}
where $B$ is the Euler Beta function. The $c_{k,q}$ coefficients are simple monomials in the $N_{k}$ which are straightforwardly computed by PM expanding $p_{r}$ and collecting powers of $r$. They are tabulated for the first few orders in~\cite{long_paper} however an all orders formula is also given by \cite{Kalin:2019rwq}.

With this general formula we can write the 0SF radial action for the scattering of electrically charged spinless bodies in Einstein-Maxwell theory, 
\begingroup
\allowdisplaybreaks
\begin{align}\label{eq:0SFradialaction}
I^{H,L}_{0,1}&=\frac{2}{(\sigma ^2-1)^{1/2}}\left(\eH \eL \sigma -2 \sigma ^2+1\right) \nonumber\\
I^{H,L}_{0,2}&=\frac{\pi}{4(\sigma^{2}-1)^{1/2}}\left(2 \eH^2 \eL^2+\eH^2(1-3\sigma^{2})-12 \eH \eL \sigma+15 \sigma ^2-3\right) \nonumber\\
I^{H,L}_{0,3}&=\frac{1}{3\left(\sigma ^2-1\right)^{5/2}}\bigg(\eH^3 \eL^3 \sigma  \left(3-2 \sigma ^2\right) +3 \eH^3 \eL \sigma  \left(4 \sigma ^4-7 \sigma ^2+3\right)\hspace{-1pt} +3 \eH^2 \eL^2 \left(8 \sigma ^4-12 \sigma ^2+3\right)\nonumber \\
&+3 \eH^2 \left(-8 \sigma ^6+16 \sigma ^4-9 \sigma ^2+1\right) +3 \eH \eL \sigma  \left(-24 \sigma ^4+40 \sigma ^2-15\right) \nonumber \\
&+64 \sigma ^6-120 \sigma ^4+60 \sigma ^2-5\bigg) \nonumber\\
I^{H,L}_{0,4}&=\frac{\pi}{64(\sigma^{2}-1)^{3/2}}\bigg(8\eH^{4}\eL^{4}-24\eH^{4}\eL^{2}(5\sigma^{2}-1)-160\eH^{3}\eL^{3}\sigma+\eH^{4}(35\sigma^{4}-30\sigma^{2}+3) \nonumber\\
&+80\eH^{3}\eL\sigma(7\sigma^{2}-3)+120\eH^{2}\eL^{2}(7\sigma^{2}-1)-10\eH^{2}(63\sigma^{4}-42\sigma^{2}+3)\nonumber\\
&-560\eH\eL\sigma(3\sigma^{2}-1)+35(33\sigma^{4}-18\sigma^{2}+1)\bigg)\,.
\end{align}%
In principle we could compute the probe radial action to arbitrary accuracy, however we'll only need this result to 4PM for our discussions below.

\section{1SF: Subleading Contributions in Small Mass-Ratio}\label{sec:1SF}
\medskip
\subsection{Effective Field Theory for Extreme Mass Ratios}
The probe motion studied above is quite simple, however going beyond probe motion is highly non-trivial. The corrections to probe motion can be computed non-perturbatively using techniques developed by the valiant efforts of the gravitational self-force program, however when one is interested in a simultaneous SF+PM expansion there is a simpler approach. Higher orders in mass ratio can be studied systematically by first deriving an effective action wherein each operator power counts with a definite power of $\lambda$, and then using this action to compute perturbatively via Feynman diagrams. 

We will now set up this effective field theory for Einstein-Maxwell theory.  We take the action \Eq{eq:action} and expand about the $\lambda=0$ solutions 
\eq{
x^\mu_j  = \bar x^\mu_j + \dx_j^\mu \qquad \textrm{and} \qquad
g_{\mu\nu} = \bar g_{\mu\nu} + \dg_{\mu\nu} \qquad \textrm{and} \qquad
A_{\mu} = \bar A_{\mu} + \dA_{\mu} \, ,
}{eq:expandFields}
where the $\lambda=0$ solutions are given in \cref{sec:0SFsolutions}. Doing this ensures the fluctuations $\dx_j,\dA,\dg$ are $\OO(\lambda)$. From the general structure of the radial action \Eq{eq:radialActionSFexpansion} it follows that to compute 1SF physics we'll need to expand the action to order $\lambda^{2}$. 

Since the light particle worldline action has an explicit power of $\mL$ it only needs to be  expanded to linear order in fluctuations. These terms describe the light particle acting as an external source for the fluctuation fields
\eq{
S_{\textrm{sources}}=-m_{L}\int\,d\tau \left(\eL\dotbarxL^{\mu}\dA(\barxL)+\frac{1}{2}\dotbarxL^{\mu}\dotbarxL^{\nu}\dg_{\mu\nu}(\barxL)\right)\,.
}{}
The remaining terms in the action must be expanded to quadratic order in fluctuations. For terms quadratic in field fluctuations we have the background field theory of photon and graviton fluctuations about a Reissner-Nordstr\"{o}m  solution,
\eq{
S_{\textrm{BF}}[\bar{g},\dg,\bar{A},\dA]&=\frac{1}{16\pi G}\int d^{4}x\,\sqrt{-g}\left[-\frac{1}{4}\dg_{\mu\nu}\bar{\nabla}^{2}\dg^{\mu\nu}+\frac{1}{8}\dg\bar{\nabla}^{2}\dg-\frac{1}{2}\dg_{\mu\nu}\dg_{\rho\sigma}\bar{R}^{\mu\rho\nu\sigma}\right.\\
&\left.-\frac{1}{2}(\dg_{\mu\rho}\dg_{\nu}^{\rho}-\dg_{\mu\nu}\dg)\bar{R}^{\mu\nu}+\frac{1}{4}(\dg_{\mu\nu}\dg^{\mu\nu}-\frac{1}{2}\dg\dg)\bar{R}-\frac{1}{2}(1-\xi_{1})F_{\mu}F^{\mu}\right]\\
&+\frac{1}{8\pi G}\int d^{4}x\,\sqrt{-g}\left[\dA_{\mu}\bar{\nabla}^{2}\dA^{\mu}-\dA_{\mu}\dA_{\nu}\bar{R}^{\mu\nu}+(1-\xi_{2})G^{2}\right]\\
&-\frac{1}{16\pi G}\int d^{4}x\,\sqrt{-g}\left[4\,\dg_{\mu\nu}\dF_{\rho}^{\nu}\bar{F}^{\mu\rho}+\dg\dF_{\mu\nu}\bar{F}^{\mu\nu}\right]\,.
}{eq:BFaction}
Here we've grouped the terms into graviton-graviton, photon-photon, and photon-graviton mixing ``Gertsenshtein'' terms. We've also added gauge fixing terms proportional to $\xi_{1},\xi_{2}$, where
\eq{
F_{\mu}=\bar{\nabla}_{\nu}\dg_{\mu}^{\nu}-\frac{1}{2}\bar{\nabla}_{\mu}\dg\qquad\textrm{and}\qquad G=\bar{\nabla}_{\mu}\dA^{\mu}\,.
}{}
Choosing $\xi_{1}=\xi_{2}=1$ puts the action into Lorenz gauge for both fields, however we will leave these gauge parameters arbitrary to allow for an additional consistency check on the final result, ie. that it is gauge independent.

Finally there are the terms involving the fluctations of the heavy body. A key observation in \cite{cheung2023effective} was that the  fluctuations of the heavy body can be exactly integrated out, leaving behind an effective \textit{recoil operator} for the fields. In Einstein-Maxwell theory this recoil operator is
\eq{
S_{\textrm{recoil}}=-\frac{\mH}{2}\int d\tau \left[\dG^{\mu}_{\textrm{H}}(\barxH)-\eH\dE^{\mu}_{\textrm{H}}(\barxH)\right]\frac{1}{\partial_{\tau}^{2}}\left[\dG_{\textrm{H}}{}_{\mu}(\barxH)-\eH\dE_{\textrm{H}}{}_{\mu}(\barxH)\right]\,,
}{eq:recoilop}
which mixes the electric field terms $\dE^{\mu}_{\textrm{H}}(\barxH)=\vH^{\alpha}\dF^{\mu}_{\alpha}(\barxH)$ and gravitational field terms $\dG^{\mu}_{\textrm{H}}(\barxH)=\vH^{\alpha}\vH^{\beta}\dG^{\mu}_{\alpha\beta}(\barxH)$, where $\dG^{\mu}_{\alpha\beta}(\barxH)$ is the difference between the full connection and its background value, $\dG^{\mu}_{\alpha\beta}(\barxH)=\Gamma^{\mu}_{\alpha\beta}-\bar{\Gamma}^{\mu}_{\alpha\beta}$. 

The 1SF effective action is then simply the action for photons and gravitons in a Reissner-Nordstr\"{o}m background, which are sourced by the light body on its probe trajectory (i.e. background-field theory), together with a recoil operator
\eq{
S_{1\textrm{SF}}=S_{\textrm{sources}}+S_{\textrm{recoil}}+S_{\textrm{BF}}[\bar{g},\dg,\bar{A},\dA]\,.
}{eq:1SFeffectiveaction}
This effective field theory is sufficient to study the conservative 1SF dynamics of a pair of charged, spinless, massive bodies in Einstein-Maxwell theory to arbitrary order in $G$. We give the associated Feynman rules in \cref{app:feynmanrules}. 

We should be clear, however, about this statement that this EFT can be used to arbitrary order in $G$---we are stating that it can be used to streamline calculations at \textit{any finite order} in $G$, not that it is an exact in $G$ description of the black hole physics. The obstruction, discussed below, is common to all EFT approaches to the gravitational two-body problem. The operational utility of \Eq{eq:1SFeffectiveaction} is that: \textit{i)} it is quadratic in the fluctuation fields $\delta g, \delta A$, and \textit{ii)} by simple Taylor expansion of known expressions for the background metric, gauge field, and test-particle trajectories, this action immediately generates simple Feynman rules for the fluctuation fields, to any order in $G$. This being said, at 4PM order and beyond the point particle effective action that we started with, \Eq{eq:action}, must be supplemented by higher-dimension operators describing the non-minimal coupling of the particles to the fields.\footnote{This is in addition to possible localized gapless degrees of freedom living on the worldline.~\cite{PhysRevD.73.104030}} These nonminimal couplings describe the finite sizes of the black holes via polarizabilities, tidal responses, etc. 

 Including non-minimal terms in the 1SF effective action allows, in principle, for systematic computations to arbitrary higher powers of $\rs/b$ for a relevant orbital length scale $b$. Given this fact, one can ask if the EFT could be used to compute observables exactly in $G$. There is an obvious practical obstruction to determining the infinite set of Wilson coefficients which identify the particle in the EFT as a black hole rather than some other lump of matter. However even if one had complete knowledge of the Wilson coefficients, this is still an \textit{effective} theory for computing observables as a series expansion in $\rs/b$ and there is no guarantee that this series converges, or even if, that it converges for even moderate values of $\rs/b$. In this sense the EFT approach presented in \cite{cheung2023effective} cannot, as presented, be used for ``full'' SF computations.

As mentioned in \Sec{sec:massratioexpansion} there are certain terms in the radial action at given PM order for which $\mL\leftrightarrow\mH$ symmetry can be used to reduce the number of independent computations.\footnote{This symmetry is only manifest in the on-shell radial action, and not necessarily in the off-shell expressions such as \Eq{eq:recoilop}.} At 2PM it will be the case that the above 1SF effective action tells us nothing new that couldn't be inferred from this symmetry. At 3PM order and beyond, the 1SF effective action will, however, describe unique physics. For example, the first diagram in \Fig{fig:2PMdiagrams} describes an incoming light body on an inertial trajectory and its static gravitational field inducing a wobble of the heavy body. In the frame of the light body this is just describing a static black hole of mass $\mL$ and the leading perturbative correction to the heavy body's geodesic equation. This clearly has an exactly analogue in the 0SF-2PM case where the black hole of mass $\mH$ is static and the light body has a perturbative geodesic equation. The diagrams in \Fig{fig:3PMdiagrams}, however have no analogue at 0SF. For example, the fourth diagram contains both heavy body and light body deviations from inertial motion.

\subsection{Radial Action}

We aim to compute the on-shell (radial) action. The required computation can phrased in a number of equivalent ways: solve the equations of motion and insert back into the action, evaluate the Gaussian path-integral, sum all tree-level Feynman diagrams, etc. Since this effective action is quadratic one could formally evaluate the path-integral, yielding an on-shell action to all orders in $G$. However it is not presently known how to invert the quadratic form in \Eq{eq:BFaction} in a closed analytic expression. Here we will instead compute perturbatively in powers of $G$. This amounts to summing over a set of tree-level Feynman diagrams with only a small number of perturbative background field operator insertions which are determined by the solutions $\barxL,\bar g,\bar A$. The expansions of the background solutions presented in \Eqs{eq:RNisotropic}{eq:probesoln} are sufficient to compute the radial action to 3PM order.

The only graphs which we can draw, to all orders in $G$, consist of a light particle sourcing a field fluctuation which propagates through the Reissner-Nordstr\"{o}m background (mixing between photon and graviton states), encountering an arbitrary number of recoil operator insertions, and eventually terminating on another light particle source vertex. The diagrams needed to compute the 1SF-2PM and 1SF-3PM radial actions are presented in \cref{fig:2PMdiagrams,fig:3PMdiagrams}. We've omitted the diagrams with two recoil operator insertions, as these don't contribute to conservative physics at this order. 

\begin{figure}
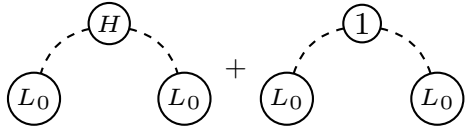

\begin{centering}
$\hbox{\HtwoPM}\mathrel{\raisebox{4ex}{$\!+\!$}}\hbox{\VtwoPM}$
\caption{Diagram structures for the 1SF-2PM radial action. The circles denote various Feynman elements. $H$ is the recoil operator, 1 is a linearized background field insertion, and $L_{0}$ is a zeroth order light particle geodesic source. Each dashed line can be either a photon or graviton propagator, and one must consider all possible identities of these lines and sum all distinct graphs with appropriate diagram symmetry factors.}\label{fig:2PMdiagrams}
\end{centering}
\end{figure}

\begin{figure}
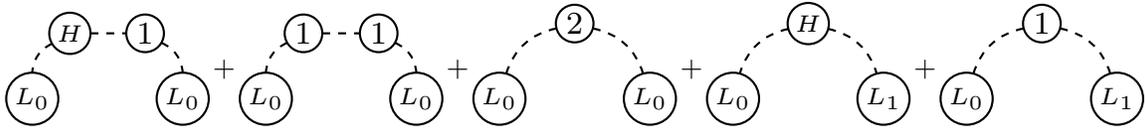

\begin{centering}
$\HVthreePM\mathrel{\raisebox{4ex}{$\!+\!$}}\VVthreePM\mathrel{\raisebox{4ex}{$\!+\!$}}\VtwothreePM\mathrel{\raisebox{4ex}{$\!+\!$}}\LzeroHLone\mathrel{\raisebox{4ex}{$\!+\!$}}\LzeroVLone$
\caption{Diagram structures for the 1SF-3PM radial action. The novel elements not found in the 2PM graphs are the $\OO(G^{2})$ background field vertex, 2, and the 1PM corrected geodesic sources $L_{1}$. Diagrams with multiple recoil insertions are not included because they do not contribute in the potential region.}
\label{fig:3PMdiagrams}
\end{centering}
\end{figure}

Since the worldlines break translation symmetry the Fourier transforms in the sources are not trivial and effectively generate Feynman loop-integrals. These integrals are all expressible as members of the families
\eq{
G^{2\textrm{PM}}(a_{1},a_{2},a_{3})=\int_{\ell}\frac{\del{\ell\cdot\vL}}{D_{1}^{a_{1}}D_{2}^{a_{2}}D_{3}^{a_{3}}}\,,
}{}
with
\eq{
\{D_{i}\}=\{(\ell\cdot\vH),\ell^{2},\,(\ell-q)^{2}\}\,,
}{} 
and
\eq{
G^{3\textrm{PM}}(a_{1},...,a_{7})=\int_{\ell_{1},\ell_{2}}\frac{\del{\ell_{1}\cdot\vL}\del{\ell_{2}\cdot\vH}}{D_{1}^{a_{1}}D_{2}^{a_{2}}D_{3}^{a_{3}}D_{4}^{a_{4}}D_{5}^{a_{5}}D_{6}^{a_{6}}D_{7}^{a_{7}}}\,,
}{}
with
\eq{
\{D_{i}\}=\{(\ell_{1}\cdot\vH),\,(\ell_{2}\cdot\vL),\ell_{1}^{2},\,\ell_{2}^{2},\,(\ell_{1}+\ell_{2}-q)^{2},\,(\ell_{1}-q)^{2},\,(\ell_{2}-q)^{2}\}\,,
}{}
where $q$ is the fourier conjugate to the impact parameter $b$. Here we've omitted the Feynman $i0$-prescriptions in the propagators as well as factors of $q^{2}$ and $\sigma$ in the numerator.  The delta functions can too be represented as linearized propagators via `reverse unitary',
\eq{
\del{\ell\cdot u}=-i\left(\frac{1}{\ell \cdot u -i0}-\frac{1}{\ell\cdot u+i0}\right),
}{}
effectively enlarging the propagator sets $\{D_{i}\}$ and allowing for well developed Feynman integral tools to be applied.

We closely follow the integration strategy imported to gravitational wave physics from collider physics by \cite{Parra-Martinez:2020dzs}. We manage the loop integrals by reducing them onto a basis of master integrals using integration by parts (IBP) identities, automated via the \texttt{LiteRed} and \texttt{FIRE6} software packages~\cite{Chetyrkin:1981qh, Laporta:2000dsw, Lee:2013mka, Smirnov:2019qkx}. The 2PM master integrals can be straightforwardly evaluated because the ``cut'' propagator (ie. the delta function) can be immediately used to localize the energy integrals and yield straightforward $d-1$-dimensional static Euclidean integrals. At 3PM this procedure no longer works because the integrals have cuts on both the heavy and light bodies' propagators. To evaluate these two-loop integrals we then use canonical differential equations to bootstrap their full $\sigma$ dependence from knowledge of their values in the static, $\sigma\rightarrow 1$ limit. We isolate the conservative physics by evaluating these static `boundary condition' integrals in the potential region. The pure basis that places our differential equations into canonical form is borrowed from~\cite{Parra-Martinez:2020dzs}. Detailed computations of the required static integrals can already be found in \cite{Parra-Martinez:2020dzs, Dlapa:2023hsl} along with a thorough discussion of the whole integration strategy, so we will not present further detail here.

Summing the contributions from these diagrams and evaluating the loop integrals using the above techniques, we arrive at the 1SF radial action for the scattering of spinless bodies in Einstein-Maxwell theory,
\eq{
I^{H,L}_{1,2}&=
\frac{\pi}{4(\sigma^{2}-1)^{1/2}}\bigg(2\eH^{2}\eL^{2}+\eL^2 \left(1-3 \sigma ^2\right)-12 \eH \eL \sigma+15 \sigma ^2-3 \bigg)\\
I^{H,L}_{1,3}&=\frac{1}{3(\sigma^{2}-1)^{5/2}}\bigg(
2\eH^3 \eL^3 \left(\sigma ^4-3 \sigma ^2+3\right)
+\eH \eL \left(\eH^2+\eL^2\right) \left(8 \sigma ^6-6 \sigma ^4-9 \sigma ^2+7\right) \\
&+6 \eH^2 \eL^2 \sigma  \left(4 \sigma ^4-4\sigma^{2}-1\right)+\eH \eL \left(-64 \sigma ^6+96 \sigma ^4-42 \sigma ^2+16\right) \\
&+\left(\eH^2+\eL^2\right)\left(-8 \sigma ^7+36 \sigma ^5-51\sigma^{3}+23 \sigma \right) +2 \sigma  \left(36 \sigma ^6-114 \sigma ^4+132\sigma^{2}-55\right)
\bigg) \\
&+\frac{\textrm{arccosh}(\sigma)}{\sigma^{2}-1}\bigg(
\eH^2 \eL^2 \left(4-8 \sigma ^2\right)+16 \eH \eL \sigma  \left(\sigma ^2-2\right) -4\left(2 \sigma ^2+1\right) \left(\eH^2+\eL^2\right)\\
&-16 \sigma ^4+48 \sigma ^2+12\bigg)\,.
}{eq:1SFradialaction}
This result passes a variety of consistency checks. The first is simply that it is independent of the gauge parameters $\xi_{1},\xi_{2}$ we had introduced. More detailed checks are that the 1SF-2PM contribution $I_{1,2}^{H,L}$ matches the 0SF-2PM contribution $I_{0,2}^{H,L}$ after swapping $H,L$ labels, and the 1SF-3PM contribution is itself symmetric under $H \leftrightarrow L$. The pure gravity limit $(\eH,\eL)\rightarrow 0$ agrees with the original computation~\cite{Bern:2019nnu, Bern:2019crd} which itself has been confirmed many times over~\cite{Kalin:2020fhe, Cheung:2020gyp, Bjerrum-Bohr:2021din, Brandhuber:2021eyq, Jakobsen:2022fcj}. Further yet, taking only $\eH\rightarrow 0$ agrees with a previously reported vector field toy self-force result~\cite{cheung2023effective}. Finally, taking $G\rightarrow0$ while holding $q_{j}=\eta_{j}m_{j}G^{1/2}$ fixed agrees with the pure electric scattering results~\cite{Bern_2022}.

With the 0SF and 1SF radial actions computed, we can utilize the aforementioned $L,H$ symmetry to report our main result, the \textit{total} radial action up to 3PM order
\eq{
\frac{i_r}{G\mH\mL}&=\log(b)I_{0,1}^{H,L}+\frac{G}{b}\big(\mH I_{0,2}^{H,L}+\mL I_{0,2}^{L,H}\big)+\frac{G^{2}}{b^{2}}(\mH^{2} I_{0,3}^{H,L}+\mH\mL I_{1,3}^{H,L}+\mL^{2} I_{0,3}^{L,H})\, ,
}{eq:totalradialaction}
with the $I^{H,L}_{n,m}(\sigma)$ functions given in \Eqs{eq:0SFradialaction}{eq:1SFradialaction}. The 3PM conservative scattering angle is then given by inserting these functions into \Eq{eq:scatteringangle}.

\section{Extremal Black Holes}\label{sec:extremal}
\medskip
It is very natural to ask if there is a simplification of the dynamics in the extremal limit. Of course Einstein-Maxwell theory has more fields than GR, but a major lesson from the modern scattering amplitudes program has been that on-shell observables can be far simpler than one might guess from the Lagrangian, especially for supersymmetric theories~\cite{ElvangHuang, Dixon:2013uaa, Cheung:2017pzi, Arkani-Hamed:2008owk}. This very idea led to Caron-Huot and Zahraee's pioneering study of classical binary dynamics of extremal black holes in $\mathcal{N}=8$ supergravity, as a potential toy model of General Relativity~\cite{Caron-Huot:2018ape}.

The specific simplicity they looked for was super-integrability of bound orbits, namely the existence of Laplace-Runge-Lenz type symmetries signalled by the absence of periapsis precession. They indeed found no precession at 1PN nor in the probe limit, and it was later demonstrated that this continues at least through 3PM~\cite{Parra-Martinez:2020dzs}. Given that Einstein-Maxwell theory is the bosonic sector of $\mathcal{N}=2$ supergravity and extremal Reissner-Nordstr\"{o}m (ERN) black holes are half-BPS solutions, it is natural to ask whether an analogous simplicity also occurs in this situation.

Given our above results for RN black holes we immediately address this question. We use the boundary-to-bound formula~\cite{Kalin:2019inp},
\eq{
\Delta\Phi(J)=\chi(J)+\chi(-J),
}{}
relating the precession angle $\Delta\Phi$ to the scattering angle $\chi$ at fixed angular momentum $J$. Since odd powers in $J$ do not contribute to the precession angle, at this order we need only look at 2PM,
\eq{
\chi^{\textrm{2PM}}=&\frac{\pi G^{2}\nu ^2 M^4 \left(4\eH^2\eL^2 -24 \eH \eL \sigma +(\eL^2+\eH^{2}) \left(1-3 \sigma ^2\right)+30 \sigma ^2-6\right)}{8 J^2 \sqrt{2 \nu  (\sigma -1)+1}}\\
&-\frac{\pi G^{2}  \sqrt{1-4 \nu } \nu ^2 M^4 \left(3 \sigma ^2-1\right) \left(\eH^2-\eL^2\right)}{8 J^2 \sqrt{2 \nu  (\sigma -1)+1}}\,.
}{}
This is non-zero for all choices of $(\eH,\eL)$, and thus the precession angle is non-zero. The non-zero precession of charged test-particle orbits
in RN backgrounds is well known~\cite{ChandrasekharBook}, however the discussion here lifts that statement to arbitrary mass ratios.

We can also ask whether the functional form of the result~\Eq{eq:1SFradialaction} sees any simplification in the extremal limit. The answer appears to be ``no''. Moreover there are no choices of $(\eH,\eL)$ for which the $\textrm{arccosh}(\sigma)$ term vanishes, so there doesn't appear to be special limits where one can decrease the transcendentality or simplify the integrals.

\subsection{Comparison with the Moduli Space Approximation}

One might also ask if there are techniques which become available in the extremal limit which simplify the calculations involved to study the dynamics. In this paper we're working with a worldline effective field theory so we do not benefit from the scattering amplitudes tools such as on-shell superspace and unitary cut constructions. There is however a very elegant Lagrangian method available for certain extremal black holes called the \textit{moduli space approximation}~\cite{Manton:1981mp}.

As previously discussed, Majumdar and Papapetrou have given a whole moduli space of exact solutions to the Einstein-Maxwell equations describing an arbitrary arrangement of \textit{static} ERN black holes. Manton's moduli space approximation states that quite generally one can compute a metric on the moduli space of BPS solitons and that the two-body dynamics is simply described by a geodesic in this space, up to $\OO(v^{2})$ and all orders in coupling. For the two-body problem this space is an $\mathbb{R}^{6}$. Translation invariance ensures that the center-of-mass coordinate $\mathbf{R}=(m_{1}\mathbf{x}_{1}+m_{2}\mathbf{x}_2)/(m_1+m_2)$ moves in a flat $\mathbb{R}^{3}$. The relative position $\mathbf{r}=\mathbf{x}_{1}-\mathbf{x}_{2}$ lives in a curved $\mathbb{R}^{3}$, and rotation invariance fixes the metric to be a function of $r=|\mathbf{r}|$ alone. Ferrell and Eardley computed this metric~\cite{Ferrell:1987gf}, finding the remarkably simple form
\eq{
ds_{MSA}^{2}=Md\mathbf{R}^{2}+\mu\left[\left(1+\frac{GM}{r}\right)^{3}-\frac{2G^{3}\mu M^{2}}{r^{3}}\right]d\mathbf{r}^{3}\,,
}{}
where $M$ and $\mu$ are the total and reduced mass respectively. 

The Ferrell-Eardley MSA Lagragian is then
\eq{
L=-M+\frac{1}{2}MV^{2}+\frac{1}{2}\mu v^{2}\left[\left(1+\frac{GM}{r}\right)^{3}-\frac{2G^{3}\mu M^{2}}{r^{3}}\right]\,.
}{}
with  $(V,v)$ the center of mass and relative velocities. We can use this along with \Eqs{eq:generalradialmomentum}{eq:generalradialaction} to immediately compute the leading post-static radial action to arbitrary order in $G$ for arbitrary mass-ratio. Up to $\OO(G^{5}v^{2})$ the result is
\eq{
i_r&=G\mH\mL v\bigg(-3\log(b)+\frac{3\pi}{2} \frac{G}{b} (\mH+\mL)+\frac{1}{8} \left(\frac{G}{b}\right)^2 \left(35(\mH^2+\mL^{2})+54 \mH \mL\right) \\
&+\frac{3\pi}{8}  \left(\frac{G}{b}\right)^3 \left(5 (\mH^3+\mL^3) +11(\mH^{2}\mL+\mH\mL^{2})\right) \\
&+\frac{21}{320} \left(\frac{G}{b}\right)^4 \left(143(\mH^{4}+\mL^{4})+412(\mH^{3}\mL+\mH\mL^{3})+538\mH^{2}\mL^{2}\right)\bigg) +\OO(v^{3}),
}{}
and of course all static contributions, $\OO(v^{0}G^{\infty})$, are zero. The overlapping terms between this and our 3PM computation, \Eqs{eq:0SFradialaction}{eq:1SFradialaction}, agree completely.\footnote{The availability of this cross-check was identified by Jones and Solon~\cite{Jones_2023}.}

\subsection{Tutti Frutti for Extremal Post-Newtonian Dynamics}
 Given that the slow-motion strong-field dynamics are readily computed it is clear that what the extremal limit affords us is a very efficient route to the post-\textit{Newtonian} dynamics. Rather than simply using the MSA as a check on our post-Minkowskian computation we can harness both the MSA and PM to efficiently compute non-overlapping terms in the PN expansion. The strategy is similar in spirit to the way we treated the mass ratio expansion; one could compute the full PM radial action and use the probe limit as a check, or one could use the simplicity of the probe limit to determine as much of the radial action as possible and perform the more difficult PM computations only where needed. This is precisely the spirit of the  ``Tutti-Frutti'' approach in GR~\cite{Bini:2019nra, Bini:2020rzn} where one computes PN coefficients using whatever means are most efficient.
 
 There is an important caveat here which distinguishes from GR. Since like-charge extremal black holes have a trivial Newtonian limit it's likely that there isn't even a regime in which one \textit{should} power count $G\sim v^{2}$. In the absence of such a regime one should regard the following discussion merely as a commentary on the orthogonality of the PM and MSA approximations, and how they together serve to map out the parameter space describing binary ERN black holes. Despite this potential issue, we will proceed to discuss the PN approximation in what follows and delay a careful litigation of its regime of validity to future work.

To describe the conservative dynamics of the two bodies from first post-Newtonian order (1PN) through 5PN there are a number of independent coefficients which require fixing.\footnote{This counting below is done assuming an isotropic gauge effective Hamiltonian, which is of course a gauge-dependent quantity. The radial action and scattering angle are gauge-independent however, and the counting of coefficients in the isotropic gauge Hamiltonian agrees with the counting of unique terms in the radial action.} At $n$PN order (for $n\leq 5$) we have the $n+1$ different terms $G^{n+1}v^{0},\,G^{n}v^{2},...,G^{1}v^{2n}$. Aside from an overall factor of $G\mH\mL$, for each of these terms each power of $G$ must multiply either $\mH$ or $\mL$. As discussed at length in \cref{sec:setup}, various terms are related under $H\leftrightarrow L$. All together this counting leaves 33 independent coefficients through 5PN. 

 Of the 33 independent coefficients, many are fixed by results which are much easier to compute than loop integrals. Through 5PN there are: 20 terms fixed by the probe limit, 5 more by the static limit, 3 more by the MSA result, and 4 more by our 3PM 1SF result. Of course there are many cross checks between the various approaches, but it is remarkable that this leaves just 1 unfixed coefficient (see \cref{fig:ERN-PN}.). This method of combining results from the various approaches scales very efficiently with PN order.  To go to 6PN would ostensibly introduce 17 new unfixed coefficients, however 11 of these are already determined by probe, static, MSA, and 3PM, and 2 more by 4PM. 
 
 Before proceeding we should address ``hereditary'' effects, such as tail terms describing radiated photons and gravitons which backscatter off the geometry and lead to conservative radiation-reaction effects. These effects are non-local in time, and spoil the counting of independent coefficients above. However, when truncating to a description of scattering orbits at fixed PN order an effectively local description can be given which resembles the standard local PN expansion up to the addition of terms logarithmic in asymptotic velocity~\cite{Bini:2017wfr, Cho:2021arx}. In GR one first encounters tail terms at 4PN order, however for ERN black holes one doesn't expect this until 6PN (hence the restriction to $n\leq5$ above). For example, at 6PN one expects to find a $G^{4}v^{6}\log(v)$ term in the radial momentum expansion in addition to the previously counted $G^{4}v^{6}$ term. At 6PN order this increases the naive count of 16 independent coefficients to 17.

 The relative suppression in PN order for tail contributions is somewhat counter-intuitive, especially given that electromagnetic dipole radiation occurs with fewer powers of $1/c$. There are two reasons for the suppression, and to se this we first utilize a radiation-zone effective theory one to estimate the order at which the tail effect would arise (see eg. \cite{Galley:2015kus}). For electromagnetic dipole radiation one expects contributions to the effective Hamiltonian scaling as $\sim G^{2}\ddot{d}^{2}$, where $d$ is the electric dipole moment of the system. For bodies with equal charge-to-mass ratio $\eH=\eL$ the leading contribution to the binary system's electric-dipole moment and mass-dipole moment are proportionally related. Since the conservative dynamics do not change the mass-dipole moment there is no dipolar radiation. The tail effect from quadrupolar radiation arises with scaling $\sim G^{2}\dddot{I}^{2}$, where $I\sim r^{i}r^{j}$. In GR one has $\dddot{I}\sim \dot{v}v\sim Gv$, and thus the tail effect arises at 4PN order, $\mathcal{O}(G^{4}v^{2})$. For extremal black holes the vanishing Newtonian limit implies a different scaling for the acceleration though, $\dot{v}\sim Gv^{2}$. One then expects tail contributions at 6PN order, $G^{4}v^{6}$, for ERN black holes. The absence of $\log(v)$ terms in \Eq{eq:PNscatteringangle} at 4PN and 5PN order, which are present in pure GR \cite{Bini:2017wfr}, supports this estimate.

The PM approximation is highly complimentary to the moduli space approximation, especially when we tackle PM order-by-order in mass ratio. The more difficult high-SF-order contributions are determined by the MSA, which at 5PN leaves only 1SF computations to be done by loop-integrals. The final contribution at 5PN, ie. the 1SF $\OO(G^{4}v^{4})$ term, is computed by three-loop Feynman integrals which can be evaluated with present-day integration technology. Evaluating this would completely determine PN physics which would have otherwise required Feynman integrals though five-loop order. For concreteness we present the conservative scattering angle through 5PN,
\begingroup
\allowdisplaybreaks
\begin{align}\label{eq:PNscatteringangle}
\chi^{1\textrm{PN}}=&\frac{3 G M}{b}\,,\nonumber\\
\chi^{2\textrm{PN}}=&\frac{3 \pi  G^2 M^2}{2 b^2}+\frac{G (6 \nu +1) M v^2}{4 b}\,,\nonumber \\
\chi^{3\textrm{PN}}=&-\frac{G^3 (16 \nu -35) M^3}{4 b^3}+\frac{3 \pi  G^2 (2 \nu +1) M^2 v^2}{8 b^2}-\frac{G \left(3 \nu ^2+2 \nu +1\right) M v^4}{8 b}\,, \nonumber\\
\chi^{4\textrm{PN}}=&-\frac{9 \pi  G^4 (4 \nu -5) M^4}{8 b^4}-\frac{G^3 \left(32 \nu ^2-34 \nu -63\right) M^3 v^2}{16 b^3}-\frac{3 \pi  G^2 \left(\nu ^2+1\right) M^2 v^4}{16 b^2}\nonumber\\
&+\frac{G \left(12 \nu ^3+10 \nu ^2+6 \nu +5\right) M v^6}{64 b}\,, \nonumber\\
\chi^{5\textrm{PN}}=&-\frac{21 G^5 (160 \nu -143) M^5}{80 b^5}-\frac{3 G^4 M^4 v^2 \pi  \left(12 \nu ^2+45 \nu -20\right)}{16 b^4}+\frac{3 G^4 \nu  M^4 v^2}{b^4}I_{1,4}^{v^{3}}\,,
\nonumber\\
&+\frac{G^3 \left(32 \nu ^3-110 \nu ^2-248 \nu -117\right) M^3 v^4}{64 b^3}+\frac{3 \pi  G^2 \left(4 \nu ^3+2 \nu ^2-2 \nu +5\right) M^2 v^6}{128 b^2}\nonumber\\
&-\frac{G \left(15 \nu ^4+16 \nu
   ^3+11 \nu ^2+6 \nu +7\right) M v^8}{128 b} \,,
\end{align}%
where $v=(\sigma^{2}-1)^{1/2}$. As mentioned, this has just one undetermined numerical coefficient, $I_{1,4}^{v^{3}}$, which is the $O(v^{3})$ term in the conservative sector 1SF-4PM radial action.

\begin{figure}[h]
\begin{center}
    \includegraphics[scale=0.35]{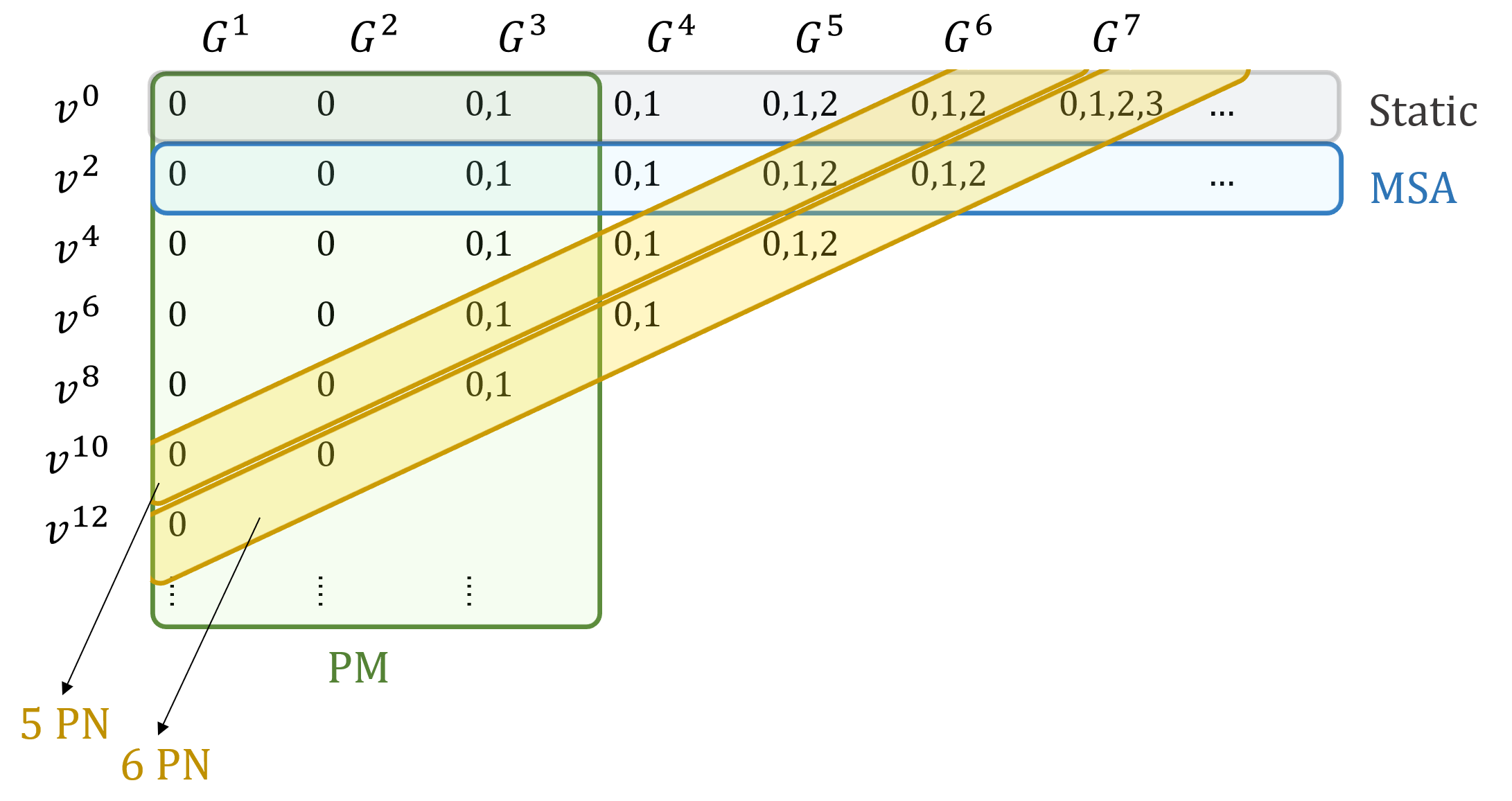}
\end{center}
    \caption{Each entry in the table indicates a term of order $v^{2n}G^{m}$ in the effective action for extremal charged black holes. Each term has independent coefficients for each of its powers of the mass-ratio, as indicated by the values in each cell. For example, $G^{1}$ and $G^{2}$ terms only have 0SF contributions while $G^{7}$ terms have contributions through 3SF order. The grey and blue rows are determined by the MSA. The green columns are determined by our new PM results. PN orders are diagonals of the table, with 5PN and 6PN highlighted.}
    \label{fig:ERN-PN}
\end{figure}

For purely electrically charged extremal black holes in Einstein-Maxwell theory we necessarily have $|\eH|=|\eL|=1$, which clearly has only two possible charge alignments, $\eH=\pm\eL$. In the aligned case we have the situation discussed above with a vanishing static limit and thus dynamics hardly resemble pure gravity. In the anti-aligned case we indeed have some resemblance to pure gravity, with eg. a non-vanishing attractive $1/r$ force and bound orbits, however in this case the MSA is completely inapplicable. Ideally one would have a situation in which the extremal black hole static forces only marginally cancel, and with the degree of cancellation determined by one or many controllable ``misalignment'' parameters. In such a case there would be bound orbits and one could hope to extract meaningful data from the PN approach described above, up to corrections controlled by powers of the misalignment parameters. While static forces contribute to powers of $1/v$ in the scattering angle which are dominant in the low-energy limit, the \textit{even} in $J$ contributions to the scattering angle that determine the precession angle are actually protected from such singular contributions. It may then be that the dominant contributions to the precession angle for slightly-misaligned charges is indeed given by the PN physics in \Eq{eq:PNscatteringangle}. We leave detailed scrutiny of this hypothesis for future work.

In supergravity theories with $\mathcal{N}>2$ one can retain purely electric charges while having a continuous space of non-trivial charge alignments, eg. the three-angles in $\mathcal{N}=8$~\cite{Caron-Huot:2018ape}. It is interesting to ask whether there are MSA-like tools for efficiently studying the case of small misalignment in such theories. In Einstein-Maxwell theory with purely electric charges the misalignment parameter can only take values $\pm1$, however we have not yet explored the more general case where the black holes have both electric and magnetic charge. In this more general case we would also introduce the desired misalignment angle. Again, it would be interesting to ask whether MSA-like tools could be developed to allow one to treat the general dyonic charge case for small misalignment angle as a perturbation away from the aligned charge case. Such a framework could genuinely realize the ideas presented in this section for using PM and MSA tools to efficiently map out the PN dynamics of extremal black holes. We will leave this for future work.

\section{Conclusions}\label{sec:conclusion}

In this paper we have computed the conservative classical scattering dynamics of two Reissner-Nordstr\"{o}m black holes at $\OO(G^{3})$. Specifically we computed the radial action, which encodes the conservative dynamics in a gauge-invariant and Lorentz-invariant manner. The computation of the radial action was performed order-by-order in mass ratio. Utilizing the symmetry under relabelling the bodies we expressed the complete result, \Eq{eq:totalradialaction}, in terms of just four independent functions of the asymptotic relative boost factor, $\sigma=\vH^{\,\,\,\mu}\vL{}_{\mu}$. These functions are the test-mass (0SF) contributions at 1PM, 2PM, and 3PM order, \Eq{eq:0SFradialaction}, and the next-to-leading in mass ratio (1SF) contribution at 3PM order, \Eq{eq:1SFradialaction}. The conservative scattering angle follows trivially from the radial action and is given in \Eq{eq:scatteringangle} in terms of the 0SF and 1SF contributions.

The perturbative computation of the 0SF radial action is a solved mechanics problem. We reviewed the general formula and applied it to evaluate the contributions through 3PM order. The computation of the 1SF radial action was performed using a recently developed effective field theory~\cite{cheung2023effective, long_paper} which systematically determines an effective action for computations accurate to a given SF order. At 1SF this effective action is just background field theory, ie. an orbiting test-body sourcing gravitons and photons in a Reissner-Nordstr\"{o}m background, supplemented with a \textit{recoil operator} to account for the backreaction onto the heavy body. Using this effective action we computed the 1SF radial action to 3PM order, completing the full 3PM computation. 

The masses and charges were kept arbitrary which allowed for an interpolation from pure gravity to purely electromagnetic scattering. The computation agrees in both these limits with known results~\cite{Bern:2019nnu, Bern:2019crd, Bern_2022}. Setting one of the electric charges to zero describes a vector field toy model for gravitational self-force, and we agree with previous PM computations for this model~\cite{long_paper}. Taking the charges to be extremal we were able to verify that the radial action has a vanishing static limit, consistent with the known \textit{exact} solutions consisting of static multi-centered extremal black holes. Furthermore we verified the leading post-static terms in our result by comparison with the radial action computed using the known effective action in the moduli space approximation.

We also discussed the possibility for combining data from post-Minkowskian calculations and the moduli space approximations to determine high orders in the post-Newtonian approximation. Combining results from the probe limit, the moduli space approximation, and the 3PM results computed here, we determined 32 of 33 independent numerical coefficients in conservative scattering angle through fifth post-Newtonian order. The post-Newtonian approximation is likely dubious for extremal black holes with like-charges, but these results may have some applicability in study of PN dynamics of dyonic extremal black holes---for which static forces do not necessarily cancel and virialized orbits can  exist.

\medskip
\begin{center} 
{\bf Acknowledgments}
\end{center}
\noindent 
We thank Clifford Cheung, Julio Parra-Martinez, Nabha Shah, and Ira Rothstein for many useful discussions and comments on the draft. J.W.G. is supported by a fellowship at the Walter Burke Institute for Theoretical Physics, a Presidential Postdoctoral Fellowship, the DOE under award number DE-SC0011632, and the Simons Foundation (Award Number 568762).

\newpage
\appendix

\section{Feynman Rules}\label{app:feynmanrules} 
The Feynman rules for the action \eqref{eq:1SFeffectiveaction} are given below, using an all ingoing momentum flow convention. We'll represent the photon propagator by solid lines,
\eq{
\mu\,\flatscalar\,\nu &= -\frac{4\pi Gi}{p^2} \left[\eta_{\mu\nu}-\left(1-\frac{1}{\xi_{2}}\right)\frac{p_{\mu}p_{\nu}}{p^{2}}\right]\,.
}{}
and the graviton propagator by wave lines
\eq{
\mu\nu\,\flatprop\,\alpha\beta &= \frac{16\pi Gi}{p^2} \bigg[\left(\eta_{\mu\alpha}\eta_{\nu\beta}+\eta_{\mu\beta}\eta_{\nu\alpha}-\eta_{\mu\nu}\eta_{\alpha\beta}\right) \\
&-\left(1-\frac{1}{\xi_{1}}\right)\left(\frac{\eta_{\beta\nu}p_{\alpha}p_{\mu}+\eta_{\alpha\nu}p_{\beta}p_{\mu}+\eta_{\beta\mu}p_{\alpha}p_{\nu}+\eta_{\alpha\mu}p_{\beta}p_{\nu}}{p^{2}}\right)\bigg]\,.
}{}
The PM expanded photon source is
\eq{
\photonsource\,\mu &=-i\mL\eL\int_{p}e^{-ipb}\left(\vL^{\mu}\,\del{\vL p}+i\OO^{\alpha\mu}(\vL,p)x_{1}{}_{\alpha}(\vL p)+\cdots\right)
}{}
and the PM expanded graviton source is
\eq{
\gravitonsource\,\mu\nu &=-i\mL\int_{p}e^{-ipb}\left(\frac{\vL^{\mu}\vL^{\nu}}{2}\,\del{\vL p}+i\OO^{\alpha\mu\nu}(\vL,p)x_{1}{}_{\alpha}(\vL p)+\cdots\right)\,,
}{}
where we've defined
\eq{
\OO^{\alpha\mu}(\vL,p)&=(\vL p)\eta^{\alpha\mu}-p^{\alpha}\vL^{\mu} \\
\OO^{\alpha\mu\nu}(\vL,p)&=\tfrac{1}{2}(\vL p)\left(\eta^{\alpha\mu}\vL^{\nu}+\eta^{\alpha\nu}\vL^{\mu}\right)-p^{\alpha}\vL^{\mu}\vL^{\nu}\,.
}{}

The recoil operator gives three Feynman vertices
\eq{
p_{1},\alpha_{1}\recoilvertAA p_{2},\alpha_{2} &= i\mH\eH^{2}\frac{\del{\vH p_{1}+\vH p_{2}}}{(\vH p_{1})(\vH p_{2})}\OO^{\mu\alpha_1}(\vH,p_{1})\OO_{\mu}^{\;\alpha_2}(\vH,p_{2})\\
p_{1},\alpha_{1}\beta_{1}\recoilvertGG p_{2},\alpha_{2}\beta_{2} &= i\mH\frac{\del{\vH p_{1}+\vH p_{2}}}{(\vH p_{1})(\vH p_{2})}\OO^{\mu\alpha_1 \beta_1}(\vH,p_{1})\OO_{\mu}^{\;\alpha_2 \beta_2}(\vH,p_{2})\\
p_{1},\alpha_{1}\recoilvertAG p_{2},\alpha_{2}\beta_{2} &= i\mH\eH\frac{\del{\vH p_{1}+\vH p_{2}}}{(\vH p_{1})(\vH p_{2})}\OO^{\mu\alpha_1}(\vH,p_{1})\OO_{\mu}^{\;\alpha_2 \beta_2}(\vH,p_{2})\,.
}{}

The Feynman rules for the background field insertions are sufficiently lengthy that we will not write them here, but they are straightforwardly computed by: taking the background field action \Eq{eq:BFaction}, stripping off the fluctuation fields $\delta g$ and $\delta A$, expanding the background fields $(\bar{g},\bar{A})$ as in \Eq{eq:RNisotropic}, and keeping the desired order in $G$.
\eq{
\backgroundvertAAone &= iG\left(\frac{d}{dG}\frac{\delta^{2}S_{\textrm{BF}}}{\delta(\delta A_{\alpha_{1}}(p_{1}))\delta(\delta A_{\alpha_{2}}(p_{2}))}\bigg|_{G=0}\right) \\
\backgroundvertGGone &= iG\left(\frac{d}{dG}\frac{\delta^{2}S_{\textrm{BF}}}{\delta(\delta g_{\alpha_{1}\beta_{1}}(p_{1}))\delta(\delta A_{\alpha_{2}\beta_{2}}(p_{2}))}\bigg|_{G=0}\right) \\
\backgroundvertAGone &= iG\left(\frac{d}{dG}\frac{\delta^{2}S_{\textrm{BF}}}{\delta(\delta A_{\alpha_{1}}(p_{1}))\delta(\delta g_{\alpha_{2}\beta_{2}}(p_{2}))}\bigg|_{G=0}\right) \\
\backgroundvertAAtwo &= i\frac{G^{2}}{2}\left(\frac{d^{2}}{dG^{2}}\frac{\delta^{2}S_{\textrm{BF}}}{\delta(\delta A_{\alpha_{1}}(p_{1}))\delta(\delta A_{\alpha_{2}}(p_{2}))}\bigg|_{G=0}\right) \\
\backgroundvertGGtwo &= i\frac{G^{2}}{2}\left(\frac{d^{2}}{dG^{2}}\frac{\delta^{2}S_{\textrm{BF}}}{\delta(\delta g_{\alpha_{1}\beta_{1}}(p_{1}))\delta(\delta A_{\alpha_{2}\beta_{2}}(p_{2}))}\bigg|_{G=0}\right) \\
\backgroundvertAGtwo &= i\frac{G^{2}}{2}\left(\frac{d^{2}}{dG^{2}}\frac{\delta^{2}S_{\textrm{BF}}}{\delta(\delta A_{\alpha_{1}}(p_{1}))\delta(\delta g_{\alpha_{2}\beta_{2}}(p_{2}))}\bigg|_{G=0}\right)\,.
}{}

\bibliography{references}

\begin{thebibliography}{140}%
\makeatletter
\providecommand \@ifxundefined [1]{%
 \@ifx{#1\undefined}
}%
\providecommand \@ifnum [1]{%
 \ifnum #1\expandafter \@firstoftwo
 \else \expandafter \@secondoftwo
 \fi
}%
\providecommand \@ifx [1]{%
 \ifx #1\expandafter \@firstoftwo
 \else \expandafter \@secondoftwo
 \fi
}%
\providecommand \natexlab [1]{#1}%
\providecommand \enquote  [1]{``#1''}%
\providecommand \bibnamefont  [1]{#1}%
\providecommand \bibfnamefont [1]{#1}%
\providecommand \citenamefont [1]{#1}%
\providecommand \href@noop [0]{\@secondoftwo}%
\providecommand \href [0]{\begingroup \@sanitize@url \@href}%
\providecommand \@href[1]{\@@startlink{#1}\@@href}%
\providecommand \@@href[1]{\endgroup#1\@@endlink}%
\providecommand \@sanitize@url [0]{\catcode `\\12\catcode `\$12\catcode `\&12\catcode `\#12\catcode `\^12\catcode `\_12\catcode `\%12\relax}%
\providecommand \@@startlink[1]{}%
\providecommand \@@endlink[0]{}%
\providecommand \url  [0]{\begingroup\@sanitize@url \@url }%
\providecommand \@url [1]{\endgroup\@href {#1}{\urlprefix }}%
\providecommand \urlprefix  [0]{URL }%
\providecommand \Eprint [0]{\href }%
\providecommand \doibase [0]{https://doi.org/}%
\providecommand \selectlanguage [0]{\@gobble}%
\providecommand \bibinfo  [0]{\@secondoftwo}%
\providecommand \bibfield  [0]{\@secondoftwo}%
\providecommand \translation [1]{[#1]}%
\providecommand \BibitemOpen [0]{}%
\providecommand \bibitemStop [0]{}%
\providecommand \bibitemNoStop [0]{.\EOS\space}%
\providecommand \EOS [0]{\spacefactor3000\relax}%
\providecommand \BibitemShut  [1]{\csname bibitem#1\endcsname}%
\let\auto@bib@innerbib\@empty
\bibitem [{\citenamefont {Abbott}\ \emph {et~al.}(2016)\citenamefont {Abbott} \emph {et~al.}}]{LIGO}%
  \BibitemOpen
  \bibfield  {author} {\bibinfo {author} {\bibfnamefont {B.~P.}\ \bibnamefont {Abbott}} \emph {et~al.} (\bibinfo {collaboration} {LIGO Scientific Collaboration and Virgo Collaboration}),\ }\bibfield  {title} {\bibinfo {title} {Observation of gravitational waves from a binary black hole merger},\ }\href {https://doi.org/10.1103/PhysRevLett.116.061102} {\bibfield  {journal} {\bibinfo  {journal} {Phys. Rev. Lett.}\ }\textbf {\bibinfo {volume} {116}},\ \bibinfo {pages} {061102} (\bibinfo {year} {2016})}\BibitemShut {NoStop}%
\bibitem [{\citenamefont {Agazie}\ \emph {et~al.}(2023)\citenamefont {Agazie} \emph {et~al.}}]{NANOGrav}%
  \BibitemOpen
  \bibfield  {author} {\bibinfo {author} {\bibfnamefont {G.}~\bibnamefont {Agazie}} \emph {et~al.},\ }\bibfield  {title} {\bibinfo {title} {The {NANOG}rav 15 yr data set: Evidence for a gravitational-wave background},\ }\href {https://doi.org/10.3847/2041-8213/acdac6} {\bibfield  {journal} {\bibinfo  {journal} {The Astrophysical Journal Letters}\ }\textbf {\bibinfo {volume} {951}},\ \bibinfo {pages} {L8} (\bibinfo {year} {2023})}\BibitemShut {NoStop}%
\bibitem [{\citenamefont {Pretorius}(2005)}]{Pretorius:2005gq}%
  \BibitemOpen
  \bibfield  {author} {\bibinfo {author} {\bibfnamefont {F.}~\bibnamefont {Pretorius}},\ }\bibfield  {title} {\bibinfo {title} {{Evolution of binary black hole spacetimes}},\ }\href {https://doi.org/10.1103/PhysRevLett.95.121101} {\bibfield  {journal} {\bibinfo  {journal} {Phys. Rev. Lett.}\ }\textbf {\bibinfo {volume} {95}},\ \bibinfo {pages} {121101} (\bibinfo {year} {2005})},\ \Eprint {https://arxiv.org/abs/gr-qc/0507014} {arXiv:gr-qc/0507014} \BibitemShut {NoStop}%
\bibitem [{\citenamefont {Lehner}\ and\ \citenamefont {Pretorius}(2014)}]{Lehner_2014}%
  \BibitemOpen
  \bibfield  {author} {\bibinfo {author} {\bibfnamefont {L.}~\bibnamefont {Lehner}}\ and\ \bibinfo {author} {\bibfnamefont {F.}~\bibnamefont {Pretorius}},\ }\bibfield  {title} {\bibinfo {title} {Numerical relativity and astrophysics},\ }\href {https://doi.org/10.1146/annurev-astro-081913-040031} {\bibfield  {journal} {\bibinfo  {journal} {Annual Review of Astronomy and Astrophysics}\ }\textbf {\bibinfo {volume} {52}},\ \bibinfo {pages} {661} (\bibinfo {year} {2014})}\BibitemShut {NoStop}%
\bibitem [{\citenamefont {Cardoso}\ \emph {et~al.}(2015)\citenamefont {Cardoso}, \citenamefont {Gualtieri}, \citenamefont {Herdeiro},\ and\ \citenamefont {Sperhake}}]{Cardoso_2015}%
  \BibitemOpen
  \bibfield  {author} {\bibinfo {author} {\bibfnamefont {V.}~\bibnamefont {Cardoso}}, \bibinfo {author} {\bibfnamefont {L.}~\bibnamefont {Gualtieri}}, \bibinfo {author} {\bibfnamefont {C.}~\bibnamefont {Herdeiro}},\ and\ \bibinfo {author} {\bibfnamefont {U.}~\bibnamefont {Sperhake}},\ }\bibfield  {title} {\bibinfo {title} {Exploring new physics frontiers through numerical relativity},\ }\bibfield  {journal} {\bibinfo  {journal} {Living Reviews in Relativity}\ }\textbf {\bibinfo {volume} {18}},\ \href {https://doi.org/10.1007/lrr-2015-1} {10.1007/lrr-2015-1} (\bibinfo {year} {2015})\BibitemShut {NoStop}%
\bibitem [{\citenamefont {Poisson}\ \emph {et~al.}(2011)\citenamefont {Poisson}, \citenamefont {Pound},\ and\ \citenamefont {Vega}}]{PoissonReview}%
  \BibitemOpen
  \bibfield  {author} {\bibinfo {author} {\bibfnamefont {E.}~\bibnamefont {Poisson}}, \bibinfo {author} {\bibfnamefont {A.}~\bibnamefont {Pound}},\ and\ \bibinfo {author} {\bibfnamefont {I.}~\bibnamefont {Vega}},\ }\bibfield  {title} {\bibinfo {title} {The motion of point particles in curved spacetime},\ }\bibfield  {journal} {\bibinfo  {journal} {Living Reviews in Relativity}\ }\textbf {\bibinfo {volume} {14}},\ \href {https://doi.org/10.12942/lrr-2011-7} {10.12942/lrr-2011-7} (\bibinfo {year} {2011})\BibitemShut {NoStop}%
\bibitem [{\citenamefont {Pound}(2015)}]{PoundReview}%
  \BibitemOpen
  \bibfield  {author} {\bibinfo {author} {\bibfnamefont {A.}~\bibnamefont {Pound}},\ }\bibfield  {title} {\bibinfo {title} {Motion of small objects in curved spacetimes: An introduction to gravitational self-force},\ }in\ \href {https://doi.org/10.1007/978-3-319-18335-0_13} {\emph {\bibinfo {booktitle} {Fundamental Theories of Physics}}}\ (\bibinfo  {publisher} {Springer International Publishing},\ \bibinfo {year} {2015})\ pp.\ \bibinfo {pages} {399--486}\BibitemShut {NoStop}%
\bibitem [{\citenamefont {Barack}\ and\ \citenamefont {Pound}(2018)}]{BarackReview}%
  \BibitemOpen
  \bibfield  {author} {\bibinfo {author} {\bibfnamefont {L.}~\bibnamefont {Barack}}\ and\ \bibinfo {author} {\bibfnamefont {A.}~\bibnamefont {Pound}},\ }\bibfield  {title} {\bibinfo {title} {Self-force and radiation reaction in general relativity},\ }\href {https://doi.org/10.1088/1361-6633/aae552} {\bibfield  {journal} {\bibinfo  {journal} {Reports on Progress in Physics}\ }\textbf {\bibinfo {volume} {82}},\ \bibinfo {pages} {016904} (\bibinfo {year} {2018})}\BibitemShut {NoStop}%
\bibitem [{\citenamefont {Buonanno}\ and\ \citenamefont {Damour}(1999)}]{Buonanno:1998gg}%
  \BibitemOpen
  \bibfield  {author} {\bibinfo {author} {\bibfnamefont {A.}~\bibnamefont {Buonanno}}\ and\ \bibinfo {author} {\bibfnamefont {T.}~\bibnamefont {Damour}},\ }\bibfield  {title} {\bibinfo {title} {{Effective one-body approach to general relativistic two-body dynamics}},\ }\href {https://doi.org/10.1103/PhysRevD.59.084006} {\bibfield  {journal} {\bibinfo  {journal} {Phys. Rev. D}\ }\textbf {\bibinfo {volume} {59}},\ \bibinfo {pages} {084006} (\bibinfo {year} {1999})},\ \Eprint {https://arxiv.org/abs/gr-qc/9811091} {arXiv:gr-qc/9811091} \BibitemShut {NoStop}%
\bibitem [{\citenamefont {Blanchet}(2014)}]{Blanchet2014}%
  \BibitemOpen
  \bibfield  {author} {\bibinfo {author} {\bibfnamefont {L.}~\bibnamefont {Blanchet}},\ }\bibfield  {title} {\bibinfo {title} {Gravitational radiation from post-newtonian sources and inspiralling compact binaries},\ }\bibfield  {journal} {\bibinfo  {journal} {Living Reviews in Relativity}\ }\textbf {\bibinfo {volume} {17}},\ \href {https://doi.org/10.12942/lrr-2014-2} {10.12942/lrr-2014-2} (\bibinfo {year} {2014})\BibitemShut {NoStop}%
\bibitem [{\citenamefont {Goldberger}\ and\ \citenamefont {Rothstein}(2006{\natexlab{a}})}]{NRGR}%
  \BibitemOpen
  \bibfield  {author} {\bibinfo {author} {\bibfnamefont {W.~D.}\ \bibnamefont {Goldberger}}\ and\ \bibinfo {author} {\bibfnamefont {I.~Z.}\ \bibnamefont {Rothstein}},\ }\bibfield  {title} {\bibinfo {title} {Effective field theory of gravity for extended objects},\ }\bibfield  {journal} {\bibinfo  {journal} {Physical Review D}\ }\textbf {\bibinfo {volume} {73}},\ \href {https://doi.org/10.1103/physrevd.73.104029} {10.1103/physrevd.73.104029} (\bibinfo {year} {2006}{\natexlab{a}})\BibitemShut {NoStop}%
\bibitem [{\citenamefont {Bertotti}\ and\ \citenamefont {Plebanski}(1960)}]{Bertotti1960}%
  \BibitemOpen
  \bibfield  {author} {\bibinfo {author} {\bibfnamefont {B.}~\bibnamefont {Bertotti}}\ and\ \bibinfo {author} {\bibfnamefont {J.}~\bibnamefont {Plebanski}},\ }\bibfield  {title} {\bibinfo {title} {Theory of gravitational perturbations in the fast motion approximation},\ }\href {https://doi.org/https://doi.org/10.1016/0003-4916(60)90132-9} {\bibfield  {journal} {\bibinfo  {journal} {Annals of Physics}\ }\textbf {\bibinfo {volume} {11}},\ \bibinfo {pages} {169} (\bibinfo {year} {1960})}\BibitemShut {NoStop}%
\bibitem [{\citenamefont {Westpfahl}\ and\ \citenamefont {Goller}(1979)}]{Westpfahl1979}%
  \BibitemOpen
  \bibfield  {author} {\bibinfo {author} {\bibfnamefont {K.}~\bibnamefont {Westpfahl}}\ and\ \bibinfo {author} {\bibfnamefont {M.}~\bibnamefont {Goller}},\ }\bibfield  {title} {\bibinfo {title} {Gravitational scattering of two relativistic particles in post-linear approximation},\ }\href {https://doi.org/https://doi.org/10.1007/BF02817047} {\bibfield  {journal} {\bibinfo  {journal} {Lett. Nuovo Cimento}\ }\textbf {\bibinfo {volume} {26}},\ \bibinfo {pages} {573} (\bibinfo {year} {1979})}\BibitemShut {NoStop}%
\bibitem [{\citenamefont {Westpfahl}(1985)}]{Westpfahl1985}%
  \BibitemOpen
  \bibfield  {author} {\bibinfo {author} {\bibfnamefont {K.}~\bibnamefont {Westpfahl}},\ }\bibfield  {title} {\bibinfo {title} {High-speed scattering of charged and uncharged particles in general relativity},\ }\href {https://doi.org/https://doi.org/10.1002/prop.2190330802} {\bibfield  {journal} {\bibinfo  {journal} {Fortschritte der Physik/Progress of Physics}\ }\textbf {\bibinfo {volume} {33}},\ \bibinfo {pages} {417} (\bibinfo {year} {1985})}\BibitemShut {NoStop}%
\bibitem [{\citenamefont {Damour}(2016)}]{Damour:2016gwp}%
  \BibitemOpen
  \bibfield  {author} {\bibinfo {author} {\bibfnamefont {T.}~\bibnamefont {Damour}},\ }\bibfield  {title} {\bibinfo {title} {{Gravitational scattering, post-Minkowskian approximation and Effective One-Body theory}},\ }\href {https://doi.org/10.1103/PhysRevD.94.104015} {\bibfield  {journal} {\bibinfo  {journal} {Phys. Rev. D}\ }\textbf {\bibinfo {volume} {94}},\ \bibinfo {pages} {104015} (\bibinfo {year} {2016})},\ \Eprint {https://arxiv.org/abs/1609.00354} {arXiv:1609.00354 [gr-qc]} \BibitemShut {NoStop}%
\bibitem [{\citenamefont {Damour}(2018)}]{Damour2018-pmeob2}%
  \BibitemOpen
  \bibfield  {author} {\bibinfo {author} {\bibfnamefont {T.}~\bibnamefont {Damour}},\ }\bibfield  {title} {\bibinfo {title} {High-energy gravitational scattering and the general relativistic two-body problem},\ }\bibfield  {journal} {\bibinfo  {journal} {Physical Review D}\ }\textbf {\bibinfo {volume} {97}},\ \href {https://doi.org/10.1103/physrevd.97.044038} {10.1103/physrevd.97.044038} (\bibinfo {year} {2018})\BibitemShut {NoStop}%
\bibitem [{\citenamefont {Cheung}\ \emph {et~al.}(2018{\natexlab{a}})\citenamefont {Cheung}, \citenamefont {Rothstein},\ and\ \citenamefont {Solon}}]{Cheung2018-2PM}%
  \BibitemOpen
  \bibfield  {author} {\bibinfo {author} {\bibfnamefont {C.}~\bibnamefont {Cheung}}, \bibinfo {author} {\bibfnamefont {I.~Z.}\ \bibnamefont {Rothstein}},\ and\ \bibinfo {author} {\bibfnamefont {M.~P.}\ \bibnamefont {Solon}},\ }\bibfield  {title} {\bibinfo {title} {From scattering amplitudes to classical potentials in the post-minkowskian expansion},\ }\bibfield  {journal} {\bibinfo  {journal} {Physical Review Letters}\ }\textbf {\bibinfo {volume} {121}},\ \href {https://doi.org/10.1103/physrevlett.121.251101} {10.1103/physrevlett.121.251101} (\bibinfo {year} {2018}{\natexlab{a}})\BibitemShut {NoStop}%
\bibitem [{\citenamefont {Barack}\ \emph {et~al.}(2010)\citenamefont {Barack}, \citenamefont {Damour},\ and\ \citenamefont {Sago}}]{Barack:2010ny}%
  \BibitemOpen
  \bibfield  {author} {\bibinfo {author} {\bibfnamefont {L.}~\bibnamefont {Barack}}, \bibinfo {author} {\bibfnamefont {T.}~\bibnamefont {Damour}},\ and\ \bibinfo {author} {\bibfnamefont {N.}~\bibnamefont {Sago}},\ }\bibfield  {title} {\bibinfo {title} {{Precession effect of the gravitational self-force in a Schwarzschild spacetime and the effective one-body formalism}},\ }\href {https://doi.org/10.1103/PhysRevD.82.084036} {\bibfield  {journal} {\bibinfo  {journal} {Phys. Rev. D}\ }\textbf {\bibinfo {volume} {82}},\ \bibinfo {pages} {084036} (\bibinfo {year} {2010})},\ \Eprint {https://arxiv.org/abs/1008.0935} {arXiv:1008.0935 [gr-qc]} \BibitemShut {NoStop}%
\bibitem [{\citenamefont {Nagar}\ and\ \citenamefont {Albanesi}(2022)}]{Nagar:2022fep}%
  \BibitemOpen
  \bibfield  {author} {\bibinfo {author} {\bibfnamefont {A.}~\bibnamefont {Nagar}}\ and\ \bibinfo {author} {\bibfnamefont {S.}~\bibnamefont {Albanesi}},\ }\bibfield  {title} {\bibinfo {title} {{Toward a gravitational self-force-informed effective-one-body waveform model for nonprecessing, eccentric, large-mass-ratio inspirals}},\ }\href {https://doi.org/10.1103/PhysRevD.106.064049} {\bibfield  {journal} {\bibinfo  {journal} {Phys. Rev. D}\ }\textbf {\bibinfo {volume} {106}},\ \bibinfo {pages} {064049} (\bibinfo {year} {2022})},\ \Eprint {https://arxiv.org/abs/2207.14002} {arXiv:2207.14002 [gr-qc]} \BibitemShut {NoStop}%
\bibitem [{\citenamefont {Detweiler}(2008)}]{Detweiler:2008ft}%
  \BibitemOpen
  \bibfield  {author} {\bibinfo {author} {\bibfnamefont {S.~L.}\ \bibnamefont {Detweiler}},\ }\bibfield  {title} {\bibinfo {title} {{A Consequence of the gravitational self-force for circular orbits of the Schwarzschild geometry}},\ }\href {https://doi.org/10.1103/PhysRevD.77.124026} {\bibfield  {journal} {\bibinfo  {journal} {Phys. Rev. D}\ }\textbf {\bibinfo {volume} {77}},\ \bibinfo {pages} {124026} (\bibinfo {year} {2008})},\ \Eprint {https://arxiv.org/abs/0804.3529} {arXiv:0804.3529 [gr-qc]} \BibitemShut {NoStop}%
\bibitem [{\citenamefont {Barack}\ and\ \citenamefont {Sago}(2011)}]{Barack:2011ed}%
  \BibitemOpen
  \bibfield  {author} {\bibinfo {author} {\bibfnamefont {L.}~\bibnamefont {Barack}}\ and\ \bibinfo {author} {\bibfnamefont {N.}~\bibnamefont {Sago}},\ }\bibfield  {title} {\bibinfo {title} {{Beyond the geodesic approximation: conservative effects of the gravitational self-force in eccentric orbits around a Schwarzschild black hole}},\ }\href {https://doi.org/10.1103/PhysRevD.83.084023} {\bibfield  {journal} {\bibinfo  {journal} {Phys. Rev. D}\ }\textbf {\bibinfo {volume} {83}},\ \bibinfo {pages} {084023} (\bibinfo {year} {2011})},\ \Eprint {https://arxiv.org/abs/1101.3331} {arXiv:1101.3331 [gr-qc]} \BibitemShut {NoStop}%
\bibitem [{\citenamefont {Bini}\ and\ \citenamefont {Damour}(2013)}]{Bini:2013zaa}%
  \BibitemOpen
  \bibfield  {author} {\bibinfo {author} {\bibfnamefont {D.}~\bibnamefont {Bini}}\ and\ \bibinfo {author} {\bibfnamefont {T.}~\bibnamefont {Damour}},\ }\bibfield  {title} {\bibinfo {title} {{Analytical determination of the two-body gravitational interaction potential at the fourth post-Newtonian approximation}},\ }\href {https://doi.org/10.1103/PhysRevD.87.121501} {\bibfield  {journal} {\bibinfo  {journal} {Phys. Rev. D}\ }\textbf {\bibinfo {volume} {87}},\ \bibinfo {pages} {121501} (\bibinfo {year} {2013})},\ \Eprint {https://arxiv.org/abs/1305.4884} {arXiv:1305.4884 [gr-qc]} \BibitemShut {NoStop}%
\bibitem [{\citenamefont {Damour}\ \emph {et~al.}(2014)\citenamefont {Damour}, \citenamefont {Guercilena}, \citenamefont {Hinder}, \citenamefont {Hopper}, \citenamefont {Nagar},\ and\ \citenamefont {Rezzolla}}]{Damour:2014afa}%
  \BibitemOpen
  \bibfield  {author} {\bibinfo {author} {\bibfnamefont {T.}~\bibnamefont {Damour}}, \bibinfo {author} {\bibfnamefont {F.}~\bibnamefont {Guercilena}}, \bibinfo {author} {\bibfnamefont {I.}~\bibnamefont {Hinder}}, \bibinfo {author} {\bibfnamefont {S.}~\bibnamefont {Hopper}}, \bibinfo {author} {\bibfnamefont {A.}~\bibnamefont {Nagar}},\ and\ \bibinfo {author} {\bibfnamefont {L.}~\bibnamefont {Rezzolla}},\ }\bibfield  {title} {\bibinfo {title} {{Strong-Field Scattering of Two Black Holes: Numerics Versus Analytics}},\ }\href {https://doi.org/10.1103/PhysRevD.89.081503} {\bibfield  {journal} {\bibinfo  {journal} {Phys. Rev. D}\ }\textbf {\bibinfo {volume} {89}},\ \bibinfo {pages} {081503} (\bibinfo {year} {2014})},\ \Eprint {https://arxiv.org/abs/1402.7307} {arXiv:1402.7307 [gr-qc]} \BibitemShut {NoStop}%
\bibitem [{\citenamefont {van~de Meent}(2017)}]{vandeMeent:2016hel}%
  \BibitemOpen
  \bibfield  {author} {\bibinfo {author} {\bibfnamefont {M.}~\bibnamefont {van~de Meent}},\ }\bibfield  {title} {\bibinfo {title} {{Self-force corrections to the periapsis advance around a spinning black hole}},\ }\href {https://doi.org/10.1103/PhysRevLett.118.011101} {\bibfield  {journal} {\bibinfo  {journal} {Phys. Rev. Lett.}\ }\textbf {\bibinfo {volume} {118}},\ \bibinfo {pages} {011101} (\bibinfo {year} {2017})},\ \Eprint {https://arxiv.org/abs/1610.03497} {arXiv:1610.03497 [gr-qc]} \BibitemShut {NoStop}%
\bibitem [{\citenamefont {Antonelli}\ \emph {et~al.}(2019)\citenamefont {Antonelli}, \citenamefont {Buonanno}, \citenamefont {Steinhoff}, \citenamefont {van~de Meent},\ and\ \citenamefont {Vines}}]{Antonelli:2019ytb}%
  \BibitemOpen
  \bibfield  {author} {\bibinfo {author} {\bibfnamefont {A.}~\bibnamefont {Antonelli}}, \bibinfo {author} {\bibfnamefont {A.}~\bibnamefont {Buonanno}}, \bibinfo {author} {\bibfnamefont {J.}~\bibnamefont {Steinhoff}}, \bibinfo {author} {\bibfnamefont {M.}~\bibnamefont {van~de Meent}},\ and\ \bibinfo {author} {\bibfnamefont {J.}~\bibnamefont {Vines}},\ }\bibfield  {title} {\bibinfo {title} {{Energetics of two-body Hamiltonians in post-Minkowskian gravity}},\ }\href {https://doi.org/10.1103/PhysRevD.99.104004} {\bibfield  {journal} {\bibinfo  {journal} {Phys. Rev. D}\ }\textbf {\bibinfo {volume} {99}},\ \bibinfo {pages} {104004} (\bibinfo {year} {2019})},\ \Eprint {https://arxiv.org/abs/1901.07102} {arXiv:1901.07102 [gr-qc]} \BibitemShut {NoStop}%
\bibitem [{\citenamefont {Bini}\ \emph {et~al.}(2019)\citenamefont {Bini}, \citenamefont {Damour},\ and\ \citenamefont {Geralico}}]{Bini:2019nra}%
  \BibitemOpen
  \bibfield  {author} {\bibinfo {author} {\bibfnamefont {D.}~\bibnamefont {Bini}}, \bibinfo {author} {\bibfnamefont {T.}~\bibnamefont {Damour}},\ and\ \bibinfo {author} {\bibfnamefont {A.}~\bibnamefont {Geralico}},\ }\bibfield  {title} {\bibinfo {title} {{Novel approach to binary dynamics: application to the fifth post-Newtonian level}},\ }\href {https://doi.org/10.1103/PhysRevLett.123.231104} {\bibfield  {journal} {\bibinfo  {journal} {Phys. Rev. Lett.}\ }\textbf {\bibinfo {volume} {123}},\ \bibinfo {pages} {231104} (\bibinfo {year} {2019})},\ \Eprint {https://arxiv.org/abs/1909.02375} {arXiv:1909.02375 [gr-qc]} \BibitemShut {NoStop}%
\bibitem [{\citenamefont {Bini}\ \emph {et~al.}(2020)\citenamefont {Bini}, \citenamefont {Damour},\ and\ \citenamefont {Geralico}}]{Bini:2020wpo}%
  \BibitemOpen
  \bibfield  {author} {\bibinfo {author} {\bibfnamefont {D.}~\bibnamefont {Bini}}, \bibinfo {author} {\bibfnamefont {T.}~\bibnamefont {Damour}},\ and\ \bibinfo {author} {\bibfnamefont {A.}~\bibnamefont {Geralico}},\ }\bibfield  {title} {\bibinfo {title} {{Binary dynamics at the fifth and fifth-and-a-half post-Newtonian orders}},\ }\href {https://doi.org/10.1103/PhysRevD.102.024062} {\bibfield  {journal} {\bibinfo  {journal} {Phys. Rev. D}\ }\textbf {\bibinfo {volume} {102}},\ \bibinfo {pages} {024062} (\bibinfo {year} {2020})},\ \Eprint {https://arxiv.org/abs/2003.11891} {arXiv:2003.11891 [gr-qc]} \BibitemShut {NoStop}%
\bibitem [{\citenamefont {Bini}\ \emph {et~al.}(2021)\citenamefont {Bini}, \citenamefont {Damour}, \citenamefont {Geralico}, \citenamefont {Laporta},\ and\ \citenamefont {Mastrolia}}]{Bini:2020rzn}%
  \BibitemOpen
  \bibfield  {author} {\bibinfo {author} {\bibfnamefont {D.}~\bibnamefont {Bini}}, \bibinfo {author} {\bibfnamefont {T.}~\bibnamefont {Damour}}, \bibinfo {author} {\bibfnamefont {A.}~\bibnamefont {Geralico}}, \bibinfo {author} {\bibfnamefont {S.}~\bibnamefont {Laporta}},\ and\ \bibinfo {author} {\bibfnamefont {P.}~\bibnamefont {Mastrolia}},\ }\bibfield  {title} {\bibinfo {title} {{Gravitational scattering at the seventh order in $G$: nonlocal contribution at the sixth post-Newtonian accuracy}},\ }\href {https://doi.org/10.1103/PhysRevD.103.044038} {\bibfield  {journal} {\bibinfo  {journal} {Phys. Rev. D}\ }\textbf {\bibinfo {volume} {103}},\ \bibinfo {pages} {044038} (\bibinfo {year} {2021})},\ \Eprint {https://arxiv.org/abs/2012.12918} {arXiv:2012.12918 [gr-qc]} \BibitemShut {NoStop}%
\bibitem [{\citenamefont {Gralla}\ and\ \citenamefont {Lobo}(2022)}]{Gralla:2021qaf}%
  \BibitemOpen
  \bibfield  {author} {\bibinfo {author} {\bibfnamefont {S.~E.}\ \bibnamefont {Gralla}}\ and\ \bibinfo {author} {\bibfnamefont {K.}~\bibnamefont {Lobo}},\ }\bibfield  {title} {\bibinfo {title} {{Self-force effects in post-Minkowskian scattering}},\ }\href {https://doi.org/10.1088/1361-6382/ac5d88} {\bibfield  {journal} {\bibinfo  {journal} {Class. Quant. Grav.}\ }\textbf {\bibinfo {volume} {39}},\ \bibinfo {pages} {095001} (\bibinfo {year} {2022})},\ \Eprint {https://arxiv.org/abs/2110.08681} {arXiv:2110.08681 [gr-qc]} \BibitemShut {NoStop}%
\bibitem [{\citenamefont {Long}\ and\ \citenamefont {Barack}(2021)}]{Long:2021ufh}%
  \BibitemOpen
  \bibfield  {author} {\bibinfo {author} {\bibfnamefont {O.}~\bibnamefont {Long}}\ and\ \bibinfo {author} {\bibfnamefont {L.}~\bibnamefont {Barack}},\ }\bibfield  {title} {\bibinfo {title} {{Time-domain metric reconstruction for hyperbolic scattering}},\ }\href {https://doi.org/10.1103/PhysRevD.104.024014} {\bibfield  {journal} {\bibinfo  {journal} {Phys. Rev. D}\ }\textbf {\bibinfo {volume} {104}},\ \bibinfo {pages} {024014} (\bibinfo {year} {2021})},\ \Eprint {https://arxiv.org/abs/2105.05630} {arXiv:2105.05630 [gr-qc]} \BibitemShut {NoStop}%
\bibitem [{\citenamefont {Khalil}\ \emph {et~al.}(2022)\citenamefont {Khalil}, \citenamefont {Buonanno}, \citenamefont {Steinhoff},\ and\ \citenamefont {Vines}}]{Khalil:2022ylj}%
  \BibitemOpen
  \bibfield  {author} {\bibinfo {author} {\bibfnamefont {M.}~\bibnamefont {Khalil}}, \bibinfo {author} {\bibfnamefont {A.}~\bibnamefont {Buonanno}}, \bibinfo {author} {\bibfnamefont {J.}~\bibnamefont {Steinhoff}},\ and\ \bibinfo {author} {\bibfnamefont {J.}~\bibnamefont {Vines}},\ }\bibfield  {title} {\bibinfo {title} {{Energetics and scattering of gravitational two-body systems at fourth post-Minkowskian order}},\ }\href {https://doi.org/10.1103/PhysRevD.106.024042} {\bibfield  {journal} {\bibinfo  {journal} {Phys. Rev. D}\ }\textbf {\bibinfo {volume} {106}},\ \bibinfo {pages} {024042} (\bibinfo {year} {2022})},\ \Eprint {https://arxiv.org/abs/2204.05047} {arXiv:2204.05047 [gr-qc]} \BibitemShut {NoStop}%
\bibitem [{\citenamefont {Barack}\ and\ \citenamefont {Long}(2022)}]{Barack:2022pde}%
  \BibitemOpen
  \bibfield  {author} {\bibinfo {author} {\bibfnamefont {L.}~\bibnamefont {Barack}}\ and\ \bibinfo {author} {\bibfnamefont {O.}~\bibnamefont {Long}},\ }\bibfield  {title} {\bibinfo {title} {{Self-force correction to the deflection angle in black-hole scattering: A scalar charge toy model}},\ }\href {https://doi.org/10.1103/PhysRevD.106.104031} {\bibfield  {journal} {\bibinfo  {journal} {Phys. Rev. D}\ }\textbf {\bibinfo {volume} {106}},\ \bibinfo {pages} {104031} (\bibinfo {year} {2022})},\ \Eprint {https://arxiv.org/abs/2209.03740} {arXiv:2209.03740 [gr-qc]} \BibitemShut {NoStop}%
\bibitem [{\citenamefont {Barack}\ \emph {et~al.}(2023)\citenamefont {Barack} \emph {et~al.}}]{Barack:2023oqp}%
  \BibitemOpen
  \bibfield  {author} {\bibinfo {author} {\bibfnamefont {L.}~\bibnamefont {Barack}} \emph {et~al.},\ }\bibfield  {title} {\bibinfo {title} {{Comparison of post-Minkowskian and self-force expansions: Scattering in a scalar charge toy model}},\ }\href {https://doi.org/10.1103/PhysRevD.108.024025} {\bibfield  {journal} {\bibinfo  {journal} {Phys. Rev. D}\ }\textbf {\bibinfo {volume} {108}},\ \bibinfo {pages} {024025} (\bibinfo {year} {2023})},\ \Eprint {https://arxiv.org/abs/2304.09200} {arXiv:2304.09200 [hep-th]} \BibitemShut {NoStop}%
\bibitem [{\citenamefont {Whittall}\ and\ \citenamefont {Barack}(2023)}]{Whittall:2023xjp}%
  \BibitemOpen
  \bibfield  {author} {\bibinfo {author} {\bibfnamefont {C.}~\bibnamefont {Whittall}}\ and\ \bibinfo {author} {\bibfnamefont {L.}~\bibnamefont {Barack}},\ }\bibfield  {title} {\bibinfo {title} {{Frequency-domain approach to self-force in hyperbolic scattering}},\ }\href {https://doi.org/10.1103/PhysRevD.108.064017} {\bibfield  {journal} {\bibinfo  {journal} {Phys. Rev. D}\ }\textbf {\bibinfo {volume} {108}},\ \bibinfo {pages} {064017} (\bibinfo {year} {2023})},\ \Eprint {https://arxiv.org/abs/2305.09724} {arXiv:2305.09724 [gr-qc]} \BibitemShut {NoStop}%
\bibitem [{\citenamefont {Galley}\ and\ \citenamefont {Hu}(2009)}]{Galley_2009}%
  \BibitemOpen
  \bibfield  {author} {\bibinfo {author} {\bibfnamefont {C.~R.}\ \bibnamefont {Galley}}\ and\ \bibinfo {author} {\bibfnamefont {B.~L.}\ \bibnamefont {Hu}},\ }\bibfield  {title} {\bibinfo {title} {Self-force on extreme mass ratio inspirals via curved spacetime effective field theory},\ }\bibfield  {journal} {\bibinfo  {journal} {Physical Review D}\ }\textbf {\bibinfo {volume} {79}},\ \href {https://doi.org/10.1103/physrevd.79.064002} {10.1103/physrevd.79.064002} (\bibinfo {year} {2009})\BibitemShut {NoStop}%
\bibitem [{\citenamefont {Elvang}\ and\ \citenamefont {Huang}(2013)}]{ElvangHuang}%
  \BibitemOpen
  \bibfield  {author} {\bibinfo {author} {\bibfnamefont {H.}~\bibnamefont {Elvang}}\ and\ \bibinfo {author} {\bibfnamefont {Y.-t.}\ \bibnamefont {Huang}},\ }\bibfield  {title} {\bibinfo {title} {{Scattering Amplitudes}},\ }\href@noop {} {\  (\bibinfo {year} {2013})},\ \Eprint {https://arxiv.org/abs/1308.1697} {arXiv:1308.1697 [hep-th]} \BibitemShut {NoStop}%
\bibitem [{\citenamefont {Dixon}(2014)}]{Dixon:2013uaa}%
  \BibitemOpen
  \bibfield  {author} {\bibinfo {author} {\bibfnamefont {L.~J.}\ \bibnamefont {Dixon}},\ }\bibfield  {title} {\bibinfo {title} {{A brief introduction to modern amplitude methods}},\ }in\ \href {https://doi.org/10.5170/CERN-2014-008.31} {\emph {\bibinfo {booktitle} {{Theoretical Advanced Study Institute in Elementary Particle Physics}: {Particle Physics: The Higgs Boson and Beyond}}}}\ (\bibinfo {year} {2014})\ pp.\ \bibinfo {pages} {31--67},\ \Eprint {https://arxiv.org/abs/1310.5353} {arXiv:1310.5353 [hep-ph]} \BibitemShut {NoStop}%
\bibitem [{\citenamefont {Cheung}(2018)}]{Cheung:2017pzi}%
  \BibitemOpen
  \bibfield  {author} {\bibinfo {author} {\bibfnamefont {C.}~\bibnamefont {Cheung}},\ }\bibinfo {title} {{TASI Lectures on Scattering Amplitudes}},\ in\ \href {https://doi.org/10.1142/9789813233348_0008} {\emph {\bibinfo {booktitle} {{Proceedings, Theoretical Advanced Study Institute in Elementary Particle Physics : Anticipating the Next Discoveries in Particle Physics (TASI 2016)}: {Boulder, CO, USA, June 6-July 1, 2016}}}},\ \bibinfo {editor} {edited by\ \bibinfo {editor} {\bibfnamefont {R.}~\bibnamefont {Essig}}\ and\ \bibinfo {editor} {\bibfnamefont {I.}~\bibnamefont {Low}}}\ (\bibinfo {year} {2018})\ pp.\ \bibinfo {pages} {571--623},\ \Eprint {https://arxiv.org/abs/1708.03872} {arXiv:1708.03872 [hep-ph]} \BibitemShut {NoStop}%
\bibitem [{\citenamefont {Chetyrkin}\ and\ \citenamefont {Tkachov}(1981)}]{Chetyrkin:1981qh}%
  \BibitemOpen
  \bibfield  {author} {\bibinfo {author} {\bibfnamefont {K.~G.}\ \bibnamefont {Chetyrkin}}\ and\ \bibinfo {author} {\bibfnamefont {F.~V.}\ \bibnamefont {Tkachov}},\ }\bibfield  {title} {\bibinfo {title} {{Integration by Parts: The Algorithm to Calculate beta Functions in 4 Loops}},\ }\href {https://doi.org/10.1016/0550-3213(81)90199-1} {\bibfield  {journal} {\bibinfo  {journal} {Nucl. Phys. B}\ }\textbf {\bibinfo {volume} {192}},\ \bibinfo {pages} {159} (\bibinfo {year} {1981})}\BibitemShut {NoStop}%
\bibitem [{\citenamefont {Henn}(2013)}]{Henn:2013pwa}%
  \BibitemOpen
  \bibfield  {author} {\bibinfo {author} {\bibfnamefont {J.~M.}\ \bibnamefont {Henn}},\ }\bibfield  {title} {\bibinfo {title} {{Multiloop integrals in dimensional regularization made simple}},\ }\href {https://doi.org/10.1103/PhysRevLett.110.251601} {\bibfield  {journal} {\bibinfo  {journal} {Phys. Rev. Lett.}\ }\textbf {\bibinfo {volume} {110}},\ \bibinfo {pages} {251601} (\bibinfo {year} {2013})},\ \Eprint {https://arxiv.org/abs/1304.1806} {arXiv:1304.1806 [hep-th]} \BibitemShut {NoStop}%
\bibitem [{\citenamefont {Kosower}\ \emph {et~al.}(2019)\citenamefont {Kosower}, \citenamefont {Maybee},\ and\ \citenamefont {O'Connell}}]{Kosower:2018adc}%
  \BibitemOpen
  \bibfield  {author} {\bibinfo {author} {\bibfnamefont {D.~A.}\ \bibnamefont {Kosower}}, \bibinfo {author} {\bibfnamefont {B.}~\bibnamefont {Maybee}},\ and\ \bibinfo {author} {\bibfnamefont {D.}~\bibnamefont {O'Connell}},\ }\bibfield  {title} {\bibinfo {title} {{Amplitudes, Observables, and Classical Scattering}},\ }\href {https://doi.org/10.1007/JHEP02(2019)137} {\bibfield  {journal} {\bibinfo  {journal} {JHEP}\ }\textbf {\bibinfo {volume} {02}},\ \bibinfo {pages} {137}},\ \Eprint {https://arxiv.org/abs/1811.10950} {arXiv:1811.10950 [hep-th]} \BibitemShut {NoStop}%
\bibitem [{\citenamefont {Bern}\ \emph {et~al.}(1994)\citenamefont {Bern}, \citenamefont {Dixon}, \citenamefont {Dunbar},\ and\ \citenamefont {Kosower}}]{Bern:1994zx}%
  \BibitemOpen
  \bibfield  {author} {\bibinfo {author} {\bibfnamefont {Z.}~\bibnamefont {Bern}}, \bibinfo {author} {\bibfnamefont {L.~J.}\ \bibnamefont {Dixon}}, \bibinfo {author} {\bibfnamefont {D.~C.}\ \bibnamefont {Dunbar}},\ and\ \bibinfo {author} {\bibfnamefont {D.~A.}\ \bibnamefont {Kosower}},\ }\bibfield  {title} {\bibinfo {title} {{One loop n point gauge theory amplitudes, unitarity and collinear limits}},\ }\href {https://doi.org/10.1016/0550-3213(94)90179-1} {\bibfield  {journal} {\bibinfo  {journal} {Nucl. Phys. B}\ }\textbf {\bibinfo {volume} {425}},\ \bibinfo {pages} {217} (\bibinfo {year} {1994})},\ \Eprint {https://arxiv.org/abs/hep-ph/9403226} {arXiv:hep-ph/9403226} \BibitemShut {NoStop}%
\bibitem [{\citenamefont {Bern}\ \emph {et~al.}(1995)\citenamefont {Bern}, \citenamefont {Dixon}, \citenamefont {Dunbar},\ and\ \citenamefont {Kosower}}]{Bern:1994cg}%
  \BibitemOpen
  \bibfield  {author} {\bibinfo {author} {\bibfnamefont {Z.}~\bibnamefont {Bern}}, \bibinfo {author} {\bibfnamefont {L.~J.}\ \bibnamefont {Dixon}}, \bibinfo {author} {\bibfnamefont {D.~C.}\ \bibnamefont {Dunbar}},\ and\ \bibinfo {author} {\bibfnamefont {D.~A.}\ \bibnamefont {Kosower}},\ }\bibfield  {title} {\bibinfo {title} {{Fusing gauge theory tree amplitudes into loop amplitudes}},\ }\href {https://doi.org/10.1016/0550-3213(94)00488-Z} {\bibfield  {journal} {\bibinfo  {journal} {Nucl. Phys. B}\ }\textbf {\bibinfo {volume} {435}},\ \bibinfo {pages} {59} (\bibinfo {year} {1995})},\ \Eprint {https://arxiv.org/abs/hep-ph/9409265} {arXiv:hep-ph/9409265} \BibitemShut {NoStop}%
\bibitem [{\citenamefont {'t~Hooft}\ and\ \citenamefont {Veltman}(1972)}]{tHooft:1972tcz}%
  \BibitemOpen
  \bibfield  {author} {\bibinfo {author} {\bibfnamefont {G.}~\bibnamefont {'t~Hooft}}\ and\ \bibinfo {author} {\bibfnamefont {M.~J.~G.}\ \bibnamefont {Veltman}},\ }\bibfield  {title} {\bibinfo {title} {{Regularization and Renormalization of Gauge Fields}},\ }\href {https://doi.org/10.1016/0550-3213(72)90279-9} {\bibfield  {journal} {\bibinfo  {journal} {Nucl. Phys. B}\ }\textbf {\bibinfo {volume} {44}},\ \bibinfo {pages} {189} (\bibinfo {year} {1972})}\BibitemShut {NoStop}%
\bibitem [{\citenamefont {Bollini}\ and\ \citenamefont {Giambiagi}(1972)}]{Bollini:1972ui}%
  \BibitemOpen
  \bibfield  {author} {\bibinfo {author} {\bibfnamefont {C.~G.}\ \bibnamefont {Bollini}}\ and\ \bibinfo {author} {\bibfnamefont {J.~J.}\ \bibnamefont {Giambiagi}},\ }\bibfield  {title} {\bibinfo {title} {{Dimensional Renormalization: The Number of Dimensions as a Regularizing Parameter}},\ }\href {https://doi.org/10.1007/BF02895558} {\bibfield  {journal} {\bibinfo  {journal} {Nuovo Cim. B}\ }\textbf {\bibinfo {volume} {12}},\ \bibinfo {pages} {20} (\bibinfo {year} {1972})}\BibitemShut {NoStop}%
\bibitem [{\citenamefont {Smirnov}(2012)}]{Smirnov:2012gma}%
  \BibitemOpen
  \bibfield  {author} {\bibinfo {author} {\bibfnamefont {V.~A.}\ \bibnamefont {Smirnov}},\ }\href {https://doi.org/10.1007/978-3-642-34886-0} {\emph {\bibinfo {title} {{Analytic tools for Feynman integrals}}}},\ Vol.\ \bibinfo {volume} {250}\ (\bibinfo {year} {2012})\BibitemShut {NoStop}%
\bibitem [{\citenamefont {Neill}\ and\ \citenamefont {Rothstein}(2013)}]{Smatrix}%
  \BibitemOpen
  \bibfield  {author} {\bibinfo {author} {\bibfnamefont {D.}~\bibnamefont {Neill}}\ and\ \bibinfo {author} {\bibfnamefont {I.~Z.}\ \bibnamefont {Rothstein}},\ }\bibfield  {title} {\bibinfo {title} {{Classical Space-Times from the {S} Matrix}},\ }\href {https://doi.org/10.1016/j.nuclphysb.2013.09.007} {\bibfield  {journal} {\bibinfo  {journal} {Nucl. Phys. B}\ }\textbf {\bibinfo {volume} {877}},\ \bibinfo {pages} {177} (\bibinfo {year} {2013})},\ \Eprint {https://arxiv.org/abs/1304.7263} {arXiv:1304.7263 [hep-th]} \BibitemShut {NoStop}%
\bibitem [{\citenamefont {Bjerrum-Bohr}\ \emph {et~al.}(2018)\citenamefont {Bjerrum-Bohr}, \citenamefont {Damgaard}, \citenamefont {Festuccia}, \citenamefont {Plant\'e},\ and\ \citenamefont {Vanhove}}]{Bjerrum-Bohr:2018xdl}%
  \BibitemOpen
  \bibfield  {author} {\bibinfo {author} {\bibfnamefont {N.~E.~J.}\ \bibnamefont {Bjerrum-Bohr}}, \bibinfo {author} {\bibfnamefont {P.~H.}\ \bibnamefont {Damgaard}}, \bibinfo {author} {\bibfnamefont {G.}~\bibnamefont {Festuccia}}, \bibinfo {author} {\bibfnamefont {L.}~\bibnamefont {Plant\'e}},\ and\ \bibinfo {author} {\bibfnamefont {P.}~\bibnamefont {Vanhove}},\ }\bibfield  {title} {\bibinfo {title} {{General Relativity from Scattering Amplitudes}},\ }\href {https://doi.org/10.1103/PhysRevLett.121.171601} {\bibfield  {journal} {\bibinfo  {journal} {Phys. Rev. Lett.}\ }\textbf {\bibinfo {volume} {121}},\ \bibinfo {pages} {171601} (\bibinfo {year} {2018})},\ \Eprint {https://arxiv.org/abs/1806.04920} {arXiv:1806.04920 [hep-th]} \BibitemShut {NoStop}%
\bibitem [{\citenamefont {Bern}\ \emph {et~al.}(2019{\natexlab{a}})\citenamefont {Bern}, \citenamefont {Cheung}, \citenamefont {Roiban}, \citenamefont {Shen}, \citenamefont {Solon},\ and\ \citenamefont {Zeng}}]{Bern:2019nnu}%
  \BibitemOpen
  \bibfield  {author} {\bibinfo {author} {\bibfnamefont {Z.}~\bibnamefont {Bern}}, \bibinfo {author} {\bibfnamefont {C.}~\bibnamefont {Cheung}}, \bibinfo {author} {\bibfnamefont {R.}~\bibnamefont {Roiban}}, \bibinfo {author} {\bibfnamefont {C.-H.}\ \bibnamefont {Shen}}, \bibinfo {author} {\bibfnamefont {M.~P.}\ \bibnamefont {Solon}},\ and\ \bibinfo {author} {\bibfnamefont {M.}~\bibnamefont {Zeng}},\ }\bibfield  {title} {\bibinfo {title} {{Scattering Amplitudes and the Conservative Hamiltonian for Binary Systems at Third Post-Minkowskian Order}},\ }\href {https://doi.org/10.1103/PhysRevLett.122.201603} {\bibfield  {journal} {\bibinfo  {journal} {Phys. Rev. Lett.}\ }\textbf {\bibinfo {volume} {122}},\ \bibinfo {pages} {201603} (\bibinfo {year} {2019}{\natexlab{a}})},\ \Eprint {https://arxiv.org/abs/1901.04424} {arXiv:1901.04424 [hep-th]} \BibitemShut {NoStop}%
\bibitem [{\citenamefont {Bern}\ \emph {et~al.}(2019{\natexlab{b}})\citenamefont {Bern}, \citenamefont {Cheung}, \citenamefont {Roiban}, \citenamefont {Shen}, \citenamefont {Solon},\ and\ \citenamefont {Zeng}}]{Bern:2019crd}%
  \BibitemOpen
  \bibfield  {author} {\bibinfo {author} {\bibfnamefont {Z.}~\bibnamefont {Bern}}, \bibinfo {author} {\bibfnamefont {C.}~\bibnamefont {Cheung}}, \bibinfo {author} {\bibfnamefont {R.}~\bibnamefont {Roiban}}, \bibinfo {author} {\bibfnamefont {C.-H.}\ \bibnamefont {Shen}}, \bibinfo {author} {\bibfnamefont {M.~P.}\ \bibnamefont {Solon}},\ and\ \bibinfo {author} {\bibfnamefont {M.}~\bibnamefont {Zeng}},\ }\bibfield  {title} {\bibinfo {title} {{Black Hole Binary Dynamics from the Double Copy and Effective Theory}},\ }\href {https://doi.org/10.1007/JHEP10(2019)206} {\bibfield  {journal} {\bibinfo  {journal} {JHEP}\ }\textbf {\bibinfo {volume} {10}},\ \bibinfo {pages} {206}},\ \Eprint {https://arxiv.org/abs/1908.01493} {arXiv:1908.01493 [hep-th]} \BibitemShut {NoStop}%
\bibitem [{\citenamefont {Bjerrum-Bohr}\ \emph {et~al.}(2020)\citenamefont {Bjerrum-Bohr}, \citenamefont {Cristofoli},\ and\ \citenamefont {Damgaard}}]{Bjerrum-Bohr:2019kec}%
  \BibitemOpen
  \bibfield  {author} {\bibinfo {author} {\bibfnamefont {N.~E.~J.}\ \bibnamefont {Bjerrum-Bohr}}, \bibinfo {author} {\bibfnamefont {A.}~\bibnamefont {Cristofoli}},\ and\ \bibinfo {author} {\bibfnamefont {P.~H.}\ \bibnamefont {Damgaard}},\ }\bibfield  {title} {\bibinfo {title} {{Post-Minkowskian Scattering Angle in Einstein Gravity}},\ }\href {https://doi.org/10.1007/JHEP08(2020)038} {\bibfield  {journal} {\bibinfo  {journal} {JHEP}\ }\textbf {\bibinfo {volume} {08}},\ \bibinfo {pages} {038}},\ \Eprint {https://arxiv.org/abs/1910.09366} {arXiv:1910.09366 [hep-th]} \BibitemShut {NoStop}%
\bibitem [{\citenamefont {K\"alin}\ and\ \citenamefont {Porto}(2020{\natexlab{a}})}]{Kalin:2019rwq}%
  \BibitemOpen
  \bibfield  {author} {\bibinfo {author} {\bibfnamefont {G.}~\bibnamefont {K\"alin}}\ and\ \bibinfo {author} {\bibfnamefont {R.~A.}\ \bibnamefont {Porto}},\ }\bibfield  {title} {\bibinfo {title} {{From Boundary Data to Bound States}},\ }\href {https://doi.org/10.1007/JHEP01(2020)072} {\bibfield  {journal} {\bibinfo  {journal} {JHEP}\ }\textbf {\bibinfo {volume} {01}},\ \bibinfo {pages} {072}},\ \Eprint {https://arxiv.org/abs/1910.03008} {arXiv:1910.03008 [hep-th]} \BibitemShut {NoStop}%
\bibitem [{\citenamefont {K\"alin}\ and\ \citenamefont {Porto}(2020{\natexlab{b}})}]{Kalin:2019inp}%
  \BibitemOpen
  \bibfield  {author} {\bibinfo {author} {\bibfnamefont {G.}~\bibnamefont {K\"alin}}\ and\ \bibinfo {author} {\bibfnamefont {R.~A.}\ \bibnamefont {Porto}},\ }\bibfield  {title} {\bibinfo {title} {{From boundary data to bound states. Part II. Scattering angle to dynamical invariants (with twist)}},\ }\href {https://doi.org/10.1007/JHEP02(2020)120} {\bibfield  {journal} {\bibinfo  {journal} {JHEP}\ }\textbf {\bibinfo {volume} {02}},\ \bibinfo {pages} {120}},\ \Eprint {https://arxiv.org/abs/1911.09130} {arXiv:1911.09130 [hep-th]} \BibitemShut {NoStop}%
\bibitem [{\citenamefont {K\"alin}\ \emph {et~al.}(2020)\citenamefont {K\"alin}, \citenamefont {Liu},\ and\ \citenamefont {Porto}}]{Kalin:2020fhe}%
  \BibitemOpen
  \bibfield  {author} {\bibinfo {author} {\bibfnamefont {G.}~\bibnamefont {K\"alin}}, \bibinfo {author} {\bibfnamefont {Z.}~\bibnamefont {Liu}},\ and\ \bibinfo {author} {\bibfnamefont {R.~A.}\ \bibnamefont {Porto}},\ }\bibfield  {title} {\bibinfo {title} {{Conservative Dynamics of Binary Systems to Third Post-Minkowskian Order from the Effective Field Theory Approach}},\ }\href {https://doi.org/10.1103/PhysRevLett.125.261103} {\bibfield  {journal} {\bibinfo  {journal} {Phys. Rev. Lett.}\ }\textbf {\bibinfo {volume} {125}},\ \bibinfo {pages} {261103} (\bibinfo {year} {2020})},\ \Eprint {https://arxiv.org/abs/2007.04977} {arXiv:2007.04977 [hep-th]} \BibitemShut {NoStop}%
\bibitem [{\citenamefont {Cheung}\ and\ \citenamefont {Solon}(2020)}]{Cheung:2020gyp}%
  \BibitemOpen
  \bibfield  {author} {\bibinfo {author} {\bibfnamefont {C.}~\bibnamefont {Cheung}}\ and\ \bibinfo {author} {\bibfnamefont {M.~P.}\ \bibnamefont {Solon}},\ }\bibfield  {title} {\bibinfo {title} {{Classical gravitational scattering at $ \mathcal{O} $(G$^{3}$) from Feynman diagrams}},\ }\href {https://doi.org/10.1007/JHEP06(2020)144} {\bibfield  {journal} {\bibinfo  {journal} {JHEP}\ }\textbf {\bibinfo {volume} {06}},\ \bibinfo {pages} {144}},\ \Eprint {https://arxiv.org/abs/2003.08351} {arXiv:2003.08351 [hep-th]} \BibitemShut {NoStop}%
\bibitem [{\citenamefont {Bjerrum-Bohr}\ \emph {et~al.}(2021{\natexlab{a}})\citenamefont {Bjerrum-Bohr}, \citenamefont {Damgaard}, \citenamefont {Plant\'e},\ and\ \citenamefont {Vanhove}}]{Bjerrum-Bohr:2021din}%
  \BibitemOpen
  \bibfield  {author} {\bibinfo {author} {\bibfnamefont {N.~E.~J.}\ \bibnamefont {Bjerrum-Bohr}}, \bibinfo {author} {\bibfnamefont {P.~H.}\ \bibnamefont {Damgaard}}, \bibinfo {author} {\bibfnamefont {L.}~\bibnamefont {Plant\'e}},\ and\ \bibinfo {author} {\bibfnamefont {P.}~\bibnamefont {Vanhove}},\ }\bibfield  {title} {\bibinfo {title} {{The amplitude for classical gravitational scattering at third Post-Minkowskian order}},\ }\href {https://doi.org/10.1007/JHEP08(2021)172} {\bibfield  {journal} {\bibinfo  {journal} {JHEP}\ }\textbf {\bibinfo {volume} {08}},\ \bibinfo {pages} {172}},\ \Eprint {https://arxiv.org/abs/2105.05218} {arXiv:2105.05218 [hep-th]} \BibitemShut {NoStop}%
\bibitem [{\citenamefont {Bern}\ \emph {et~al.}(2021)\citenamefont {Bern}, \citenamefont {Parra-Martinez}, \citenamefont {Roiban}, \citenamefont {Ruf}, \citenamefont {Shen}, \citenamefont {Solon},\ and\ \citenamefont {Zeng}}]{Bern:2021dqo}%
  \BibitemOpen
  \bibfield  {author} {\bibinfo {author} {\bibfnamefont {Z.}~\bibnamefont {Bern}}, \bibinfo {author} {\bibfnamefont {J.}~\bibnamefont {Parra-Martinez}}, \bibinfo {author} {\bibfnamefont {R.}~\bibnamefont {Roiban}}, \bibinfo {author} {\bibfnamefont {M.~S.}\ \bibnamefont {Ruf}}, \bibinfo {author} {\bibfnamefont {C.-H.}\ \bibnamefont {Shen}}, \bibinfo {author} {\bibfnamefont {M.~P.}\ \bibnamefont {Solon}},\ and\ \bibinfo {author} {\bibfnamefont {M.}~\bibnamefont {Zeng}},\ }\bibfield  {title} {\bibinfo {title} {{Scattering Amplitudes and Conservative Binary Dynamics at ${\cal O}(G^4)$}},\ }\href {https://doi.org/10.1103/PhysRevLett.126.171601} {\bibfield  {journal} {\bibinfo  {journal} {Phys. Rev. Lett.}\ }\textbf {\bibinfo {volume} {126}},\ \bibinfo {pages} {171601} (\bibinfo {year} {2021})},\ \Eprint {https://arxiv.org/abs/2101.07254} {arXiv:2101.07254 [hep-th]} \BibitemShut {NoStop}%
\bibitem [{\citenamefont {Bern}\ \emph {et~al.}(2022{\natexlab{a}})\citenamefont {Bern}, \citenamefont {Parra-Martinez}, \citenamefont {Roiban}, \citenamefont {Ruf}, \citenamefont {Shen}, \citenamefont {Solon},\ and\ \citenamefont {Zeng}}]{Bern:2021yeh}%
  \BibitemOpen
  \bibfield  {author} {\bibinfo {author} {\bibfnamefont {Z.}~\bibnamefont {Bern}}, \bibinfo {author} {\bibfnamefont {J.}~\bibnamefont {Parra-Martinez}}, \bibinfo {author} {\bibfnamefont {R.}~\bibnamefont {Roiban}}, \bibinfo {author} {\bibfnamefont {M.~S.}\ \bibnamefont {Ruf}}, \bibinfo {author} {\bibfnamefont {C.-H.}\ \bibnamefont {Shen}}, \bibinfo {author} {\bibfnamefont {M.~P.}\ \bibnamefont {Solon}},\ and\ \bibinfo {author} {\bibfnamefont {M.}~\bibnamefont {Zeng}},\ }\bibfield  {title} {\bibinfo {title} {{Scattering Amplitudes, the Tail Effect, and Conservative Binary Dynamics at O(G4)}},\ }\href {https://doi.org/10.1103/PhysRevLett.128.161103} {\bibfield  {journal} {\bibinfo  {journal} {Phys. Rev. Lett.}\ }\textbf {\bibinfo {volume} {128}},\ \bibinfo {pages} {161103} (\bibinfo {year} {2022}{\natexlab{a}})},\ \Eprint {https://arxiv.org/abs/2112.10750} {arXiv:2112.10750 [hep-th]} \BibitemShut {NoStop}%
\bibitem [{\citenamefont {Herrmann}\ \emph {et~al.}(2021{\natexlab{a}})\citenamefont {Herrmann}, \citenamefont {Parra-Martinez}, \citenamefont {Ruf},\ and\ \citenamefont {Zeng}}]{Herrmann:2021tct}%
  \BibitemOpen
  \bibfield  {author} {\bibinfo {author} {\bibfnamefont {E.}~\bibnamefont {Herrmann}}, \bibinfo {author} {\bibfnamefont {J.}~\bibnamefont {Parra-Martinez}}, \bibinfo {author} {\bibfnamefont {M.~S.}\ \bibnamefont {Ruf}},\ and\ \bibinfo {author} {\bibfnamefont {M.}~\bibnamefont {Zeng}},\ }\bibfield  {title} {\bibinfo {title} {{Radiative classical gravitational observables at $ \mathcal{O} $(G$^{3}$) from scattering amplitudes}},\ }\href {https://doi.org/10.1007/JHEP10(2021)148} {\bibfield  {journal} {\bibinfo  {journal} {JHEP}\ }\textbf {\bibinfo {volume} {10}},\ \bibinfo {pages} {148}},\ \Eprint {https://arxiv.org/abs/2104.03957} {arXiv:2104.03957 [hep-th]} \BibitemShut {NoStop}%
\bibitem [{\citenamefont {Di~Vecchia}\ \emph {et~al.}(2021{\natexlab{a}})\citenamefont {Di~Vecchia}, \citenamefont {Heissenberg}, \citenamefont {Russo},\ and\ \citenamefont {Veneziano}}]{DiVecchia:2021bdo}%
  \BibitemOpen
  \bibfield  {author} {\bibinfo {author} {\bibfnamefont {P.}~\bibnamefont {Di~Vecchia}}, \bibinfo {author} {\bibfnamefont {C.}~\bibnamefont {Heissenberg}}, \bibinfo {author} {\bibfnamefont {R.}~\bibnamefont {Russo}},\ and\ \bibinfo {author} {\bibfnamefont {G.}~\bibnamefont {Veneziano}},\ }\bibfield  {title} {\bibinfo {title} {{The eikonal approach to gravitational scattering and radiation at $ \mathcal{O} $(G$^{3}$)}},\ }\href {https://doi.org/10.1007/JHEP07(2021)169} {\bibfield  {journal} {\bibinfo  {journal} {JHEP}\ }\textbf {\bibinfo {volume} {07}},\ \bibinfo {pages} {169}},\ \Eprint {https://arxiv.org/abs/2104.03256} {arXiv:2104.03256 [hep-th]} \BibitemShut {NoStop}%
\bibitem [{\citenamefont {Herrmann}\ \emph {et~al.}(2021{\natexlab{b}})\citenamefont {Herrmann}, \citenamefont {Parra-Martinez}, \citenamefont {Ruf},\ and\ \citenamefont {Zeng}}]{Herrmann:2021lqe}%
  \BibitemOpen
  \bibfield  {author} {\bibinfo {author} {\bibfnamefont {E.}~\bibnamefont {Herrmann}}, \bibinfo {author} {\bibfnamefont {J.}~\bibnamefont {Parra-Martinez}}, \bibinfo {author} {\bibfnamefont {M.~S.}\ \bibnamefont {Ruf}},\ and\ \bibinfo {author} {\bibfnamefont {M.}~\bibnamefont {Zeng}},\ }\bibfield  {title} {\bibinfo {title} {{Gravitational Bremsstrahlung from Reverse Unitarity}},\ }\href {https://doi.org/10.1103/PhysRevLett.126.201602} {\bibfield  {journal} {\bibinfo  {journal} {Phys. Rev. Lett.}\ }\textbf {\bibinfo {volume} {126}},\ \bibinfo {pages} {201602} (\bibinfo {year} {2021}{\natexlab{b}})},\ \Eprint {https://arxiv.org/abs/2101.07255} {arXiv:2101.07255 [hep-th]} \BibitemShut {NoStop}%
\bibitem [{\citenamefont {Di~Vecchia}\ \emph {et~al.}(2021{\natexlab{b}})\citenamefont {Di~Vecchia}, \citenamefont {Heissenberg}, \citenamefont {Russo},\ and\ \citenamefont {Veneziano}}]{DiVecchia:2021ndb}%
  \BibitemOpen
  \bibfield  {author} {\bibinfo {author} {\bibfnamefont {P.}~\bibnamefont {Di~Vecchia}}, \bibinfo {author} {\bibfnamefont {C.}~\bibnamefont {Heissenberg}}, \bibinfo {author} {\bibfnamefont {R.}~\bibnamefont {Russo}},\ and\ \bibinfo {author} {\bibfnamefont {G.}~\bibnamefont {Veneziano}},\ }\bibfield  {title} {\bibinfo {title} {{Radiation Reaction from Soft Theorems}},\ }\href {https://doi.org/10.1016/j.physletb.2021.136379} {\bibfield  {journal} {\bibinfo  {journal} {Phys. Lett. B}\ }\textbf {\bibinfo {volume} {818}},\ \bibinfo {pages} {136379} (\bibinfo {year} {2021}{\natexlab{b}})},\ \Eprint {https://arxiv.org/abs/2101.05772} {arXiv:2101.05772 [hep-th]} \BibitemShut {NoStop}%
\bibitem [{\citenamefont {Jakobsen}\ \emph {et~al.}(2021)\citenamefont {Jakobsen}, \citenamefont {Mogull}, \citenamefont {Plefka},\ and\ \citenamefont {Steinhoff}}]{Jakobsen:2021smu}%
  \BibitemOpen
  \bibfield  {author} {\bibinfo {author} {\bibfnamefont {G.~U.}\ \bibnamefont {Jakobsen}}, \bibinfo {author} {\bibfnamefont {G.}~\bibnamefont {Mogull}}, \bibinfo {author} {\bibfnamefont {J.}~\bibnamefont {Plefka}},\ and\ \bibinfo {author} {\bibfnamefont {J.}~\bibnamefont {Steinhoff}},\ }\bibfield  {title} {\bibinfo {title} {{Classical Gravitational Bremsstrahlung from a Worldline Quantum Field Theory}},\ }\href {https://doi.org/10.1103/PhysRevLett.126.201103} {\bibfield  {journal} {\bibinfo  {journal} {Phys. Rev. Lett.}\ }\textbf {\bibinfo {volume} {126}},\ \bibinfo {pages} {201103} (\bibinfo {year} {2021})},\ \Eprint {https://arxiv.org/abs/2101.12688} {arXiv:2101.12688 [gr-qc]} \BibitemShut {NoStop}%
\bibitem [{\citenamefont {Mougiakakos}\ \emph {et~al.}(2021)\citenamefont {Mougiakakos}, \citenamefont {Riva},\ and\ \citenamefont {Vernizzi}}]{Mougiakakos:2021ckm}%
  \BibitemOpen
  \bibfield  {author} {\bibinfo {author} {\bibfnamefont {S.}~\bibnamefont {Mougiakakos}}, \bibinfo {author} {\bibfnamefont {M.~M.}\ \bibnamefont {Riva}},\ and\ \bibinfo {author} {\bibfnamefont {F.}~\bibnamefont {Vernizzi}},\ }\bibfield  {title} {\bibinfo {title} {{Gravitational Bremsstrahlung in the post-Minkowskian effective field theory}},\ }\href {https://doi.org/10.1103/PhysRevD.104.024041} {\bibfield  {journal} {\bibinfo  {journal} {Phys. Rev. D}\ }\textbf {\bibinfo {volume} {104}},\ \bibinfo {pages} {024041} (\bibinfo {year} {2021})},\ \Eprint {https://arxiv.org/abs/2102.08339} {arXiv:2102.08339 [gr-qc]} \BibitemShut {NoStop}%
\bibitem [{\citenamefont {Brandhuber}\ \emph {et~al.}(2021)\citenamefont {Brandhuber}, \citenamefont {Chen}, \citenamefont {Travaglini},\ and\ \citenamefont {Wen}}]{Brandhuber:2021eyq}%
  \BibitemOpen
  \bibfield  {author} {\bibinfo {author} {\bibfnamefont {A.}~\bibnamefont {Brandhuber}}, \bibinfo {author} {\bibfnamefont {G.}~\bibnamefont {Chen}}, \bibinfo {author} {\bibfnamefont {G.}~\bibnamefont {Travaglini}},\ and\ \bibinfo {author} {\bibfnamefont {C.}~\bibnamefont {Wen}},\ }\bibfield  {title} {\bibinfo {title} {{Classical gravitational scattering from a gauge-invariant double copy}},\ }\href {https://doi.org/10.1007/JHEP10(2021)118} {\bibfield  {journal} {\bibinfo  {journal} {JHEP}\ }\textbf {\bibinfo {volume} {10}},\ \bibinfo {pages} {118}},\ \Eprint {https://arxiv.org/abs/2108.04216} {arXiv:2108.04216 [hep-th]} \BibitemShut {NoStop}%
\bibitem [{\citenamefont {Manohar}\ \emph {et~al.}(2022)\citenamefont {Manohar}, \citenamefont {Ridgway},\ and\ \citenamefont {Shen}}]{Manohar:2022dea}%
  \BibitemOpen
  \bibfield  {author} {\bibinfo {author} {\bibfnamefont {A.~V.}\ \bibnamefont {Manohar}}, \bibinfo {author} {\bibfnamefont {A.~K.}\ \bibnamefont {Ridgway}},\ and\ \bibinfo {author} {\bibfnamefont {C.-H.}\ \bibnamefont {Shen}},\ }\bibfield  {title} {\bibinfo {title} {{Radiated Angular Momentum and Dissipative Effects in Classical Scattering}},\ }\href {https://doi.org/10.1103/PhysRevLett.129.121601} {\bibfield  {journal} {\bibinfo  {journal} {Phys. Rev. Lett.}\ }\textbf {\bibinfo {volume} {129}},\ \bibinfo {pages} {121601} (\bibinfo {year} {2022})},\ \Eprint {https://arxiv.org/abs/2203.04283} {arXiv:2203.04283 [hep-th]} \BibitemShut {NoStop}%
\bibitem [{\citenamefont {Bjerrum-Bohr}\ \emph {et~al.}(2022)\citenamefont {Bjerrum-Bohr}, \citenamefont {Damgaard}, \citenamefont {Plante},\ and\ \citenamefont {Vanhove}}]{Bjerrum-Bohr:2022blt}%
  \BibitemOpen
  \bibfield  {author} {\bibinfo {author} {\bibfnamefont {N.~E.~J.}\ \bibnamefont {Bjerrum-Bohr}}, \bibinfo {author} {\bibfnamefont {P.~H.}\ \bibnamefont {Damgaard}}, \bibinfo {author} {\bibfnamefont {L.}~\bibnamefont {Plante}},\ and\ \bibinfo {author} {\bibfnamefont {P.}~\bibnamefont {Vanhove}},\ }\bibfield  {title} {\bibinfo {title} {{The SAGEX review on scattering amplitudes Chapter 13: Post-Minkowskian expansion from scattering amplitudes}},\ }\href {https://doi.org/10.1088/1751-8121/ac7a78} {\bibfield  {journal} {\bibinfo  {journal} {J. Phys. A}\ }\textbf {\bibinfo {volume} {55}},\ \bibinfo {pages} {443014} (\bibinfo {year} {2022})},\ \Eprint {https://arxiv.org/abs/2203.13024} {arXiv:2203.13024 [hep-th]} \BibitemShut {NoStop}%
\bibitem [{\citenamefont {K\"alin}\ and\ \citenamefont {Porto}(2020{\natexlab{c}})}]{Kalin:2020mvi}%
  \BibitemOpen
  \bibfield  {author} {\bibinfo {author} {\bibfnamefont {G.}~\bibnamefont {K\"alin}}\ and\ \bibinfo {author} {\bibfnamefont {R.~A.}\ \bibnamefont {Porto}},\ }\bibfield  {title} {\bibinfo {title} {{Post-Minkowskian Effective Field Theory for Conservative Binary Dynamics}},\ }\href {https://doi.org/10.1007/JHEP11(2020)106} {\bibfield  {journal} {\bibinfo  {journal} {JHEP}\ }\textbf {\bibinfo {volume} {11}},\ \bibinfo {pages} {106}},\ \Eprint {https://arxiv.org/abs/2006.01184} {arXiv:2006.01184 [hep-th]} \BibitemShut {NoStop}%
\bibitem [{\citenamefont {Dlapa}\ \emph {et~al.}(2022{\natexlab{a}})\citenamefont {Dlapa}, \citenamefont {K\"alin}, \citenamefont {Liu},\ and\ \citenamefont {Porto}}]{Dlapa:2021npj}%
  \BibitemOpen
  \bibfield  {author} {\bibinfo {author} {\bibfnamefont {C.}~\bibnamefont {Dlapa}}, \bibinfo {author} {\bibfnamefont {G.}~\bibnamefont {K\"alin}}, \bibinfo {author} {\bibfnamefont {Z.}~\bibnamefont {Liu}},\ and\ \bibinfo {author} {\bibfnamefont {R.~A.}\ \bibnamefont {Porto}},\ }\bibfield  {title} {\bibinfo {title} {{Dynamics of binary systems to fourth Post-Minkowskian order from the effective field theory approach}},\ }\href {https://doi.org/10.1016/j.physletb.2022.137203} {\bibfield  {journal} {\bibinfo  {journal} {Phys. Lett. B}\ }\textbf {\bibinfo {volume} {831}},\ \bibinfo {pages} {137203} (\bibinfo {year} {2022}{\natexlab{a}})},\ \Eprint {https://arxiv.org/abs/2106.08276} {arXiv:2106.08276 [hep-th]} \BibitemShut {NoStop}%
\bibitem [{\citenamefont {Dlapa}\ \emph {et~al.}(2022{\natexlab{b}})\citenamefont {Dlapa}, \citenamefont {K\"alin}, \citenamefont {Liu},\ and\ \citenamefont {Porto}}]{Dlapa:2021vgp}%
  \BibitemOpen
  \bibfield  {author} {\bibinfo {author} {\bibfnamefont {C.}~\bibnamefont {Dlapa}}, \bibinfo {author} {\bibfnamefont {G.}~\bibnamefont {K\"alin}}, \bibinfo {author} {\bibfnamefont {Z.}~\bibnamefont {Liu}},\ and\ \bibinfo {author} {\bibfnamefont {R.~A.}\ \bibnamefont {Porto}},\ }\bibfield  {title} {\bibinfo {title} {{Conservative Dynamics of Binary Systems at Fourth Post-Minkowskian Order in the Large-Eccentricity Expansion}},\ }\href {https://doi.org/10.1103/PhysRevLett.128.161104} {\bibfield  {journal} {\bibinfo  {journal} {Phys. Rev. Lett.}\ }\textbf {\bibinfo {volume} {128}},\ \bibinfo {pages} {161104} (\bibinfo {year} {2022}{\natexlab{b}})},\ \Eprint {https://arxiv.org/abs/2112.11296} {arXiv:2112.11296 [hep-th]} \BibitemShut {NoStop}%
\bibitem [{\citenamefont {Jakobsen}\ and\ \citenamefont {Mogull}(2022)}]{Jakobsen:2022fcj}%
  \BibitemOpen
  \bibfield  {author} {\bibinfo {author} {\bibfnamefont {G.~U.}\ \bibnamefont {Jakobsen}}\ and\ \bibinfo {author} {\bibfnamefont {G.}~\bibnamefont {Mogull}},\ }\bibfield  {title} {\bibinfo {title} {{Conservative and Radiative Dynamics of Spinning Bodies at Third Post-Minkowskian Order Using Worldline Quantum Field Theory}},\ }\href {https://doi.org/10.1103/PhysRevLett.128.141102} {\bibfield  {journal} {\bibinfo  {journal} {Phys. Rev. Lett.}\ }\textbf {\bibinfo {volume} {128}},\ \bibinfo {pages} {141102} (\bibinfo {year} {2022})},\ \Eprint {https://arxiv.org/abs/2201.07778} {arXiv:2201.07778 [hep-th]} \BibitemShut {NoStop}%
\bibitem [{\citenamefont {Jakobsen}\ \emph {et~al.}(2023{\natexlab{a}})\citenamefont {Jakobsen}, \citenamefont {Mogull}, \citenamefont {Plefka}, \citenamefont {Sauer},\ and\ \citenamefont {Xu}}]{Jakobsen:2023ndj}%
  \BibitemOpen
  \bibfield  {author} {\bibinfo {author} {\bibfnamefont {G.~U.}\ \bibnamefont {Jakobsen}}, \bibinfo {author} {\bibfnamefont {G.}~\bibnamefont {Mogull}}, \bibinfo {author} {\bibfnamefont {J.}~\bibnamefont {Plefka}}, \bibinfo {author} {\bibfnamefont {B.}~\bibnamefont {Sauer}},\ and\ \bibinfo {author} {\bibfnamefont {Y.}~\bibnamefont {Xu}},\ }\bibfield  {title} {\bibinfo {title} {{Conservative Scattering of Spinning Black Holes at Fourth Post-Minkowskian Order}},\ }\href {https://doi.org/10.1103/PhysRevLett.131.151401} {\bibfield  {journal} {\bibinfo  {journal} {Phys. Rev. Lett.}\ }\textbf {\bibinfo {volume} {131}},\ \bibinfo {pages} {151401} (\bibinfo {year} {2023}{\natexlab{a}})},\ \Eprint {https://arxiv.org/abs/2306.01714} {arXiv:2306.01714 [hep-th]} \BibitemShut {NoStop}%
\bibitem [{\citenamefont {Dlapa}\ \emph {et~al.}(2023{\natexlab{a}})\citenamefont {Dlapa}, \citenamefont {K\"alin}, \citenamefont {Liu}, \citenamefont {Neef},\ and\ \citenamefont {Porto}}]{Dlapa:2022lmu}%
  \BibitemOpen
  \bibfield  {author} {\bibinfo {author} {\bibfnamefont {C.}~\bibnamefont {Dlapa}}, \bibinfo {author} {\bibfnamefont {G.}~\bibnamefont {K\"alin}}, \bibinfo {author} {\bibfnamefont {Z.}~\bibnamefont {Liu}}, \bibinfo {author} {\bibfnamefont {J.}~\bibnamefont {Neef}},\ and\ \bibinfo {author} {\bibfnamefont {R.~A.}\ \bibnamefont {Porto}},\ }\bibfield  {title} {\bibinfo {title} {{Radiation Reaction and Gravitational Waves at Fourth Post-Minkowskian Order}},\ }\href {https://doi.org/10.1103/PhysRevLett.130.101401} {\bibfield  {journal} {\bibinfo  {journal} {Phys. Rev. Lett.}\ }\textbf {\bibinfo {volume} {130}},\ \bibinfo {pages} {101401} (\bibinfo {year} {2023}{\natexlab{a}})},\ \Eprint {https://arxiv.org/abs/2210.05541} {arXiv:2210.05541 [hep-th]} \BibitemShut {NoStop}%
\bibitem [{\citenamefont {Dlapa}\ \emph {et~al.}(2023{\natexlab{b}})\citenamefont {Dlapa}, \citenamefont {K\"alin}, \citenamefont {Liu}, \citenamefont {Neef},\ and\ \citenamefont {Porto}}]{dlapa2023-letter}%
  \BibitemOpen
  \bibfield  {author} {\bibinfo {author} {\bibfnamefont {C.}~\bibnamefont {Dlapa}}, \bibinfo {author} {\bibfnamefont {G.}~\bibnamefont {K\"alin}}, \bibinfo {author} {\bibfnamefont {Z.}~\bibnamefont {Liu}}, \bibinfo {author} {\bibfnamefont {J.}~\bibnamefont {Neef}},\ and\ \bibinfo {author} {\bibfnamefont {R.~A.}\ \bibnamefont {Porto}},\ }\bibfield  {title} {\bibinfo {title} {Radiation reaction and gravitational waves at fourth post-minkowskian order},\ }\href {https://doi.org/10.1103/PhysRevLett.130.101401} {\bibfield  {journal} {\bibinfo  {journal} {Phys. Rev. Lett.}\ }\textbf {\bibinfo {volume} {130}},\ \bibinfo {pages} {101401} (\bibinfo {year} {2023}{\natexlab{b}})}\BibitemShut {NoStop}%
\bibitem [{\citenamefont {Mogull}\ \emph {et~al.}(2021)\citenamefont {Mogull}, \citenamefont {Plefka},\ and\ \citenamefont {Steinhoff}}]{Mogull:2020sak}%
  \BibitemOpen
  \bibfield  {author} {\bibinfo {author} {\bibfnamefont {G.}~\bibnamefont {Mogull}}, \bibinfo {author} {\bibfnamefont {J.}~\bibnamefont {Plefka}},\ and\ \bibinfo {author} {\bibfnamefont {J.}~\bibnamefont {Steinhoff}},\ }\bibfield  {title} {\bibinfo {title} {{Classical black hole scattering from a worldline quantum field theory}},\ }\href {https://doi.org/10.1007/JHEP02(2021)048} {\bibfield  {journal} {\bibinfo  {journal} {JHEP}\ }\textbf {\bibinfo {volume} {02}},\ \bibinfo {pages} {048}},\ \Eprint {https://arxiv.org/abs/2010.02865} {arXiv:2010.02865 [hep-th]} \BibitemShut {NoStop}%
\bibitem [{\citenamefont {Jakobsen}\ \emph {et~al.}(2023{\natexlab{b}})\citenamefont {Jakobsen}, \citenamefont {Mogull}, \citenamefont {Plefka},\ and\ \citenamefont {Sauer}}]{jakobsen2023dissipative}%
  \BibitemOpen
  \bibfield  {author} {\bibinfo {author} {\bibfnamefont {G.~U.}\ \bibnamefont {Jakobsen}}, \bibinfo {author} {\bibfnamefont {G.}~\bibnamefont {Mogull}}, \bibinfo {author} {\bibfnamefont {J.}~\bibnamefont {Plefka}},\ and\ \bibinfo {author} {\bibfnamefont {B.}~\bibnamefont {Sauer}},\ }\href@noop {} {\bibinfo {title} {Dissipative scattering of spinning black holes at fourth post-minkowskian order}} (\bibinfo {year} {2023}{\natexlab{b}}),\ \Eprint {https://arxiv.org/abs/2308.11514} {arXiv:2308.11514 [hep-th]} \BibitemShut {NoStop}%
\bibitem [{\citenamefont {Di~Vecchia}\ \emph {et~al.}(2023)\citenamefont {Di~Vecchia}, \citenamefont {Heissenberg}, \citenamefont {Russo},\ and\ \citenamefont {Veneziano}}]{DiVecchia:2023frv}%
  \BibitemOpen
  \bibfield  {author} {\bibinfo {author} {\bibfnamefont {P.}~\bibnamefont {Di~Vecchia}}, \bibinfo {author} {\bibfnamefont {C.}~\bibnamefont {Heissenberg}}, \bibinfo {author} {\bibfnamefont {R.}~\bibnamefont {Russo}},\ and\ \bibinfo {author} {\bibfnamefont {G.}~\bibnamefont {Veneziano}},\ }\bibfield  {title} {\bibinfo {title} {{The gravitational eikonal: from particle, string and brane collisions to black-hole encounters}},\ }\href@noop {} {\  (\bibinfo {year} {2023})},\ \Eprint {https://arxiv.org/abs/2306.16488} {arXiv:2306.16488 [hep-th]} \BibitemShut {NoStop}%
\bibitem [{\citenamefont {Damgaard}\ \emph {et~al.}(2023)\citenamefont {Damgaard}, \citenamefont {Hansen}, \citenamefont {Plant\'e},\ and\ \citenamefont {Vanhove}}]{Damgaard:2023ttc}%
  \BibitemOpen
  \bibfield  {author} {\bibinfo {author} {\bibfnamefont {P.~H.}\ \bibnamefont {Damgaard}}, \bibinfo {author} {\bibfnamefont {E.~R.}\ \bibnamefont {Hansen}}, \bibinfo {author} {\bibfnamefont {L.}~\bibnamefont {Plant\'e}},\ and\ \bibinfo {author} {\bibfnamefont {P.}~\bibnamefont {Vanhove}},\ }\bibfield  {title} {\bibinfo {title} {{Classical observables from the exponential representation of the gravitational S-matrix}},\ }\href {https://doi.org/10.1007/JHEP09(2023)183} {\bibfield  {journal} {\bibinfo  {journal} {JHEP}\ }\textbf {\bibinfo {volume} {09}},\ \bibinfo {pages} {183}},\ \Eprint {https://arxiv.org/abs/2307.04746} {arXiv:2307.04746 [hep-th]} \BibitemShut {NoStop}%
\bibitem [{\citenamefont {Mason}\ and\ \citenamefont {Skinner}(2009)}]{Mason_2009}%
  \BibitemOpen
  \bibfield  {author} {\bibinfo {author} {\bibfnamefont {L.}~\bibnamefont {Mason}}\ and\ \bibinfo {author} {\bibfnamefont {D.}~\bibnamefont {Skinner}},\ }\bibfield  {title} {\bibinfo {title} {Gravity, twistors and the {MHV} formalism},\ }\href {https://doi.org/10.1007/s00220-009-0972-4} {\bibfield  {journal} {\bibinfo  {journal} {Communications in Mathematical Physics}\ }\textbf {\bibinfo {volume} {294}},\ \bibinfo {pages} {827} (\bibinfo {year} {2009})}\BibitemShut {NoStop}%
\bibitem [{\citenamefont {Cheung}\ \emph {et~al.}(2021)\citenamefont {Cheung}, \citenamefont {Shah},\ and\ \citenamefont {Solon}}]{Cheung_2021}%
  \BibitemOpen
  \bibfield  {author} {\bibinfo {author} {\bibfnamefont {C.}~\bibnamefont {Cheung}}, \bibinfo {author} {\bibfnamefont {N.}~\bibnamefont {Shah}},\ and\ \bibinfo {author} {\bibfnamefont {M.~P.}\ \bibnamefont {Solon}},\ }\bibfield  {title} {\bibinfo {title} {Mining the geodesic equation for scattering data},\ }\bibfield  {journal} {\bibinfo  {journal} {Physical Review D}\ }\textbf {\bibinfo {volume} {103}},\ \href {https://doi.org/10.1103/physrevd.103.024030} {10.1103/physrevd.103.024030} (\bibinfo {year} {2021})\BibitemShut {NoStop}%
\bibitem [{\citenamefont {Bautista}\ \emph {et~al.}(2023{\natexlab{a}})\citenamefont {Bautista}, \citenamefont {Guevara}, \citenamefont {Kavanagh},\ and\ \citenamefont {Vines}}]{bautista2023scattering-1}%
  \BibitemOpen
  \bibfield  {author} {\bibinfo {author} {\bibfnamefont {Y.~F.}\ \bibnamefont {Bautista}}, \bibinfo {author} {\bibfnamefont {A.}~\bibnamefont {Guevara}}, \bibinfo {author} {\bibfnamefont {C.}~\bibnamefont {Kavanagh}},\ and\ \bibinfo {author} {\bibfnamefont {J.}~\bibnamefont {Vines}},\ }\href@noop {} {\bibinfo {title} {Scattering in black hole backgrounds and higher-spin amplitudes: Part i}} (\bibinfo {year} {2023}{\natexlab{a}}),\ \Eprint {https://arxiv.org/abs/2107.10179} {arXiv:2107.10179 [hep-th]} \BibitemShut {NoStop}%
\bibitem [{\citenamefont {Bautista}\ \emph {et~al.}(2023{\natexlab{b}})\citenamefont {Bautista}, \citenamefont {Guevara}, \citenamefont {Kavanagh},\ and\ \citenamefont {Vines}}]{bautista2023scattering-2}%
  \BibitemOpen
  \bibfield  {author} {\bibinfo {author} {\bibfnamefont {Y.~F.}\ \bibnamefont {Bautista}}, \bibinfo {author} {\bibfnamefont {A.}~\bibnamefont {Guevara}}, \bibinfo {author} {\bibfnamefont {C.}~\bibnamefont {Kavanagh}},\ and\ \bibinfo {author} {\bibfnamefont {J.}~\bibnamefont {Vines}},\ }\href@noop {} {\bibinfo {title} {Scattering in black hole backgrounds and higher-spin amplitudes: Part ii}} (\bibinfo {year} {2023}{\natexlab{b}}),\ \Eprint {https://arxiv.org/abs/2212.07965} {arXiv:2212.07965 [hep-th]} \BibitemShut {NoStop}%
\bibitem [{\citenamefont {Adamo}\ \emph {et~al.}(2023{\natexlab{a}})\citenamefont {Adamo}, \citenamefont {Cristofoli}, \citenamefont {Ilderton},\ and\ \citenamefont {Klisch}}]{Adamo_2023}%
  \BibitemOpen
  \bibfield  {author} {\bibinfo {author} {\bibfnamefont {T.}~\bibnamefont {Adamo}}, \bibinfo {author} {\bibfnamefont {A.}~\bibnamefont {Cristofoli}}, \bibinfo {author} {\bibfnamefont {A.}~\bibnamefont {Ilderton}},\ and\ \bibinfo {author} {\bibfnamefont {S.}~\bibnamefont {Klisch}},\ }\bibfield  {title} {\bibinfo {title} {All order gravitational waveforms from scattering amplitudes},\ }\bibfield  {journal} {\bibinfo  {journal} {Physical Review Letters}\ }\textbf {\bibinfo {volume} {131}},\ \href {https://doi.org/10.1103/physrevlett.131.011601} {10.1103/physrevlett.131.011601} (\bibinfo {year} {2023}{\natexlab{a}})\BibitemShut {NoStop}%
\bibitem [{\citenamefont {Adamo}\ \emph {et~al.}(2023{\natexlab{b}})\citenamefont {Adamo}, \citenamefont {Cristofoli}, \citenamefont {Ilderton},\ and\ \citenamefont {Klisch}}]{Adamo:2023cfp}%
  \BibitemOpen
  \bibfield  {author} {\bibinfo {author} {\bibfnamefont {T.}~\bibnamefont {Adamo}}, \bibinfo {author} {\bibfnamefont {A.}~\bibnamefont {Cristofoli}}, \bibinfo {author} {\bibfnamefont {A.}~\bibnamefont {Ilderton}},\ and\ \bibinfo {author} {\bibfnamefont {S.}~\bibnamefont {Klisch}},\ }\bibfield  {title} {\bibinfo {title} {{Scattering amplitudes for self-force}},\ }\href@noop {} {\  (\bibinfo {year} {2023}{\natexlab{b}})},\ \Eprint {https://arxiv.org/abs/2307.00431} {arXiv:2307.00431 [hep-th]} \BibitemShut {NoStop}%
\bibitem [{\citenamefont {Cheung}\ \emph {et~al.}(2024)\citenamefont {Cheung}, \citenamefont {Parra-Martinez}, \citenamefont {Rothstein}, \citenamefont {Shah},\ and\ \citenamefont {Wilson-Gerow}}]{cheung2023effective}%
  \BibitemOpen
  \bibfield  {author} {\bibinfo {author} {\bibfnamefont {C.}~\bibnamefont {Cheung}}, \bibinfo {author} {\bibfnamefont {J.}~\bibnamefont {Parra-Martinez}}, \bibinfo {author} {\bibfnamefont {I.~Z.}\ \bibnamefont {Rothstein}}, \bibinfo {author} {\bibfnamefont {N.}~\bibnamefont {Shah}},\ and\ \bibinfo {author} {\bibfnamefont {J.}~\bibnamefont {Wilson-Gerow}},\ }\bibfield  {title} {\bibinfo {title} {Effective field theory for extreme mass ratio binaries},\ }\href {https://doi.org/10.1103/PhysRevLett.132.091402} {\bibfield  {journal} {\bibinfo  {journal} {Phys. Rev. Lett.}\ }\textbf {\bibinfo {volume} {132}},\ \bibinfo {pages} {091402} (\bibinfo {year} {2024})}\BibitemShut {NoStop}%
\bibitem [{\citenamefont {Cheung}\ \emph {et~al.}()\citenamefont {Cheung}, \citenamefont {Parra-Martinez}, \citenamefont {Rothstein}, \citenamefont {Shah},\ and\ \citenamefont {Wilson-Gerow}}]{long_paper}%
  \BibitemOpen
  \bibfield  {author} {\bibinfo {author} {\bibfnamefont {C.}~\bibnamefont {Cheung}}, \bibinfo {author} {\bibfnamefont {J.}~\bibnamefont {Parra-Martinez}}, \bibinfo {author} {\bibfnamefont {I.~Z.}\ \bibnamefont {Rothstein}}, \bibinfo {author} {\bibfnamefont {N.}~\bibnamefont {Shah}},\ and\ \bibinfo {author} {\bibfnamefont {J.}~\bibnamefont {Wilson-Gerow}},\ }\href@noop {} {\bibinfo  {journal} {in preparation}\ }\BibitemShut {NoStop}%
\bibitem [{\citenamefont {Kosmopoulos}\ and\ \citenamefont {Solon}(2023)}]{kosmopoulos2023gravitational}%
  \BibitemOpen
\bibfield  {journal} {  }\bibfield  {author} {\bibinfo {author} {\bibfnamefont {D.}~\bibnamefont {Kosmopoulos}}\ and\ \bibinfo {author} {\bibfnamefont {M.~P.}\ \bibnamefont {Solon}},\ }\href@noop {} {\bibinfo {title} {Gravitational self force from scattering amplitudes in curved space}} (\bibinfo {year} {2023}),\ \Eprint {https://arxiv.org/abs/2308.15304} {arXiv:2308.15304 [hep-th]} \BibitemShut {NoStop}%
\bibitem [{\citenamefont {Jones}(2023)}]{jones2023classical}%
  \BibitemOpen
  \bibfield  {author} {\bibinfo {author} {\bibfnamefont {C.~R.~T.}\ \bibnamefont {Jones}},\ }\href@noop {} {\bibinfo {title} {Classical dynamics of vortex solitons from perturbative scattering amplitudes}} (\bibinfo {year} {2023}),\ \Eprint {https://arxiv.org/abs/2305.08902} {arXiv:2305.08902 [hep-th]} \BibitemShut {NoStop}%
\bibitem [{\citenamefont {Bern}\ \emph {et~al.}(2022{\natexlab{b}})\citenamefont {Bern}, \citenamefont {Gatica}, \citenamefont {Herrmann}, \citenamefont {Luna},\ and\ \citenamefont {Zeng}}]{Bern_2022}%
  \BibitemOpen
  \bibfield  {author} {\bibinfo {author} {\bibfnamefont {Z.}~\bibnamefont {Bern}}, \bibinfo {author} {\bibfnamefont {J.~P.}\ \bibnamefont {Gatica}}, \bibinfo {author} {\bibfnamefont {E.}~\bibnamefont {Herrmann}}, \bibinfo {author} {\bibfnamefont {A.}~\bibnamefont {Luna}},\ and\ \bibinfo {author} {\bibfnamefont {M.}~\bibnamefont {Zeng}},\ }\bibfield  {title} {\bibinfo {title} {Scalar {QED} as a toy model for higher-order effects in classical gravitational scattering},\ }\bibfield  {journal} {\bibinfo  {journal} {Journal of High Energy Physics}\ }\textbf {\bibinfo {volume} {2022}},\ \href {https://doi.org/10.1007/jhep08(2022)131} {10.1007/jhep08(2022)131} (\bibinfo {year} {2022}{\natexlab{b}})\BibitemShut {NoStop}%
\bibitem [{\citenamefont {Bern}\ \emph {et~al.}(2023)\citenamefont {Bern}, \citenamefont {Herrmann}, \citenamefont {Roiban}, \citenamefont {Ruf}, \citenamefont {Smirnov}, \citenamefont {Smirnov},\ and\ \citenamefont {Zeng}}]{bern2023conservative}%
  \BibitemOpen
  \bibfield  {author} {\bibinfo {author} {\bibfnamefont {Z.}~\bibnamefont {Bern}}, \bibinfo {author} {\bibfnamefont {E.}~\bibnamefont {Herrmann}}, \bibinfo {author} {\bibfnamefont {R.}~\bibnamefont {Roiban}}, \bibinfo {author} {\bibfnamefont {M.~S.}\ \bibnamefont {Ruf}}, \bibinfo {author} {\bibfnamefont {A.~V.}\ \bibnamefont {Smirnov}}, \bibinfo {author} {\bibfnamefont {V.~A.}\ \bibnamefont {Smirnov}},\ and\ \bibinfo {author} {\bibfnamefont {M.}~\bibnamefont {Zeng}},\ }\href@noop {} {\bibinfo {title} {Conservative binary dynamics at order $o(\alpha^5)$ in electrodynamics}} (\bibinfo {year} {2023}),\ \Eprint {https://arxiv.org/abs/2305.08981} {arXiv:2305.08981 [hep-th]} \BibitemShut {NoStop}%
\bibitem [{\citenamefont {Gibbons}(1975)}]{gibbons1975vacuum}%
  \BibitemOpen
  \bibfield  {author} {\bibinfo {author} {\bibfnamefont {G.~W.}\ \bibnamefont {Gibbons}},\ }\bibfield  {title} {\bibinfo {title} {Vacuum polarization and the spontaneous loss of charge by black holes},\ }\href@noop {} {\bibfield  {journal} {\bibinfo  {journal} {Communications in Mathematical Physics}\ }\textbf {\bibinfo {volume} {44}},\ \bibinfo {pages} {245} (\bibinfo {year} {1975})}\BibitemShut {NoStop}%
\bibitem [{\citenamefont {Blandford}\ and\ \citenamefont {Znajek}(1977)}]{10.1093/mnras/179.3.433}%
  \BibitemOpen
  \bibfield  {author} {\bibinfo {author} {\bibfnamefont {R.~D.}\ \bibnamefont {Blandford}}\ and\ \bibinfo {author} {\bibfnamefont {R.~L.}\ \bibnamefont {Znajek}},\ }\bibfield  {title} {\bibinfo {title} {{Electromagnetic extraction of energy from Kerr black holes}},\ }\href {https://doi.org/10.1093/mnras/179.3.433} {\bibfield  {journal} {\bibinfo  {journal} {Monthly Notices of the Royal Astronomical Society}\ }\textbf {\bibinfo {volume} {179}},\ \bibinfo {pages} {433} (\bibinfo {year} {1977})},\ \Eprint {https://arxiv.org/abs/https://academic.oup.com/mnras/article-pdf/179/3/433/9333653/mnras179-0433.pdf} {https://academic.oup.com/mnras/article-pdf/179/3/433/9333653/mnras179-0433.pdf} \BibitemShut {NoStop}%
\bibitem [{\citenamefont {Strominger}\ and\ \citenamefont {Vafa}(1996)}]{Strominger_1996}%
  \BibitemOpen
  \bibfield  {author} {\bibinfo {author} {\bibfnamefont {A.}~\bibnamefont {Strominger}}\ and\ \bibinfo {author} {\bibfnamefont {C.}~\bibnamefont {Vafa}},\ }\bibfield  {title} {\bibinfo {title} {Microscopic origin of the bekenstein-hawking entropy},\ }\href {https://doi.org/10.1016/0370-2693(96)00345-0} {\bibfield  {journal} {\bibinfo  {journal} {Physics Letters B}\ }\textbf {\bibinfo {volume} {379}},\ \bibinfo {pages} {99} (\bibinfo {year} {1996})}\BibitemShut {NoStop}%
\bibitem [{\citenamefont {Arkani-Hamed}\ \emph {et~al.}(2007)\citenamefont {Arkani-Hamed}, \citenamefont {Motl}, \citenamefont {Nicolis},\ and\ \citenamefont {Vafa}}]{Arkani-Hamed_2007}%
  \BibitemOpen
  \bibfield  {author} {\bibinfo {author} {\bibfnamefont {N.}~\bibnamefont {Arkani-Hamed}}, \bibinfo {author} {\bibfnamefont {L.}~\bibnamefont {Motl}}, \bibinfo {author} {\bibfnamefont {A.}~\bibnamefont {Nicolis}},\ and\ \bibinfo {author} {\bibfnamefont {C.}~\bibnamefont {Vafa}},\ }\bibfield  {title} {\bibinfo {title} {The string landscape, black holes and gravity as the weakest force},\ }\href {https://doi.org/10.1088/1126-6708/2007/06/060} {\bibfield  {journal} {\bibinfo  {journal} {Journal of High Energy Physics}\ }\textbf {\bibinfo {volume} {2007}},\ \bibinfo {pages} {060} (\bibinfo {year} {2007})}\BibitemShut {NoStop}%
\bibitem [{\citenamefont {Cheung}\ \emph {et~al.}(2018{\natexlab{b}})\citenamefont {Cheung}, \citenamefont {Liu},\ and\ \citenamefont {Remmen}}]{Cheung_2018}%
  \BibitemOpen
  \bibfield  {author} {\bibinfo {author} {\bibfnamefont {C.}~\bibnamefont {Cheung}}, \bibinfo {author} {\bibfnamefont {J.}~\bibnamefont {Liu}},\ and\ \bibinfo {author} {\bibfnamefont {G.~N.}\ \bibnamefont {Remmen}},\ }\bibfield  {title} {\bibinfo {title} {Proof of the weak gravity conjecture from black hole entropy},\ }\bibfield  {journal} {\bibinfo  {journal} {Journal of High Energy Physics}\ }\textbf {\bibinfo {volume} {2018}},\ \href {https://doi.org/10.1007/jhep10(2018)004} {10.1007/jhep10(2018)004} (\bibinfo {year} {2018}{\natexlab{b}})\BibitemShut {NoStop}%
\bibitem [{\citenamefont {Gibbons}\ and\ \citenamefont {Hull}(1982)}]{GIBBONS1982190}%
  \BibitemOpen
  \bibfield  {author} {\bibinfo {author} {\bibfnamefont {G.}~\bibnamefont {Gibbons}}\ and\ \bibinfo {author} {\bibfnamefont {C.}~\bibnamefont {Hull}},\ }\bibfield  {title} {\bibinfo {title} {A bogomolny bound for general relativity and solitons in n=2 supergravity},\ }\href {https://doi.org/https://doi.org/10.1016/0370-2693(82)90751-1} {\bibfield  {journal} {\bibinfo  {journal} {Physics Letters B}\ }\textbf {\bibinfo {volume} {109}},\ \bibinfo {pages} {190} (\bibinfo {year} {1982})}\BibitemShut {NoStop}%
\bibitem [{\citenamefont {Kallosh}\ \emph {et~al.}(1992)\citenamefont {Kallosh}, \citenamefont {Linde}, \citenamefont {Ort\'{\i}n}, \citenamefont {Peet},\ and\ \citenamefont {Van~Proeyen}}]{PhysRevD.46.5278}%
  \BibitemOpen
  \bibfield  {author} {\bibinfo {author} {\bibfnamefont {R.}~\bibnamefont {Kallosh}}, \bibinfo {author} {\bibfnamefont {A.}~\bibnamefont {Linde}}, \bibinfo {author} {\bibfnamefont {T.}~\bibnamefont {Ort\'{\i}n}}, \bibinfo {author} {\bibfnamefont {A.}~\bibnamefont {Peet}},\ and\ \bibinfo {author} {\bibfnamefont {A.}~\bibnamefont {Van~Proeyen}},\ }\bibfield  {title} {\bibinfo {title} {Supersymmetry as a cosmic censor},\ }\href {https://doi.org/10.1103/PhysRevD.46.5278} {\bibfield  {journal} {\bibinfo  {journal} {Phys. Rev. D}\ }\textbf {\bibinfo {volume} {46}},\ \bibinfo {pages} {5278} (\bibinfo {year} {1992})}\BibitemShut {NoStop}%
\bibitem [{\citenamefont {Papapetrou}(1945)}]{papapetrou_1945}%
  \BibitemOpen
  \bibfield  {author} {\bibinfo {author} {\bibfnamefont {A.}~\bibnamefont {Papapetrou}},\ }\bibfield  {title} {\bibinfo {title} {A static solution of the equations of the gravitational field for an arbitary charge-distribution},\ }\href {http://www.jstor.org/stable/20488481} {\bibfield  {journal} {\bibinfo  {journal} {Proceedings of the Royal Irish Academy. Section A: Mathematical and Physical Sciences}\ }\textbf {\bibinfo {volume} {51}},\ \bibinfo {pages} {191} (\bibinfo {year} {1945})}\BibitemShut {NoStop}%
\bibitem [{\citenamefont {Majumdar}(1947)}]{PhysRev.72.390}%
  \BibitemOpen
  \bibfield  {author} {\bibinfo {author} {\bibfnamefont {S.~D.}\ \bibnamefont {Majumdar}},\ }\bibfield  {title} {\bibinfo {title} {A class of exact solutions of einstein's field equations},\ }\href {https://doi.org/10.1103/PhysRev.72.390} {\bibfield  {journal} {\bibinfo  {journal} {Phys. Rev.}\ }\textbf {\bibinfo {volume} {72}},\ \bibinfo {pages} {390} (\bibinfo {year} {1947})}\BibitemShut {NoStop}%
\bibitem [{\citenamefont {Hartle}\ and\ \citenamefont {Hawking}(1972)}]{hartle1972solutions}%
  \BibitemOpen
  \bibfield  {author} {\bibinfo {author} {\bibfnamefont {J.~B.}\ \bibnamefont {Hartle}}\ and\ \bibinfo {author} {\bibfnamefont {S.~W.}\ \bibnamefont {Hawking}},\ }\bibfield  {title} {\bibinfo {title} {Solutions of the einstein-maxwell equations with many black holes},\ }\href@noop {} {\bibfield  {journal} {\bibinfo  {journal} {Communications in Mathematical Physics}\ }\textbf {\bibinfo {volume} {26}},\ \bibinfo {pages} {87} (\bibinfo {year} {1972})}\BibitemShut {NoStop}%
\bibitem [{\citenamefont {Caron-Huot}\ and\ \citenamefont {Zahraee}(2019)}]{Caron-Huot:2018ape}%
  \BibitemOpen
  \bibfield  {author} {\bibinfo {author} {\bibfnamefont {S.}~\bibnamefont {Caron-Huot}}\ and\ \bibinfo {author} {\bibfnamefont {Z.}~\bibnamefont {Zahraee}},\ }\bibfield  {title} {\bibinfo {title} {{Integrability of Black Hole Orbits in Maximal Supergravity}},\ }\href {https://doi.org/10.1007/JHEP07(2019)179} {\bibfield  {journal} {\bibinfo  {journal} {JHEP}\ }\textbf {\bibinfo {volume} {07}},\ \bibinfo {pages} {179}},\ \Eprint {https://arxiv.org/abs/1810.04694} {arXiv:1810.04694 [hep-th]} \BibitemShut {NoStop}%
\bibitem [{\citenamefont {Liu}\ \emph {et~al.}(2020)\citenamefont {Liu}, \citenamefont {Christiansen}, \citenamefont {Guo}, \citenamefont {Cai},\ and\ \citenamefont {Kim}}]{PhysRevD.102.103520}%
  \BibitemOpen
  \bibfield  {author} {\bibinfo {author} {\bibfnamefont {L.}~\bibnamefont {Liu}}, \bibinfo {author} {\bibfnamefont {O.}~\bibnamefont {Christiansen}}, \bibinfo {author} {\bibfnamefont {Z.-K.}\ \bibnamefont {Guo}}, \bibinfo {author} {\bibfnamefont {R.-G.}\ \bibnamefont {Cai}},\ and\ \bibinfo {author} {\bibfnamefont {S.~P.}\ \bibnamefont {Kim}},\ }\bibfield  {title} {\bibinfo {title} {Gravitational and electromagnetic radiation from binary black holes with electric and magnetic charges: Circular orbits on a cone},\ }\href {https://doi.org/10.1103/PhysRevD.102.103520} {\bibfield  {journal} {\bibinfo  {journal} {Phys. Rev. D}\ }\textbf {\bibinfo {volume} {102}},\ \bibinfo {pages} {103520} (\bibinfo {year} {2020})}\BibitemShut {NoStop}%
\bibitem [{\citenamefont {Zi}\ \emph {et~al.}(2023)\citenamefont {Zi}, \citenamefont {Zhou}, \citenamefont {Wang}, \citenamefont {Li}, \citenamefont {Zhang},\ and\ \citenamefont {Chen}}]{PhysRevD.107.023005}%
  \BibitemOpen
  \bibfield  {author} {\bibinfo {author} {\bibfnamefont {T.}~\bibnamefont {Zi}}, \bibinfo {author} {\bibfnamefont {Z.}~\bibnamefont {Zhou}}, \bibinfo {author} {\bibfnamefont {H.-T.}\ \bibnamefont {Wang}}, \bibinfo {author} {\bibfnamefont {P.-C.}\ \bibnamefont {Li}}, \bibinfo {author} {\bibfnamefont {J.-d.}\ \bibnamefont {Zhang}},\ and\ \bibinfo {author} {\bibfnamefont {B.}~\bibnamefont {Chen}},\ }\bibfield  {title} {\bibinfo {title} {Analytic kludge waveforms for extreme-mass-ratio inspirals of a charged object around a kerr-newman black hole},\ }\href {https://doi.org/10.1103/PhysRevD.107.023005} {\bibfield  {journal} {\bibinfo  {journal} {Phys. Rev. D}\ }\textbf {\bibinfo {volume} {107}},\ \bibinfo {pages} {023005} (\bibinfo {year} {2023})}\BibitemShut {NoStop}%
\bibitem [{\citenamefont {Benavides-Gallego}\ and\ \citenamefont {Han}(2023)}]{Benavides_Gallego_2023}%
  \BibitemOpen
  \bibfield  {author} {\bibinfo {author} {\bibfnamefont {C.~A.}\ \bibnamefont {Benavides-Gallego}}\ and\ \bibinfo {author} {\bibfnamefont {W.-B.}\ \bibnamefont {Han}},\ }\bibfield  {title} {\bibinfo {title} {Gravitational waves and electromagnetic radiation from charged black hole binaries},\ }\href {https://doi.org/10.3390/sym15020537} {\bibfield  {journal} {\bibinfo  {journal} {Symmetry}\ }\textbf {\bibinfo {volume} {15}},\ \bibinfo {pages} {537} (\bibinfo {year} {2023})}\BibitemShut {NoStop}%
\bibitem [{\citenamefont {Juli{\'{e}}}(2018)}]{Juli__2018}%
  \BibitemOpen
  \bibfield  {author} {\bibinfo {author} {\bibfnamefont {F.-L.}\ \bibnamefont {Juli{\'{e}}}},\ }\bibfield  {title} {\bibinfo {title} {On the motion of hairy black holes in einstein-maxwell-dilaton theories},\ }\href {https://doi.org/10.1088/1475-7516/2018/01/026} {\bibfield  {journal} {\bibinfo  {journal} {Journal of Cosmology and Astroparticle Physics}\ }\textbf {\bibinfo {volume} {2018}}\bibinfo  {number} { (01)},\ \bibinfo {pages} {026}}\BibitemShut {NoStop}%
\bibitem [{\citenamefont {Khalil}\ \emph {et~al.}(2018)\citenamefont {Khalil}, \citenamefont {Sennett}, \citenamefont {Steinhoff}, \citenamefont {Vines},\ and\ \citenamefont {Buonanno}}]{Khalil_2018}%
  \BibitemOpen
\bibfield  {number} {  }\bibfield  {author} {\bibinfo {author} {\bibfnamefont {M.}~\bibnamefont {Khalil}}, \bibinfo {author} {\bibfnamefont {N.}~\bibnamefont {Sennett}}, \bibinfo {author} {\bibfnamefont {J.}~\bibnamefont {Steinhoff}}, \bibinfo {author} {\bibfnamefont {J.}~\bibnamefont {Vines}},\ and\ \bibinfo {author} {\bibfnamefont {A.}~\bibnamefont {Buonanno}},\ }\bibfield  {title} {\bibinfo {title} {Hairy binary black holes in einstein-maxwell-dilaton theory and their effective-one-body description},\ }\bibfield  {journal} {\bibinfo  {journal} {Physical Review D}\ }\textbf {\bibinfo {volume} {98}},\ \href {https://doi.org/10.1103/physrevd.98.104010} {10.1103/physrevd.98.104010} (\bibinfo {year} {2018})\BibitemShut {NoStop}%
\bibitem [{\citenamefont {Bjerrum-Bohr}(2002)}]{Bjerrum-Bohr:2002aqa}%
  \BibitemOpen
  \bibfield  {author} {\bibinfo {author} {\bibfnamefont {N.~E.~J.}\ \bibnamefont {Bjerrum-Bohr}},\ }\bibfield  {title} {\bibinfo {title} {{Leading quantum gravitational corrections to scalar QED}},\ }\href {https://doi.org/10.1103/PhysRevD.66.084023} {\bibfield  {journal} {\bibinfo  {journal} {Phys. Rev. D}\ }\textbf {\bibinfo {volume} {66}},\ \bibinfo {pages} {084023} (\bibinfo {year} {2002})},\ \Eprint {https://arxiv.org/abs/hep-th/0206236} {arXiv:hep-th/0206236} \BibitemShut {NoStop}%
\bibitem [{\citenamefont {Butt}(2006)}]{Butt:2006gv}%
  \BibitemOpen
  \bibfield  {author} {\bibinfo {author} {\bibfnamefont {M.~S.}\ \bibnamefont {Butt}},\ }\bibfield  {title} {\bibinfo {title} {{Leading quantum gravitational corrections to QED}},\ }\href {https://doi.org/10.1103/PhysRevD.74.125007} {\bibfield  {journal} {\bibinfo  {journal} {Phys. Rev. D}\ }\textbf {\bibinfo {volume} {74}},\ \bibinfo {pages} {125007} (\bibinfo {year} {2006})},\ \Eprint {https://arxiv.org/abs/gr-qc/0605137} {arXiv:gr-qc/0605137} \BibitemShut {NoStop}%
\bibitem [{\citenamefont {Faller}(2008)}]{Faller:2007sy}%
  \BibitemOpen
  \bibfield  {author} {\bibinfo {author} {\bibfnamefont {S.}~\bibnamefont {Faller}},\ }\bibfield  {title} {\bibinfo {title} {{Effective Field Theory of Gravity: Leading Quantum Gravitational Corrections to Newtons and Coulombs Law}},\ }\href {https://doi.org/10.1103/PhysRevD.77.124039} {\bibfield  {journal} {\bibinfo  {journal} {Phys. Rev. D}\ }\textbf {\bibinfo {volume} {77}},\ \bibinfo {pages} {124039} (\bibinfo {year} {2008})},\ \Eprint {https://arxiv.org/abs/0708.1701} {arXiv:0708.1701 [hep-th]} \BibitemShut {NoStop}%
\bibitem [{\citenamefont {Holstein}\ and\ \citenamefont {Ross}(2008)}]{Holstein:2008sy}%
  \BibitemOpen
  \bibfield  {author} {\bibinfo {author} {\bibfnamefont {B.~R.}\ \bibnamefont {Holstein}}\ and\ \bibinfo {author} {\bibfnamefont {A.}~\bibnamefont {Ross}},\ }\bibfield  {title} {\bibinfo {title} {{Long Distance Effects in Mixed Electromagnetic-Gravitational Scattering}},\ }\href@noop {} {\  (\bibinfo {year} {2008})},\ \Eprint {https://arxiv.org/abs/0802.0717} {arXiv:0802.0717 [hep-ph]} \BibitemShut {NoStop}%
\bibitem [{\citenamefont {Gupta}(2022)}]{gupta2022binary}%
  \BibitemOpen
  \bibfield  {author} {\bibinfo {author} {\bibfnamefont {P.~K.}\ \bibnamefont {Gupta}},\ }\href@noop {} {\bibinfo {title} {Binary dynamics from einstein-maxwell theory at second post-newtonian order using effective field theory}} (\bibinfo {year} {2022}),\ \Eprint {https://arxiv.org/abs/2205.11591} {arXiv:2205.11591 [gr-qc]} \BibitemShut {NoStop}%
\bibitem [{\citenamefont {Zilh\~ao}\ \emph {et~al.}(2012)\citenamefont {Zilh\~ao}, \citenamefont {Cardoso}, \citenamefont {Herdeiro}, \citenamefont {Lehner},\ and\ \citenamefont {Sperhake}}]{PhysRevD.85.124062}%
  \BibitemOpen
  \bibfield  {author} {\bibinfo {author} {\bibfnamefont {M.}~\bibnamefont {Zilh\~ao}}, \bibinfo {author} {\bibfnamefont {V.}~\bibnamefont {Cardoso}}, \bibinfo {author} {\bibfnamefont {C.}~\bibnamefont {Herdeiro}}, \bibinfo {author} {\bibfnamefont {L.}~\bibnamefont {Lehner}},\ and\ \bibinfo {author} {\bibfnamefont {U.}~\bibnamefont {Sperhake}},\ }\bibfield  {title} {\bibinfo {title} {Collisions of charged black holes},\ }\href {https://doi.org/10.1103/PhysRevD.85.124062} {\bibfield  {journal} {\bibinfo  {journal} {Phys. Rev. D}\ }\textbf {\bibinfo {volume} {85}},\ \bibinfo {pages} {124062} (\bibinfo {year} {2012})}\BibitemShut {NoStop}%
\bibitem [{\citenamefont {Bozzola}\ and\ \citenamefont {Paschalidis}(2019)}]{PhysRevD.99.104044}%
  \BibitemOpen
  \bibfield  {author} {\bibinfo {author} {\bibfnamefont {G.}~\bibnamefont {Bozzola}}\ and\ \bibinfo {author} {\bibfnamefont {V.}~\bibnamefont {Paschalidis}},\ }\bibfield  {title} {\bibinfo {title} {Initial data for general relativistic simulations of multiple electrically charged black holes with linear and angular momenta},\ }\href {https://doi.org/10.1103/PhysRevD.99.104044} {\bibfield  {journal} {\bibinfo  {journal} {Phys. Rev. D}\ }\textbf {\bibinfo {volume} {99}},\ \bibinfo {pages} {104044} (\bibinfo {year} {2019})}\BibitemShut {NoStop}%
\bibitem [{\citenamefont {Bozzola}\ and\ \citenamefont {Paschalidis}(2021)}]{PhysRevD.104.044004}%
  \BibitemOpen
  \bibfield  {author} {\bibinfo {author} {\bibfnamefont {G.}~\bibnamefont {Bozzola}}\ and\ \bibinfo {author} {\bibfnamefont {V.}~\bibnamefont {Paschalidis}},\ }\bibfield  {title} {\bibinfo {title} {Numerical-relativity simulations of the quasicircular inspiral and merger of nonspinning, charged black holes: Methods and comparison with approximate approaches},\ }\href {https://doi.org/10.1103/PhysRevD.104.044004} {\bibfield  {journal} {\bibinfo  {journal} {Phys. Rev. D}\ }\textbf {\bibinfo {volume} {104}},\ \bibinfo {pages} {044004} (\bibinfo {year} {2021})}\BibitemShut {NoStop}%
\bibitem [{\citenamefont {Parra-Martinez}\ \emph {et~al.}(2020)\citenamefont {Parra-Martinez}, \citenamefont {Ruf},\ and\ \citenamefont {Zeng}}]{Parra-Martinez:2020dzs}%
  \BibitemOpen
  \bibfield  {author} {\bibinfo {author} {\bibfnamefont {J.}~\bibnamefont {Parra-Martinez}}, \bibinfo {author} {\bibfnamefont {M.~S.}\ \bibnamefont {Ruf}},\ and\ \bibinfo {author} {\bibfnamefont {M.}~\bibnamefont {Zeng}},\ }\bibfield  {title} {\bibinfo {title} {{Extremal black hole scattering at $\mathcal{O}(G^3)$: graviton dominance, eikonal exponentiation, and differential equations}},\ }\href {https://doi.org/10.1007/JHEP11(2020)023} {\bibfield  {journal} {\bibinfo  {journal} {JHEP}\ }\textbf {\bibinfo {volume} {11}},\ \bibinfo {pages} {023}},\ \Eprint {https://arxiv.org/abs/2005.04236} {arXiv:2005.04236 [hep-th]} \BibitemShut {NoStop}%
\bibitem [{\citenamefont {Di~Vecchia}\ \emph {et~al.}(2020)\citenamefont {Di~Vecchia}, \citenamefont {Heissenberg}, \citenamefont {Russo},\ and\ \citenamefont {Veneziano}}]{DiVecchia:2020ymx}%
  \BibitemOpen
  \bibfield  {author} {\bibinfo {author} {\bibfnamefont {P.}~\bibnamefont {Di~Vecchia}}, \bibinfo {author} {\bibfnamefont {C.}~\bibnamefont {Heissenberg}}, \bibinfo {author} {\bibfnamefont {R.}~\bibnamefont {Russo}},\ and\ \bibinfo {author} {\bibfnamefont {G.}~\bibnamefont {Veneziano}},\ }\bibfield  {title} {\bibinfo {title} {{Universality of ultra-relativistic gravitational scattering}},\ }\href {https://doi.org/10.1016/j.physletb.2020.135924} {\bibfield  {journal} {\bibinfo  {journal} {Phys. Lett. B}\ }\textbf {\bibinfo {volume} {811}},\ \bibinfo {pages} {135924} (\bibinfo {year} {2020})},\ \Eprint {https://arxiv.org/abs/2008.12743} {arXiv:2008.12743 [hep-th]} \BibitemShut {NoStop}%
\bibitem [{\citenamefont {Bjerrum-Bohr}\ \emph {et~al.}(2021{\natexlab{b}})\citenamefont {Bjerrum-Bohr}, \citenamefont {Damgaard}, \citenamefont {Plant\'e},\ and\ \citenamefont {Vanhove}}]{Bjerrum-Bohr:2021vuf}%
  \BibitemOpen
  \bibfield  {author} {\bibinfo {author} {\bibfnamefont {N.~E.~J.}\ \bibnamefont {Bjerrum-Bohr}}, \bibinfo {author} {\bibfnamefont {P.~H.}\ \bibnamefont {Damgaard}}, \bibinfo {author} {\bibfnamefont {L.}~\bibnamefont {Plant\'e}},\ and\ \bibinfo {author} {\bibfnamefont {P.}~\bibnamefont {Vanhove}},\ }\bibfield  {title} {\bibinfo {title} {{Classical gravity from loop amplitudes}},\ }\href {https://doi.org/10.1103/PhysRevD.104.026009} {\bibfield  {journal} {\bibinfo  {journal} {Phys. Rev. D}\ }\textbf {\bibinfo {volume} {104}},\ \bibinfo {pages} {026009} (\bibinfo {year} {2021}{\natexlab{b}})},\ \Eprint {https://arxiv.org/abs/2104.04510} {arXiv:2104.04510 [hep-th]} \BibitemShut {NoStop}%
\bibitem [{\citenamefont {Jones}\ and\ \citenamefont {Solon}(2023)}]{Jones_2023}%
  \BibitemOpen
  \bibfield  {author} {\bibinfo {author} {\bibfnamefont {C.~R.~T.}\ \bibnamefont {Jones}}\ and\ \bibinfo {author} {\bibfnamefont {M.}~\bibnamefont {Solon}},\ }\bibfield  {title} {\bibinfo {title} {Scattering amplitudes and n-body post-minkowskian hamiltonians in general relativity and beyond},\ }\bibfield  {journal} {\bibinfo  {journal} {Journal of High Energy Physics}\ }\textbf {\bibinfo {volume} {2023}},\ \href {https://doi.org/10.1007/jhep02(2023)105} {10.1007/jhep02(2023)105} (\bibinfo {year} {2023})\BibitemShut {NoStop}%
\bibitem [{\citenamefont {Denef}(2000{\natexlab{a}})}]{Denef_2000}%
  \BibitemOpen
  \bibfield  {author} {\bibinfo {author} {\bibfnamefont {F.}~\bibnamefont {Denef}},\ }\bibfield  {title} {\bibinfo {title} {Supergravity flows and d-brane stability},\ }\href {https://doi.org/10.1088/1126-6708/2000/08/050} {\bibfield  {journal} {\bibinfo  {journal} {Journal of High Energy Physics}\ }\textbf {\bibinfo {volume} {2000}},\ \bibinfo {pages} {050} (\bibinfo {year} {2000}{\natexlab{a}})}\BibitemShut {NoStop}%
\bibitem [{\citenamefont {Denef}(2000{\natexlab{b}})}]{denef2000correspondence}%
  \BibitemOpen
  \bibfield  {author} {\bibinfo {author} {\bibfnamefont {F.}~\bibnamefont {Denef}},\ }\href@noop {} {\bibinfo {title} {On the correspondence between d-branes and stationary supergravity solutions of type ii calabi-yau compactifications}} (\bibinfo {year} {2000}{\natexlab{b}}),\ \Eprint {https://arxiv.org/abs/hep-th/0010222} {arXiv:hep-th/0010222 [hep-th]} \BibitemShut {NoStop}%
\bibitem [{\citenamefont {Bates}\ and\ \citenamefont {Denef}(2011)}]{Bates_2011}%
  \BibitemOpen
  \bibfield  {author} {\bibinfo {author} {\bibfnamefont {B.}~\bibnamefont {Bates}}\ and\ \bibinfo {author} {\bibfnamefont {F.}~\bibnamefont {Denef}},\ }\bibfield  {title} {\bibinfo {title} {Exact solutions for supersymmetric stationary black hole composites},\ }\bibfield  {journal} {\bibinfo  {journal} {Journal of High Energy Physics}\ }\textbf {\bibinfo {volume} {2011}},\ \href {https://doi.org/10.1007/jhep11(2011)127} {10.1007/jhep11(2011)127} (\bibinfo {year} {2011})\BibitemShut {NoStop}%
\bibitem [{\citenamefont {Manton}(1982)}]{Manton:1981mp}%
  \BibitemOpen
  \bibfield  {author} {\bibinfo {author} {\bibfnamefont {N.~S.}\ \bibnamefont {Manton}},\ }\bibfield  {title} {\bibinfo {title} {{A Remark on the Scattering of BPS Monopoles}},\ }\href {https://doi.org/10.1016/0370-2693(82)90950-9} {\bibfield  {journal} {\bibinfo  {journal} {Phys. Lett. B}\ }\textbf {\bibinfo {volume} {110}},\ \bibinfo {pages} {54} (\bibinfo {year} {1982})}\BibitemShut {NoStop}%
\bibitem [{\citenamefont {Gibbons}\ and\ \citenamefont {Ruback}(1986)}]{PhysRevLett.57.1492}%
  \BibitemOpen
  \bibfield  {author} {\bibinfo {author} {\bibfnamefont {G.~W.}\ \bibnamefont {Gibbons}}\ and\ \bibinfo {author} {\bibfnamefont {P.~J.}\ \bibnamefont {Ruback}},\ }\bibfield  {title} {\bibinfo {title} {Motion of extreme reissner-nordstrom black holes in the low-velocity limit},\ }\href {https://doi.org/10.1103/PhysRevLett.57.1492} {\bibfield  {journal} {\bibinfo  {journal} {Phys. Rev. Lett.}\ }\textbf {\bibinfo {volume} {57}},\ \bibinfo {pages} {1492} (\bibinfo {year} {1986})}\BibitemShut {NoStop}%
\bibitem [{\citenamefont {Ferrell}\ and\ \citenamefont {Eardley}(1987)}]{Ferrell:1987gf}%
  \BibitemOpen
  \bibfield  {author} {\bibinfo {author} {\bibfnamefont {R.~C.}\ \bibnamefont {Ferrell}}\ and\ \bibinfo {author} {\bibfnamefont {D.~M.}\ \bibnamefont {Eardley}},\ }\bibfield  {title} {\bibinfo {title} {{Slow motion scattering and coalescence of maximally charged black holes}},\ }\href {https://doi.org/10.1103/PhysRevLett.59.1617} {\bibfield  {journal} {\bibinfo  {journal} {Phys. Rev. Lett.}\ }\textbf {\bibinfo {volume} {59}},\ \bibinfo {pages} {1617} (\bibinfo {year} {1987})}\BibitemShut {NoStop}%
\bibitem [{\citenamefont {Camps}\ \emph {et~al.}(2017)\citenamefont {Camps}, \citenamefont {Hadar},\ and\ \citenamefont {Manton}}]{Camps:2017gxz}%
  \BibitemOpen
  \bibfield  {author} {\bibinfo {author} {\bibfnamefont {J.}~\bibnamefont {Camps}}, \bibinfo {author} {\bibfnamefont {S.}~\bibnamefont {Hadar}},\ and\ \bibinfo {author} {\bibfnamefont {N.~S.}\ \bibnamefont {Manton}},\ }\bibfield  {title} {\bibinfo {title} {{Exact Gravitational Wave Signatures from Colliding Extreme Black Holes}},\ }\href {https://doi.org/10.1103/PhysRevD.96.061501} {\bibfield  {journal} {\bibinfo  {journal} {Phys. Rev. D}\ }\textbf {\bibinfo {volume} {96}},\ \bibinfo {pages} {061501} (\bibinfo {year} {2017})},\ \Eprint {https://arxiv.org/abs/1704.08520} {arXiv:1704.08520 [gr-qc]} \BibitemShut {NoStop}%
\bibitem [{\citenamefont {Damour}(2020)}]{Damour:2019lcq}%
  \BibitemOpen
  \bibfield  {author} {\bibinfo {author} {\bibfnamefont {T.}~\bibnamefont {Damour}},\ }\bibfield  {title} {\bibinfo {title} {{Classical and quantum scattering in post-Minkowskian gravity}},\ }\href {https://doi.org/10.1103/PhysRevD.102.024060} {\bibfield  {journal} {\bibinfo  {journal} {Phys. Rev. D}\ }\textbf {\bibinfo {volume} {102}},\ \bibinfo {pages} {024060} (\bibinfo {year} {2020})},\ \Eprint {https://arxiv.org/abs/1912.02139} {arXiv:1912.02139 [gr-qc]} \BibitemShut {NoStop}%
\bibitem [{\citenamefont {Chandrasekhar}(1998)}]{ChandrasekharBook}%
  \BibitemOpen
  \bibfield  {author} {\bibinfo {author} {\bibfnamefont {S.}~\bibnamefont {Chandrasekhar}},\ }\href@noop {} {\emph {\bibinfo {title} {The mathematical theory of black holes}}},\ Vol.~\bibinfo {volume} {69}\ (\bibinfo  {publisher} {Oxford university press},\ \bibinfo {year} {1998})\BibitemShut {NoStop}%
\bibitem [{\citenamefont {Duff}(1973)}]{duff}%
  \BibitemOpen
  \bibfield  {author} {\bibinfo {author} {\bibfnamefont {M.~J.}\ \bibnamefont {Duff}},\ }\bibfield  {title} {\bibinfo {title} {{Quantum Tree Graphs and the Schwarzschild Solution}},\ }\href {https://doi.org/10.1103/PhysRevD.7.2317} {\bibfield  {journal} {\bibinfo  {journal} {Phys. Rev. D}\ }\textbf {\bibinfo {volume} {7}},\ \bibinfo {pages} {2317} (\bibinfo {year} {1973})}\BibitemShut {NoStop}%
\bibitem [{\citenamefont {Mougiakakos}\ and\ \citenamefont {Vanhove}(2021)}]{DdimSchw}%
  \BibitemOpen
  \bibfield  {author} {\bibinfo {author} {\bibfnamefont {S.}~\bibnamefont {Mougiakakos}}\ and\ \bibinfo {author} {\bibfnamefont {P.}~\bibnamefont {Vanhove}},\ }\bibfield  {title} {\bibinfo {title} {{Schwarzschild-Tangherlini metric from scattering amplitudes in various dimensions}},\ }\href {https://doi.org/10.1103/PhysRevD.103.026001} {\bibfield  {journal} {\bibinfo  {journal} {Phys. Rev. D}\ }\textbf {\bibinfo {volume} {103}},\ \bibinfo {pages} {026001} (\bibinfo {year} {2021})},\ \Eprint {https://arxiv.org/abs/2010.08882} {arXiv:2010.08882 [hep-th]} \BibitemShut {NoStop}%
\bibitem [{\citenamefont {D'Onofrio}\ \emph {et~al.}(2022)\citenamefont {D'Onofrio}, \citenamefont {Fragomeno}, \citenamefont {Gambino},\ and\ \citenamefont {Riccioni}}]{DOnofrio:2022cvn}%
  \BibitemOpen
  \bibfield  {author} {\bibinfo {author} {\bibfnamefont {S.}~\bibnamefont {D'Onofrio}}, \bibinfo {author} {\bibfnamefont {F.}~\bibnamefont {Fragomeno}}, \bibinfo {author} {\bibfnamefont {C.}~\bibnamefont {Gambino}},\ and\ \bibinfo {author} {\bibfnamefont {F.}~\bibnamefont {Riccioni}},\ }\bibfield  {title} {\bibinfo {title} {{The Reissner-Nordstr\"om-Tangherlini solution from scattering amplitudes of charged scalars}},\ }\href {https://doi.org/10.1007/JHEP09(2022)013} {\bibfield  {journal} {\bibinfo  {journal} {JHEP}\ }\textbf {\bibinfo {volume} {09}},\ \bibinfo {pages} {013}},\ \Eprint {https://arxiv.org/abs/2207.05841} {arXiv:2207.05841 [hep-th]} \BibitemShut {NoStop}%
\bibitem [{\citenamefont {Damour}\ and\ \citenamefont {Schaefer}(1988)}]{Damour:1988mr}%
  \BibitemOpen
  \bibfield  {author} {\bibinfo {author} {\bibfnamefont {T.}~\bibnamefont {Damour}}\ and\ \bibinfo {author} {\bibfnamefont {G.}~\bibnamefont {Schaefer}},\ }\bibfield  {title} {\bibinfo {title} {{Higher Order Relativistic Periastron Advances and Binary Pulsars}},\ }\href {https://doi.org/10.1007/BF02828697} {\bibfield  {journal} {\bibinfo  {journal} {Nuovo Cim. B}\ }\textbf {\bibinfo {volume} {101}},\ \bibinfo {pages} {127} (\bibinfo {year} {1988})}\BibitemShut {NoStop}%
\bibitem [{\citenamefont {Goldberger}\ and\ \citenamefont {Rothstein}(2006{\natexlab{b}})}]{PhysRevD.73.104030}%
  \BibitemOpen
  \bibfield  {author} {\bibinfo {author} {\bibfnamefont {W.~D.}\ \bibnamefont {Goldberger}}\ and\ \bibinfo {author} {\bibfnamefont {I.~Z.}\ \bibnamefont {Rothstein}},\ }\bibfield  {title} {\bibinfo {title} {Dissipative effects in the worldline approach to black hole dynamics},\ }\href {https://doi.org/10.1103/PhysRevD.73.104030} {\bibfield  {journal} {\bibinfo  {journal} {Phys. Rev. D}\ }\textbf {\bibinfo {volume} {73}},\ \bibinfo {pages} {104030} (\bibinfo {year} {2006}{\natexlab{b}})}\BibitemShut {NoStop}%
\bibitem [{\citenamefont {Laporta}(2000)}]{Laporta:2000dsw}%
  \BibitemOpen
  \bibfield  {author} {\bibinfo {author} {\bibfnamefont {S.}~\bibnamefont {Laporta}},\ }\bibfield  {title} {\bibinfo {title} {{High precision calculation of multiloop Feynman integrals by difference equations}},\ }\href {https://doi.org/10.1142/S0217751X00002159} {\bibfield  {journal} {\bibinfo  {journal} {Int. J. Mod. Phys. A}\ }\textbf {\bibinfo {volume} {15}},\ \bibinfo {pages} {5087} (\bibinfo {year} {2000})},\ \Eprint {https://arxiv.org/abs/hep-ph/0102033} {arXiv:hep-ph/0102033} \BibitemShut {NoStop}%
\bibitem [{\citenamefont {Lee}(2014)}]{Lee:2013mka}%
  \BibitemOpen
  \bibfield  {author} {\bibinfo {author} {\bibfnamefont {R.~N.}\ \bibnamefont {Lee}},\ }\bibfield  {title} {\bibinfo {title} {{LiteRed 1.4: a powerful tool for reduction of multiloop integrals}},\ }\href {https://doi.org/10.1088/1742-6596/523/1/012059} {\bibfield  {journal} {\bibinfo  {journal} {J. Phys. Conf. Ser.}\ }\textbf {\bibinfo {volume} {523}},\ \bibinfo {pages} {012059} (\bibinfo {year} {2014})},\ \Eprint {https://arxiv.org/abs/1310.1145} {arXiv:1310.1145 [hep-ph]} \BibitemShut {NoStop}%
\bibitem [{\citenamefont {Smirnov}\ and\ \citenamefont {Chuharev}(2020)}]{Smirnov:2019qkx}%
  \BibitemOpen
  \bibfield  {author} {\bibinfo {author} {\bibfnamefont {A.~V.}\ \bibnamefont {Smirnov}}\ and\ \bibinfo {author} {\bibfnamefont {F.~S.}\ \bibnamefont {Chuharev}},\ }\bibfield  {title} {\bibinfo {title} {{FIRE6: Feynman Integral REduction with Modular Arithmetic}},\ }\href {https://doi.org/10.1016/j.cpc.2019.106877} {\bibfield  {journal} {\bibinfo  {journal} {Comput. Phys. Commun.}\ }\textbf {\bibinfo {volume} {247}},\ \bibinfo {pages} {106877} (\bibinfo {year} {2020})},\ \Eprint {https://arxiv.org/abs/1901.07808} {arXiv:1901.07808 [hep-ph]} \BibitemShut {NoStop}%
\bibitem [{\citenamefont {Dlapa}\ \emph {et~al.}(2023{\natexlab{c}})\citenamefont {Dlapa}, \citenamefont {K\"alin}, \citenamefont {Liu},\ and\ \citenamefont {Porto}}]{Dlapa:2023hsl}%
  \BibitemOpen
  \bibfield  {author} {\bibinfo {author} {\bibfnamefont {C.}~\bibnamefont {Dlapa}}, \bibinfo {author} {\bibfnamefont {G.}~\bibnamefont {K\"alin}}, \bibinfo {author} {\bibfnamefont {Z.}~\bibnamefont {Liu}},\ and\ \bibinfo {author} {\bibfnamefont {R.~A.}\ \bibnamefont {Porto}},\ }\bibfield  {title} {\bibinfo {title} {{Bootstrapping the relativistic two-body problem}},\ }\href {https://doi.org/10.1007/JHEP08(2023)109} {\bibfield  {journal} {\bibinfo  {journal} {JHEP}\ }\textbf {\bibinfo {volume} {08}},\ \bibinfo {pages} {109}},\ \Eprint {https://arxiv.org/abs/2304.01275} {arXiv:2304.01275 [hep-th]} \BibitemShut {NoStop}%
\bibitem [{\citenamefont {Arkani-Hamed}\ \emph {et~al.}(2010)\citenamefont {Arkani-Hamed}, \citenamefont {Cachazo},\ and\ \citenamefont {Kaplan}}]{Arkani-Hamed:2008owk}%
  \BibitemOpen
  \bibfield  {author} {\bibinfo {author} {\bibfnamefont {N.}~\bibnamefont {Arkani-Hamed}}, \bibinfo {author} {\bibfnamefont {F.}~\bibnamefont {Cachazo}},\ and\ \bibinfo {author} {\bibfnamefont {J.}~\bibnamefont {Kaplan}},\ }\bibfield  {title} {\bibinfo {title} {{What is the Simplest Quantum Field Theory?}},\ }\href {https://doi.org/10.1007/JHEP09(2010)016} {\bibfield  {journal} {\bibinfo  {journal} {JHEP}\ }\textbf {\bibinfo {volume} {09}},\ \bibinfo {pages} {016}},\ \Eprint {https://arxiv.org/abs/0808.1446} {arXiv:0808.1446 [hep-th]} \BibitemShut {NoStop}%
\bibitem [{\citenamefont {Bini}\ and\ \citenamefont {Damour}(2017)}]{Bini:2017wfr}%
  \BibitemOpen
  \bibfield  {author} {\bibinfo {author} {\bibfnamefont {D.}~\bibnamefont {Bini}}\ and\ \bibinfo {author} {\bibfnamefont {T.}~\bibnamefont {Damour}},\ }\bibfield  {title} {\bibinfo {title} {{Gravitational scattering of two black holes at the fourth post-Newtonian approximation}},\ }\href {https://doi.org/10.1103/PhysRevD.96.064021} {\bibfield  {journal} {\bibinfo  {journal} {Phys. Rev. D}\ }\textbf {\bibinfo {volume} {96}},\ \bibinfo {pages} {064021} (\bibinfo {year} {2017})},\ \Eprint {https://arxiv.org/abs/1706.06877} {arXiv:1706.06877 [gr-qc]} \BibitemShut {NoStop}%
\bibitem [{\citenamefont {Cho}\ \emph {et~al.}(2022)\citenamefont {Cho}, \citenamefont {K\"alin},\ and\ \citenamefont {Porto}}]{Cho:2021arx}%
  \BibitemOpen
  \bibfield  {author} {\bibinfo {author} {\bibfnamefont {G.}~\bibnamefont {Cho}}, \bibinfo {author} {\bibfnamefont {G.}~\bibnamefont {K\"alin}},\ and\ \bibinfo {author} {\bibfnamefont {R.~A.}\ \bibnamefont {Porto}},\ }\bibfield  {title} {\bibinfo {title} {{From boundary data to bound states. Part III. Radiative effects}},\ }\href {https://doi.org/10.1007/JHEP04(2022)154} {\bibfield  {journal} {\bibinfo  {journal} {JHEP}\ }\textbf {\bibinfo {volume} {04}},\ \bibinfo {pages} {154}},\ \bibinfo {note} {[Erratum: JHEP 07, 002 (2022)]},\ \Eprint {https://arxiv.org/abs/2112.03976} {arXiv:2112.03976 [hep-th]} \BibitemShut {NoStop}%
\bibitem [{\citenamefont {Galley}\ \emph {et~al.}(2016)\citenamefont {Galley}, \citenamefont {Leibovich}, \citenamefont {Porto},\ and\ \citenamefont {Ross}}]{Galley:2015kus}%
  \BibitemOpen
  \bibfield  {author} {\bibinfo {author} {\bibfnamefont {C.~R.}\ \bibnamefont {Galley}}, \bibinfo {author} {\bibfnamefont {A.~K.}\ \bibnamefont {Leibovich}}, \bibinfo {author} {\bibfnamefont {R.~A.}\ \bibnamefont {Porto}},\ and\ \bibinfo {author} {\bibfnamefont {A.}~\bibnamefont {Ross}},\ }\bibfield  {title} {\bibinfo {title} {{Tail effect in gravitational radiation reaction: Time nonlocality and renormalization group evolution}},\ }\href {https://doi.org/10.1103/PhysRevD.93.124010} {\bibfield  {journal} {\bibinfo  {journal} {Phys. Rev. D}\ }\textbf {\bibinfo {volume} {93}},\ \bibinfo {pages} {124010} (\bibinfo {year} {2016})},\ \Eprint {https://arxiv.org/abs/1511.07379} {arXiv:1511.07379 [gr-qc]} \BibitemShut {NoStop}%
\end{thebibliography}%

\end{document}